\documentclass[twocolumn]{aastex62}
\usepackage{lineno}

\usepackage{rotating}
\usepackage{natbib}
\usepackage{makecell}
\usepackage{amsmath}
\usepackage{subfigure}
\usepackage{tabularx}
\setcitestyle{sort,comma}
\usepackage{afterpage}
\usepackage{csvsimple}

\usepackage{unicode-math}

\usepackage[none]{hyphenat}

\graphicspath{{./}{figures/}}

\shorttitle{Uniform Reinterpretation of Rocky Exoplanet Secondary Eclipse Observations}

\begin{document}



\title{\large{Uniform Reinterpretation of Rocky Exoplanet Secondary Eclipse Observations and the Impact of Stellar and Orbital Uncertainties}}

\correspondingauthor{Christopher Monaghan}
\email{christopherm@ucla.edu}

\author[0009-0005-9152-9480]{Christopher Monaghan} 
\affil{Department of Earth, Planetary, and Space Sciences, University of California, Los Angeles, Los Angeles, CA, USA}
\affil{Department of Physics and Trottier Institute for Research on Exoplanets, Universit\'{e} de Montr\'{e}al, Montreal, QC, Canada}

\author[0000-0001-5578-1498]{Bj\"{o}rn Benneke}
\affil{Department of Earth, Planetary, and Space Sciences, University of California, Los Angeles, Los Angeles, CA, USA}
\affil{Department of Physics and Trottier Institute for Research on Exoplanets, Universit\'{e} de Montr\'{e}al, Montreal, QC, Canada}

\author[0000-0001-5848-6750]{Nicholas J. Connors} 
\affil{Department of Physics and Trottier Institute for Research on Exoplanets, Universit\'{e} de Montr\'{e}al, Montreal, QC, Canada}

\author[0000-0002-2195-735X]{Louis-Philippe Coulombe} 
\affil{Department of Physics and Trottier Institute for Research on Exoplanets, Universit\'{e} de Montr\'{e}al, Montreal, QC, Canada}
\affil{Plan\'{e}tarium de Montr\'{e}al, Espace pour la Vie, Montr\'{e}al, QC, Canada}

\author[0000-0001-6809-3520]{Pierre-Alexis Roy} 
\affil{Department of Earth, Planetary, and Space Sciences, University of California, Los Angeles, Los Angeles, CA, USA}
\affil{Department of Physics and Trottier Institute for Research on Exoplanets, Universit\'{e} de Montr\'{e}al, Montreal, QC, Canada}

\begin{abstract}


Secondary eclipse observations are a powerful way to investigate whether or not a rocky exoplanet hosts an atmosphere, as an atmospheric presence would transport heat to the nightside and render the dayside colder than anticipated. The interpretation of the secondary eclipse observations relies, however, on models based on imperfect knowledge of the host star properties and the system parameters. Any uncertainties in such astrophysical variables will propagate into both atmospheric and bare-rock models, potentially leading to poorly constrained results and erroneous conclusions. In this work, we introduce a framework to efficiently account for the stellar and orbital uncertainties when modeling the emission spectra of rocky exoplanets, and demonstrate its use by reanalyzing the current suite of rocky exoplanets with published eclipse observations. Our analysis reveals notable uncertainty in the predicted eclipse depth even for a simple dark ($A_{\mathrm{B}}$ = 0) bare rock as a result of the finite precision of the system’s parameters and treatment of the host star’s flux. In some cases, the model uncertainty is comparable to the observational uncertainty, further complicating our capability to constrain an atmospheric presence from secondary-eclipse observations. From our modeling schematic, we derive a linear correlation between the model uncertainty and the error in $R_{\mathrm{p}}/R_{\mathrm{*}}$, $ a_{\mathrm{p}}/R_{\mathrm{*}}$, and $T_{\mathrm{*}}$, therefore enabling a more robust compositional analysis in future studies. The model uncertainty serves as a fundamental precision limit to surface analyses, and must be mitigated to strongly constrain the composition of exoplanets in future eclipse observations. 

\end{abstract}



\section{Introduction} \label{sec:intro}

\subsection{Characterization of Terrestrial Exoplanets with Emission Spectroscopy}\label{sec:emissionspectroscopy}

A critical area of research in the modern era of astrophysics is the search for atmospheres on rocky exoplanets. The recent success of the \textit{James Webb Space Telescope} (JWST) has further accelerated the search for terrestrial exoplanets with atmospheric signatures, particularly in the context of potential habitability \citep{kreidberg2025lookrockyexoplanetsjwst}. Transmission spectro- scopy has long been the primary method of atmospheric characterization for both terrestrial planets and gas giants \citep{Kreidberg_2018, Deming_Louie_Sheets_2018}. However, for smaller worlds with weaker signals, the Transit Light Source (TLS) effect may inject contaminants into the transit data that dominate the observed spectra, potentially leading to false detections and systematic errors \citep[e.g.,][]{Rackham_Apai_Giampapa_2018, Rackham_Apai_Giampapa_2019, Rackham_Espinoza_Berdyugina_Korhonen_MacDonald_Montet_Morris_Oshagh_Shapiro_Unruh_etal._2023, Rackham_deWit_2024}. 


\begin{table*}[]
    \scriptsize
    \centering
    {
    \begin{tabular}{>{\bfseries}l|c c | c  c c |c}
    \hline
    \hline
    
        \textbf{Planet} & $R_{\mathrm{p}}/R_{\mathrm{*}}$ & $a_{\mathrm{p}}/R_{\mathrm{*}}$ &  $T_{\mathrm{*}}$ & log(g)$_{\mathrm{*}}$ & [M/H]$_{\mathrm{*}}$ & $T_{\mathrm{max}}$ \\ 
         &  &  & K & log$_{10}$(cm/s$^{2}$) & dex & K \\
    
    \hline
    \hline
TOI-561\,b    & 0.0152 $\pm$ 0.0003 & 2.69 $\pm$ 0.02  & 5372 $\pm$ 70  & 4.50 $\pm$ 0.12 & -0.40 $\pm$ 0.12 & 2947 $\pm$ 45 \\
K2-141\,b     & 0.0203 $\pm$ 0.0009 & 2.36 $\pm$ 0.07  & 4570 $\pm$ 100 & 4.62 $\pm$ 0.03 & 0.00 $\pm$ 0.06  & 2689 $\pm$ 71 \\
55 Cnc\,e     & 0.0182 $\pm$ 0.0002 & 3.52 $\pm$ 0.01  & 5214 $\pm$ 53  & 4.38 $\pm$ 0.05 & 0.37 $\pm$ 0.02  & 2560 $\pm$ 37 \\
TOI-431\,b    & 0.0160 $\pm$ 0.0007 & 3.32 $\pm$ 0.13  & 4850 $\pm$ 75  & 4.60 $\pm$ 0.06 & 0.20 $\pm$ 0.05  & 2404 $\pm$ 56 \\
GJ 367\,b     & 0.0140 $\pm$ 0.0003 & 3.33 $\pm$ 0.16  & 3522 $\pm$ 70  & 4.78 $\pm$ 0.03 & -0.01 $\pm$ 0.12 & 1744 $\pm$ 54 \\
TOI-1685\,b   & 0.0296 $\pm$ 0.0006 & 5.37 $\pm$ 0.22  & 3575 $\pm$ 53  & 4.78 $\pm$ 0.03 & 0.30 $\pm$ 0.10  & 1394 $\pm$ 35 \\
GJ 1252\,b    & 0.0277 $\pm$ 0.0011 & 5.03 $\pm$ 0.27  & 3458 $\pm$ 140 & 4.83 $\pm$ 0.05 & 0.10 $\pm$ 0.10  & 1392 $\pm$ 67 \\
LTT 3780\,b   & 0.0321 $\pm$ 0.0007 & 6.79 $\pm$ 0.29  & 3358 $\pm$ 92  & 4.85 $\pm$ 0.11 & 0.06 $\pm$ 0.11  & 1167 $\pm$ 39 \\
LHS 3844\,b   & 0.0635 $\pm$ 0.0009 & 7.11 $\pm$ 0.03  & 3036 $\pm$ 77  & 5.06 $\pm$ 0.01 & - & 1031 $\pm$ 33 \\
GJ 486\,b     & 0.0372 $\pm$ 0.0001 & 11.38 $\pm$ 0.15 & 3317 $\pm$ 37  & 4.91 $\pm$ 0.01 & -0.15 $\pm$ 0.13 & 889 $\pm$ 12  \\
TOI-1468\,b   & 0.0346 $\pm$ 0.0010 & 12.20 $\pm$ 0.35 & 3376 $\pm$ 45  & 4.85 $\pm$ 0.03 & -0.01 $\pm$ 0.22 & 873 $\pm$ 16  \\
LHS 1478\,b   & 0.0439 $\pm$ 0.0015 & 16.35 $\pm$ 0.55 & 3415 $\pm$ 82  & 5.02 $\pm$ 0.03 & -0.38 $\pm$ 0.29 & 763 $\pm$ 22  \\
GJ 1132\,b    & 0.0494 $\pm$ 0.0001 & 15.26 $\pm$ 0.59 & 3229 $\pm$ 78  & 5.04 $\pm$ 0.03 & -0.17 $\pm$ 0.15 & 746 $\pm$ 22  \\
GJ 3929\,b    & 0.0318 $\pm$ 0.0007 & 17.05 $\pm$ 0.43 & 3384 $\pm$ 88  & 4.89 $\pm$ 0.05 & -0.02 $\pm$ 0.12 & 743 $\pm$ 23  \\
LTT 1445 A\,b & 0.0454 $\pm$ 0.0012 & 30.23 $\pm$ 2.35 & 3340 $\pm$ 150 & 4.98 $\pm$ 0.07 & -0.34 $\pm$ 0.09 & 549 $\pm$ 31  \\
LHS 1140\,c   & 0.0528 $\pm$ 0.0013 & 26.88 $\pm$ 0.62 & 3096 $\pm$ 48  & 5.04 $\pm$ 0.02 & -0.15 $\pm$ 0.09 & 539 $\pm$ 10  \\
TRAPPIST-1\,b & 0.0859 $\pm$ 0.0004 & 20.84 $\pm$ 0.16 & 2566 $\pm$ 26  & 5.24 $\pm$ 0.01 & 0.05 $\pm$ 0.09  & 508 $\pm$ 6   \\
TRAPPIST-1\,c & 0.0844 $\pm$ 0.0004 & 28.55 $\pm$ 0.21 & 2566 $\pm$ 26  & 5.24 $\pm$ 0.01 & 0.05 $\pm$ 0.09  & 434 $\pm$ 5   \\

\hline
\hline
    \end{tabular} }
    \caption{Properties and associated error bars of rocky exoplanets ($R_{\mathrm{p}}$
    $\leq$ 2R$_{\oplus}$) with emission-based observations, sorted by the planet's maximum dayside temperature. $T_{\mathrm{max}}$ is derived for each planet using Equation \ref{eq:tmax}. Astrophysical parameters of each planet are set to match those used in the most recent emission-based study, and are sourced from the following for each target: TOI-561\,b \citep{teske2025thickvolatileatmosphereultrahot}, K2-141\,b \citep{Bonomo_Dumusque_Massa_Mortier_Bongiolatti_Malavolta_Sozzetti_Buchhave_Damasso_Haywood_etal._2023}, 55 Cnc\,e \citep{Hu_Bello-Arufe_Zhang_Paragas_Zilinskas_Van_Buchem_Bess_Patel_Ito_Damiano_et_al._2024}, TOI-431\,b \citep{Osborn_2021}, GJ 367\,b \citep{GoffoGJ367}, TOI-1685\,b \citep{2024ApJ...971L..12B}, GJ 1252\,b \citep{Crossfield_Malik_Hill_Kane_Foley_Polanski_Coria_Brande_Zhang_Wienke_et_al._2022}, LTT 3780\,b \citep{2024A&A...682A..66B}, LHS 3844\,b \citep{Vanderspek_2019}, GJ 486\,b \citep{Mansfield_Xue_Zhang_Mahajan_Ih_Koll_Bean_Coy_Eastman_Kempton_et_al._2024}, TOI-1468\,b \citep{Meier_Vald_s_2025} LHS 1478\,b \citep{august2024hotrockssurveyi}, GJ 1132\,b \citep{Xue_Bean_Zhang_Mahajan_Ih_Eastman_Lunine_Mansfield_Coy_Kempton_et_al._2024}, GJ 3929\,b \citep{Xue_Zhang_Coy_Brady_Ji_Bean_Radica_Seifahrt_Sturmer_Luque_etal._2025}, LTT 1445 A\,b \citep{2023AJ....166..171P}, LHS 1140\,c \citep{2024ApJ...960L...3C} TRAPPIST-1\,b \citep{agol2021refiningtransittimingphotometric}, \& TRAPPIST-1\,c \citep{agol2021refiningtransittimingphotometric}.}
    \label{tab:planets_in_study}
\end{table*}

Thermal emission measurements of a planet's surface flux during a secondary eclipse have emerged as an efficient method for detecting atmospheres on tidally locked rocky worlds \citep{2018haex.bookE..40A, 2009ApJ...703.1884S, Mansfield_Kite_Hu_Koll_Malik_Bean_Kempton_2019, Koll_2019}. Both photometric and spectroscopic observations of a planet's occultation can be used to constrain a planet's atmospheric presence and composition. Secondary eclipse observations directly measure the planet's occultation depth, which encodes the surface flux of both the host star and the planet ($f_{\mathrm{*}}$ and $f_{\mathrm{p}}$, respectively). Across the bolometric wavelength space, the eclipse depth can be modeled as:

\begin{equation}
    (F_{\mathrm{p}}/F_{\mathrm{*}})(\lambda) = \left(\frac{R_{\mathrm{p}}}{R_{\mathrm{*}}}\right)^{2} \frac{f_{\mathrm{p}}(\lambda)}{f_{\mathrm{*}}(\lambda)}
    \label{eq:eclipsedepth}
\end{equation}

Where $R_{\mathrm{*}}$ and $R_{\mathrm{p}}$ are the radii of the star and planet, respectively. The eclipse depth can be used to infer the planet's dayside brightness temperature $T_{\mathrm{d}}$, which represents the temperature a blackbody must be in order to reproduce the observed intensity of light. The brightness temperature can be calculated from the planet's eclipse depth by assuming it radiates as a blackbody, and finding the value of $T_{\mathrm{d}}$ that reproduces the measured eclipse depth:

\begin{equation}
    f_{\mathrm{p}}(\lambda) = \pi B_{\lambda}(\lambda, T_{\mathrm{d}}) = \pi\frac{2hc^{2}}{\lambda^{5}}\frac{1}{e^{\frac{hc}{\lambda k_{B}T_{\mathrm{d}}}}-1}
    \label{eq:blackbody_flux}
\end{equation}

A tidally locked world with a thick atmosphere that redistributes heat to the nightside hemisphere will exhibit a colder dayside than a bare planet. Thus, observing a shallow eclipse depth consistent with a cooler dayside may therefore be indicative of atmospheric recirculation. The presence of an atmosphere is most often investigated by comparing our observations to those expected of a 'dark bare rock' that absorbs and reradiates the incident stellar flux. One notable parameter used to quantify the planet's consistency to a dark bare rock is that of the brightness temperature ratio $\mathcal{R}$:

\begin{equation}
    \mathcal{R} = T_{\mathrm{d}}/T_{\mathrm{d,max}}
\end{equation}

where $T_{\mathrm{d,max}}$ represents the maximum dayside bright- ness temperature consistent with the emissions of a dark bare rock model. The temperature of such models is often calculated following::

\begin{equation}
    T_{\mathrm{d}} = T_{\mathrm{*}} \sqrt{\frac{R_{\mathrm{*}}}{a_{\mathrm{p}}}}f^{\frac{1}{4}}(1-A_{\mathrm{B}})^{\frac{1}{4}}
    \label{eq:tmax}
\end{equation}

where $T_{\mathrm{*}}$ is the effective stellar temperature, $a_{\mathrm{p}}$ is the planet's semi-major axis, $f$ the heat redistribution factor and $A_{\mathrm{B}}$ the planet's bond albedo, which represents the total fraction of energy reflected by the planet's surface \citep{48d96982-3471-3bab-9b6f-f8b21bafd6fb}. For a dark bare rock, we generally assume that $f = \frac{2}{3}$ and $A_{\mathrm{B}}=0$ \citep{2008ApJS..179..484H, 48d96982-3471-3bab-9b6f-f8b21bafd6fb}.

An overarching goal of rocky exoplanet emission observations is to determine whether or not a planet may have an atmosphere by comparing the observed eclipse depth and corresponding brightness temperature to a suite of eclipse depth models for a variety of surface and atmospheric compositions. Planets observed with $\mathcal{R} \approx 1$ are assumed to be consistent with a bare rock without a significant atmospheric presence, while planets with $\mathcal{R} < 1$ may host either redistributive atmospheres or highly reflective surfaces. For the highly-irradiated worlds most amenable to emission-based studies, however, the albedos of bare rock planets are thought to be lowered significantly by space weathering from both stellar winds and micrometeor impacts \citep{2015aste.book..597B, 2019A&A...627A..43D,   lyu2024superearthlhs3844btidallylocked, Coy_Ih_Kite_Koll_Tenthoff_Bean_WeinerMansfield_Zhang_Xue_Kempton_etal._2025}. Thus, particularly low values of $\mathcal{R}$ are most often attributed to the presence of an atmosphere.

\subsection{Existing \& Future Eclipse Observations}\label{sec:eclobservations}

There are currently 18 rocky exoplanets ($R_{\mathrm{p}} \leq$ 2R$_{\oplus}$) with published secondary eclipse observations (Table \ref{tab:planets_in_study}). Although many of these studies have observed eclipse depths consistent with the null hypothesis of a dark bare rock, a number of promising atmosphere-hosting candidates have emerged. In particular, highly-irradiated lava worlds have been observed to host dayside tempera- tures far lower than anticipated, potentially indicating the presence of volatile-rich atmospheres fueled by volcanic outgassing \citep[e.g.,][]{Hu_Bello-Arufe_Zhang_Paragas_Zilinskas_Van_Buchem_Bess_Patel_Ito_Damiano_et_al._2024, Patel_Brandeker_Kitzmann_de_la_Roche_Bello-Arufe_Heng_Valdes_Persson_Zhang_Demory_et_al._2024, teske2025thickvolatileatmosphereultrahot, Monaghan_Roy_Benneke_Crossfield_Coulombe_Piaulet-Ghorayeb_Kreidberg_Dressing_Kane_Dragomir_etal._2025}. Additionally, a number of observations performed by the Mid InfraRed Instrument (MIRI) on the JWST may be consistent with both thin and inverted atmospheric models \citep[e.g.,][]{august2024hotrockssurveyi, Zieba_Kreidberg_Ducrot_Gillon_Morley_Schaefer_Tamburo_Koll_Lyu_Acuna_et_al._2023, Ducrot_Lagage_Min_Gillon_Bell_Tremblin_Greene_Dyrek_Bouwman_Waters_etal._2025}.

The variety of results from the current suite of rocky world eclipse studies has reinforced the need for further observational investment. A number of ongoing and upcoming JWST programs will further study the formation, composition, and survivability of rocky exoplanet atmospheres under different conditions through secondary eclipse and phase curve observations, the most notable of which being the 500-hour Rocky Worlds DDT \citep{Redfield_Batalha_Benneke_Biller_Espinoza_France_Konopacky_Kreidberg_Rauscher_Sing_2024}. This survey aims to investigate the presence of atmospheres on 9 key terrestrial exoplanets surrounding M dwarfs, in order to investigate the nature of the "cosmic shoreline", a driving hypothesis that aims to describe atmospheric retention on rocky worlds \citep{Zahnle_2017, Pass_Charbonneau_Vanderburg_2025, Ji_Chatterjee_Coy_Kite_2025, bertathompson20253dcosmicshorelinenurturing, menigallardo2025empiricaldeterminationcosmicshoreline}. Additional large-scale observations performed by the JWST to study the retention of atmospheres on rocky exoplanets includes the Hot Rocks Survey (GO 3730, PI: Diamond-Lowe), LAVALAMPS (GO 4818, PI: Mansfield), and GO 8864 (PI: Dang).

In light of these major time investments, it is important to ensure that our modeling schematic is sufficient enough to robustly characterize a planet's surface composition and atmospheric properties. While much attention has been dedicated to the mitigation of stellar contamination and uncertainty in transmission based studied, the uncertainties inherent in emission spectro- scopy have only recently been investigated \citep[e.g.,][]{Ducrot_Lagage_Min_Gillon_Bell_Tremblin_Greene_Dyrek_Bouwman_Waters_etal._2025, Coy_Ih_Kite_Koll_Tenthoff_Bean_WeinerMansfield_Zhang_Xue_Kempton_etal._2025, fauchez2025stellarmodelslimitexoplanet}. Deriving properties from the planet's observed eclipse depth requires the development of planetary emission models, which are generated using the astrophysical parameters observed in the extrasolar system alongside a stellar spectrum to approximate the host star's spectral energy distribution (SED). Any discrepancies between the true and assumed parameters will propagate as systematic errors in the generated emission models, leading to poor constraints on the planet's composition and, more worryingly, an underestimation of the results' uncertainties. Such discrepancies have undoubtedly been injected into the forward models of previously published results, and will impact the robustness of future observations if left uncorrected. Thus, it is critical to investigate how parameter uncertainty propagates into our forward modeling routines.

Here, we present a uniform reinterpretation of the existing suite of secondary eclipse observations performed on rocky exoplanets to further investigate sources of uncertainty injected by model inaccuracies and systematic imprecision. We begin by outlining various points of concern in existing schematics for modeling the emissions from rocky exoplanets in Section \ref{sec:problemswithmodels}. With this knowledge, we present our implemented modeling schematic in Section \ref{sec:methods}, which produces highly-precise bare rock models and enables our investigation of model uncertainty. We perform and analyze the results of our uniform reanalysis in Section \ref{sec:results}. Finally, in Section \ref{sec:Discussion}, we further discuss our findings in the context of spectroscopy, individual parameter dependence, and future eclipse observations.

\begin{figure*}[t!]
\centering
\includegraphics[width=\linewidth,trim={0 0 0 0cm},clip]{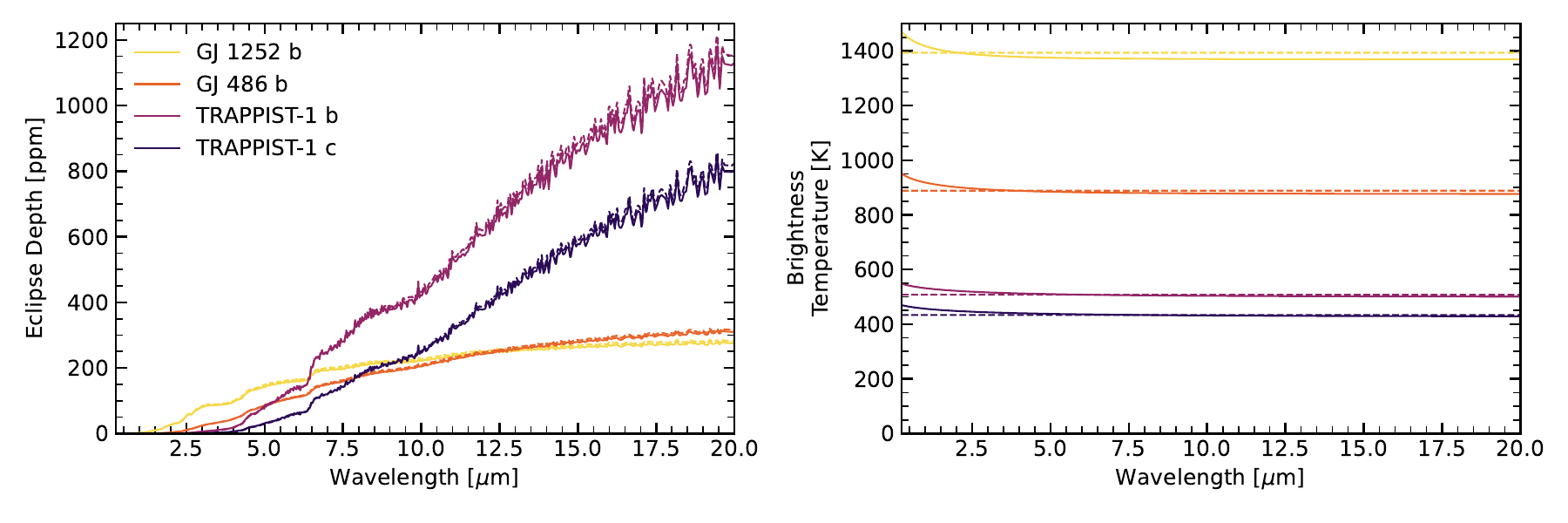}
\caption{Comparison between a uniform temperature blackbody and a dark bare rock with a dayside temperature gradient for a selection of rocky exoplanets with published thermal emission observations as modeled by our modeling framework \texttt{JESTER} (Section \ref{sec:jester}). The dashed lines represent the results for a uniform temperature blackbody. A linearly interpolated SPHINX stellar model is used for each planet's host star \citep{Iyer_Line_Muirhead_Fortney_Gharib-Nezhad_2023, Iyer_Line_Muirhead_Fortney_Gharib-Nezhad_2024}.}

\label{fig:uniformvsgradientmodelbarerock}
\end{figure*}

\section{Model Simplifications \& Inaccuracies}\label{sec:problemswithmodels}

Developing practical forward models relies on accurate knowledge of the planetary and stellar parameters in order to simulate the instrument response to various surface and atmospheric compositions. Results from such analyses, however, may be nullified if the produced models do not accurately reflect the true conditions on the planet. In this section, we present a number of factors in existing modeling schematics that may limit the accuracy of our forward models. We focus on the treatment of bare surfaces, although our conclusions can be easily extended to atmosphere models, which are also subject to a number of uncertainties that are difficult to quantify numerically \citep[e.g.,][]{Nixon_Welbanks_McGill_Kempton_2024}.

\subsection{Uniform Blackbody Models}\label{sec:uniformmodels}

The most simple form of forward model used for a dark bare rock is that of a uniform temperature blackbody that radiates at $T_{\mathrm{d,max}}$ (Equation \ref{eq:tmax}). The observed eclipse depth and corresponding brightness temperature may then be compared to the model results in order to determine if the planet is consistent with a bare rock composition. While these models serve as useful approximations, a bare rock world without heat recirculation should instead host a dayside temperature gradient that peaks at the substellar point before decreasing towards the terminator line of the planet, beyond which the planet's temperature should be consistent with 0K \citep[e.g.,][]{2008ApJS..179..484H, COWAN2011ApJ...726...82C}. Negating the impact of subsurface temperature variations \citep[e.g.,][]{lyu2025impactsubsurfacetemperaturegradients}, a temperature gradient should be present on all bare rocks regardless of the planet's mineralogical composition. For a dark bare rock, the assumption of a uniform dayside temperature results in an overestimation of the planet's mid-infrared eclipse depth (Figure \ref{fig:uniformvsgradientmodelbarerock}). Although this offset is often quite small, the error injected by the assumption of a one-dimensional dayside becomes far more difficult to predict when modeling bare rock surfaces with a nonzero albedo as a result of wavelength-dependent features in both $f_{\mathrm{*}}$ and $f_{\mathrm{p}}$ having various impacts at different temperatures.

A number of studies have begun implementing the use of non-uniform temperature models when simulating the emissions from atmosphereless planets. For example, \citet{Hammond_Guimond_Lichtenberg_Nicholls_Fisher_Luque_Meier_Taylor_Changeat_Dang_et_al._2024} and \citet{Zieba_Kreidberg_Ducrot_Gillon_Morley_Schaefer_Tamburo_Koll_Lyu_Acuna_et_al._2023} locally solve for the planet's energy balance by separating the dayside hemisphere into individual points on a grid, while \citet{Hu_Ehlmann_Seager_2012}, \citet{lyu2024superearthlhs3844btidallylocked}, \citet{First_Mishra_Gazel_Lewis_Letai_Hanssen_2025}, and \citet{Ducrot_Lagage_Min_Gillon_Bell_Tremblin_Greene_Dyrek_Bouwman_Waters_etal._2025} solve for the planet's energy balance by integrating across the planet's latitude and longitude. Alternatively, while the models presented by \citet{Paragas_Knutson_Hu_Ehlmann_Alemanno_Helbert_Maturilli_Zhang_Iyer_Rossman_2025} assume a uniform dayside temperature, they implement a one-dimensional correction factor to account for how the absorbed stellar energy is redistributed across the planet by scaling the correction to match the 2D models from \citet{Hu_Ehlmann_Seager_2012}. Although these reinterpretations of the dayside are far more realistic than the assumption of a uniform temperature, a consistent approach has not yet been applied to every planet listed in Table \ref{tab:planets_in_study}, and slight offsets between each model's interpretation have previously been noted \citep[e.g.,][]{lyu2024superearthlhs3844btidallylocked, Paragas_Knutson_Hu_Ehlmann_Alemanno_Helbert_Maturilli_Zhang_Iyer_Rossman_2025}.


\begin{figure*}[t!]
\centering
\includegraphics[width=\linewidth]{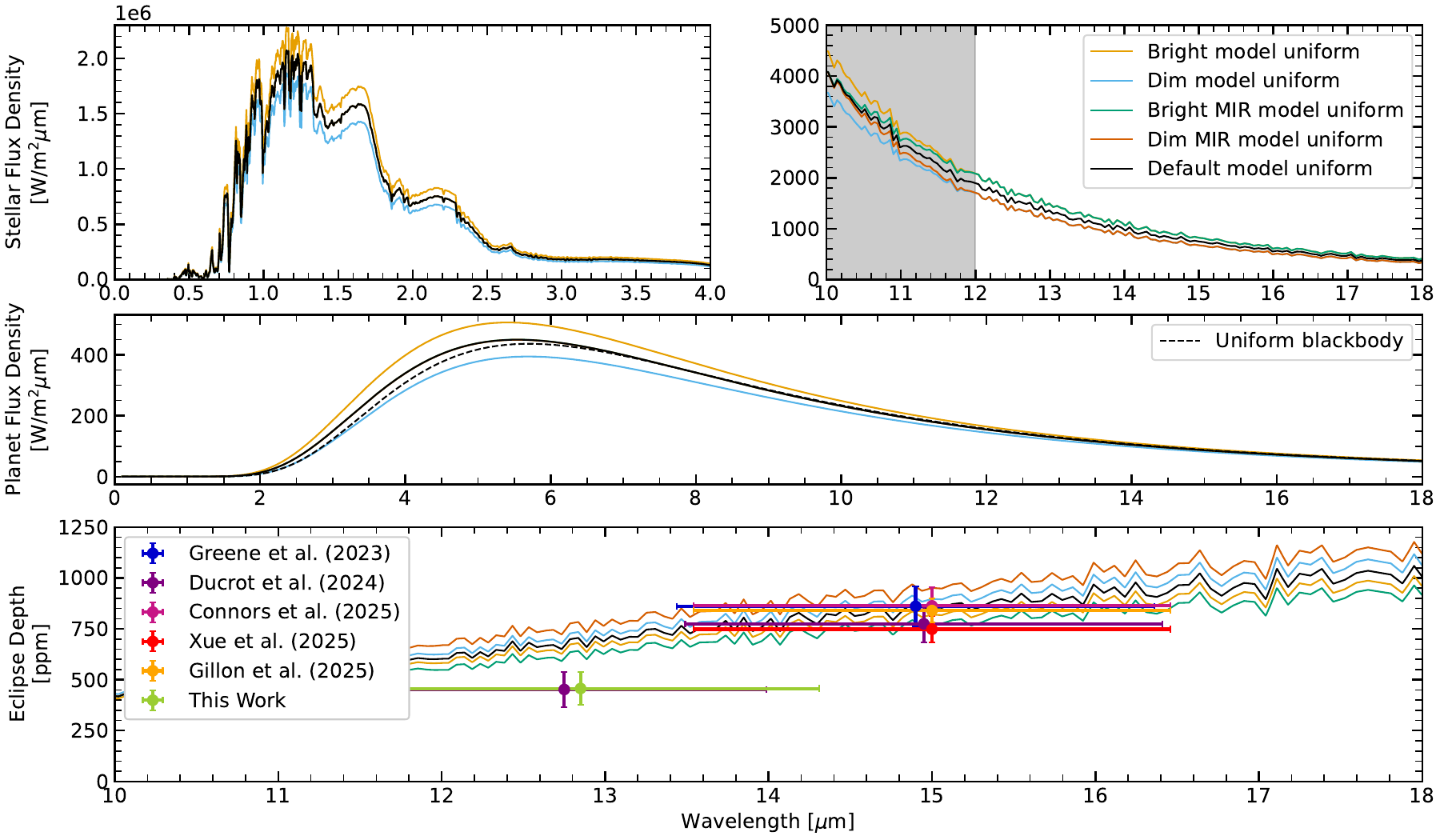}
\caption{Impact of slight offsets to the stellar spectra on a dark bare rock's flux density and eclipse depth, as simulated for TRAPPIST-1\,b using our modeling framework \texttt{JESTER} (Section \ref{sec:jester}). The top panels indicate the surface flux densities of the modeled host stars at NIR and MIR wavelengths. Different iterations of a SPHINX model ($T_{\mathrm{eff}} = 2600$K, log(g) = 5.0, [M/H] = 0.0, C/O = 0.5) are used to generate five different dark bare rock spectra \citep{Iyer_Line_Muirhead_Fortney_Gharib-Nezhad_2023, Iyer_Line_Muirhead_Fortney_Gharib-Nezhad_2024}. The models apply a 10\% increase or decrease to the star's surface flux density either across all wavelengths or for $\lambda > 12\mu$m, with a small transition region between 10$\mu$m and 12$\mu$m. The middle panel shows the simulated surface flux densities of TRAPPIST-1\,b as a dark bare rock while the bottom panel shows the resulting models for the planet's eclipse depth, alongside previously reported values from JWST observations \citep{Greene_Bell_Ducrot_Dyrek_Lagage_Fortney_2023,Ducrot_Lagage_Min_Gillon_Bell_Tremblin_Greene_Dyrek_Bouwman_Waters_etal._2025, Connors_Monaghan_Benneke_Dang_2025, Xue_Zhang_Coy_Brady_Ji_Bean_Radica_Seifahrt_Sturmer_Luque_etal._2025, gillon2025jwstthermalphasecurves}.}
\label{fig:stellarspectra}
\end{figure*}

\subsection{Spectral Energy Distribution of the Host Star}\label{sec:stellarspectrum}

One major concern when producing forward models is the accuracy of the host star's flux. In absence of interior heating, the planet's surface flux is composed of incident flux from the host star that is either reflected or absorbed and re-radiated. A stellar model is often used to approximate the SED of the host star, and is required when generating forward models across the bolometric wavelength regime. Two common stellar grids used in modeling are SPHINX and PHOENIX \citep{Iyer_Line_Muirhead_Fortney_Gharib-Nezhad_2023, Iyer_Line_Muirhead_Fortney_Gharib-Nezhad_2024, 2013A&A...553A...6H}. Although the frameworks tend to generate similar models for most spectral types, they diverge for cooler, late-type M-dwarf stars due to the use of different opacity sources when modeling the stellar spectrum \citep{fauchez2025stellarmodelslimitexoplanet}. Regardless of the model used, if $f_{\mathrm{*}}$ does not accurately reflect the true flux emitted by the host star, the simulated eclipse depth spectra will be modeled incorrectly, regardless of the treatment of the planet's surface flux \citep{fauchez2025stellarmodelslimitexoplanet}. These discrepancies between the modeled flux and the true flux have a number of potential causes, including the presence of clouds, stellar faculae, and molecular absorption \citep[e.g.,][]{Tsuji_2002, Rajpurohit_Allard_Rajpurohit_Sharma_Teixeira_Mousis_Kamlesh_2019, Miles_Biller_Patapis_Worthen_Rickman_Hoch_Skemer_Perrin_Whiteford_Chen_etal._2023}. The error injected into forward models is also nonlinear in nature, as the modeled SED affects both $f_{\mathrm{*}}$ and the incident energy on the planet's surface, thus modifying $f_{\mathrm{p}}$ as well (Figure \ref{fig:stellarspectra}).

To mitigate the effects of a poorly matching stellar model, a number of studies have performed absolute flux calibration by measuring the true flux of the host star during the planet's occultation and scaling the stellar model such that the simulated bandpass response of the model matches the observed flux \citep[e.g.,][]{Greene_Bell_Ducrot_Dyrek_Lagage_Fortney_2023, Ducrot_Lagage_Min_Gillon_Bell_Tremblin_Greene_Dyrek_Bouwman_Waters_etal._2025, Zieba_Kreidberg_Ducrot_Gillon_Morley_Schaefer_Tamburo_Koll_Lyu_Acuna_et_al._2023, august2024hotrockssurveyi, 2025hotrocksiii, xue2025jwstrockyworldsddt}. Although this calibration step may lead to more accurately modeled eclipse depths within the bandpass of the instrument used for calibration, the validity of the stellar model outside of the instrument's throughput is unknown, and thus scaling the stellar model across all wavelengths may introduce further uncertainty. Assuming the NIR spectrum of the host star is well modeled by a synthetic spectrum \citep[e.g.,][]{fauchez2025stellarmodelslimitexoplanet}, it may be more reliable to use the uncalibrated stellar model to calculate the planet's surface flux before using the scaled stellar model to calculate $(F_{\mathrm{p}}/F_{\mathrm{*}})$ in Equation \ref{eq:eclipsedepth}, as an offset in $f_{\mathrm{*}}$ at longer wavelengths will not affect $f_{\mathrm{p}}$ significantly, but will modify the local eclipse depths.

\citet{fauchez2025stellarmodelslimitexoplanet} have previously provided a thorough investigation regarding how stellar models heavily limit studies in emission spectroscopy. In particular, they argue for the collection of the stellar SED from 5-28 $\mu$m using MIRI MRS. These observations would only be marginally more expensive than eclipse observations, and would avoid the introduction of uncertainty associated with the absolute calibration step outlined above. While MRS observations would no doubt provide information critical to measuring the true stellar flux, it is also necessary to characterize the stellar flux shortward of 5 $\mu$m, as a significant majority of the host star's flux is emitted at shorter wavelengths. Therefore, much of the planet's reflected and emitted flux is governed by the host star's shortwave emissions, necessitating the precise characterization of the star's entire SED (Figure \ref{fig:stellarspectra}).


\subsection{Precision of Astrophysical Parameters} \label{sec:astroparams}


Imprecise knowledge of the planetary system's para- meters may also inject uncertainties by resulting in the generation of models that do not accurately depict the planet's conditions. Our knowledge of a system's physical parameters are limited by a number of factors, including the precision of existing instrumentation and photon noise. Parameters derived from such observations may also rely on limited equations or models. These uncertainties may result in the production of forward models that do not accurately represent the system in question, resulting in poor constraints on the planet's true composition. The uncertainty injected by such astrophysical parameters should serve as a fundamental limit to the precision of forward models. 

The impact of parameter uncertainty on the precision of model emission spectra has not yet been investigated in depth. From Equation \ref{eq:eclipsedepth}, it's clear that the uncertainty in $R_{\mathrm{p}}/{R_{\mathrm{*}}}$ will impact the precision of toy emission spectra \citep{xue2025jwstrockyworldsddt}. However, uncertainty in the parameters used to derive both $f_{\mathrm{p}}(\lambda)$ and $f_{\mathrm{*}}(\lambda)$ may also inject significant uncertainty into our forward models. Recent analyses of rocky exoplanet emission observations \citep[e.g.,][]{Zhang_Hu_Inglis_Dai_Bean_Knutson_Lam_Goffo_Gandolfi_2024, Xue_Bean_Zhang_Mahajan_Ih_Eastman_Lunine_Mansfield_Coy_Kempton_et_al._2024, Mansfield_Xue_Zhang_Mahajan_Ih_Koll_Bean_Coy_Eastman_Kempton_et_al._2024, Lin_Daylan_2026} have used the uncertainty in $T_{\mathrm{*}}$ and $a_{\mathrm{p}}/R_{\mathrm{*}}$ to derive $T_{\mathrm{d,max}}$ and $\Delta T_{\mathrm{d,max}}$ using Equation \ref{eq:tmax}, therefore facilitating a more robust analysis of the planet's brightness temperature ratio. Furthermore, \citet{Coy_Ih_Kite_Koll_Tenthoff_Bean_WeinerMansfield_Zhang_Xue_Kempton_etal._2025} recently derived the $\mathcal{R}$ values of rocky exoplanets orbiting M dwarf stars by sampling stellar models from the SPHINX stellar grid \citep{Iyer_Line_Muirhead_Fortney_Gharib-Nezhad_2023, Iyer_Line_Muirhead_Fortney_Gharib-Nezhad_2024} using the reported uncertainty in $T_{\mathrm{*}}$, log(g)$_{\mathrm{*}}$, and [M/H]$_{*}$. While these methods have enabled a more thorough analysis of model-based uncertainty in toy emission spectra and the derivation of $\mathcal{R}$, a uniform approach has not yet been implemented for planetary models with a non-uniform dayside temperature.

We use the preceding issues with forward modeling to govern our reanalysis of rocky exoplanet eclipse observations. In particular, our goal is to model the emission spectra of each planet such that a non-uniform dayside temperature is considered, and the impact of both physical and stellar uncertainty on the model emission spectra is measured. Using these model emission spectra, we can investigate the measured eclipse depths from each planet to determine whether they may be consistent with an atmospheric presence. 

\section{Methods}\label{sec:methods}

\subsection{Modeling the Flux from a Bare Rock Exoplanet}\label{sec:jester}

A bare rock without heat recirculation or interior heating will produce a surface flux in energy balance with the incident flux from the host star. \texttt{JESTER} (\textit{\textbf{J}ESTER: an \textbf{E}xoplanet \textbf{S}urface \textbf{T}ool for calculating the \textbf{E}missions from \textbf{R}ocky worlds}) is employed to produce our bare rock forward models by simulating the emission spectra of each planet in our analysis using a two-dimensional energy balance equation solved numerically \citep{Monaghan_Roy_Benneke_Crossfield_Coulombe_Piaulet-Ghorayeb_Kreidberg_Dressing_Kane_Dragomir_etal._2025, Connors_Monaghan_Benneke_Dang_2025}. \texttt{JESTER} can produce forward models for eclipse spectra of different compositions using the corresponding wavelength-dependent single-scattering albedo $w(\lambda)$ of the selected surface composition. The single-scattering albedo is converted into more relevant quantities following Hapke theory \citep{Hapke_2012}. The directional-hemispheric reflectance $r_{\mathrm{dh}}(\lambda)$ represents the total fraction of light scattered in all upward-going directions by a collimated light source, and can be calculated as:

\begin{equation}
r_{\mathrm{dh}}(\lambda) = \frac{1 - \gamma(\lambda)}{1 + 2\gamma(\lambda)\mu_{0}}
\label{eq:rdh}
\end{equation}

where $\gamma(\lambda) = \sqrt{1-w(\lambda)}$ and $\mu_{0} = \mathrm{cos}(\theta)$. $\theta$ represents the local solar zenith angle at a chosen point along the planet's surface. Additionally, the directional emissivity $\epsilon_{\mathrm{d}}(\lambda)$ measures the total fraction of emitted radiation in a single direction, and may be calculated following:

\begin{equation}
    \epsilon_{\mathrm{d}}(\lambda) = 1-r_{\mathrm{dh}}(\lambda)=\gamma(\lambda)\frac{1+2\mu_{0}}{1+2\gamma(\lambda)\mu_{0}}\label{eq:ed}
\end{equation}

The hemispherical emissivity $\epsilon_{\mathrm{h}}(\lambda)$ represents the hemispherical average of $\epsilon_{\mathrm{d}}$, and can be calculated with:

\begin{equation}
    \epsilon_{\mathrm{h}}(\lambda) = \frac{2\gamma(\lambda)}{1+\gamma(\lambda)}\left(1+\frac{1-\gamma(\lambda)}{6(1+\gamma(\lambda))}\right) \label{eq:eh}  
\end{equation}



\texttt{JESTER} uses these conversions to model the surface of a bare rock exoplanet using existing surface models, enabling us to probe the mineralogical surface composi- tion of a planet from its emission spectrum. \citep[e.g.,][]{Hammond_Guimond_Lichtenberg_Nicholls_Fisher_Luque_Meier_Taylor_Changeat_Dang_et_al._2024, Paragas_Knutson_Hu_Ehlmann_Alemanno_Helbert_Maturilli_Zhang_Iyer_Rossman_2025}.

It should be noted that many surface models that provide values of $\omega(\lambda)$ do not extend the entirety of the bolometric wavelength space. For example, the models provided by \citet{Paragas_Knutson_Hu_Ehlmann_Alemanno_Helbert_Maturilli_Zhang_Iyer_Rossman_2025} only report the $r_{\mathrm{dh}}(\lambda)$ of each model for $0.4$ $\mu$m $< 20$ $\mu$m. Simply ignoring the wavelengths outside of the included wavelength interval will result in an artificially inflated dayside temperature to account for the 'missing' flux. Thus, in order to more accurately account for the energy balance of a bare rock planet, we assume a uniform value of $w(\lambda)$ for $\lambda > 20$ $\mu$m and for $\lambda < 0.4$ $\mu$m using the closest value included within the parameter space for each. This modification prevents artificially inflating the planet's dayside temperature when producing forward models of different surfaces, and is employed to simulate the various surface compositions shown in Figure \ref{fig:surfacecom}. 

Ignoring the effects of tidal or interior heating, \texttt{JESTER} accounts for the temperature gradient across the planet's surface by separating the dayside hemisphere into a number of annular regions centered around the substellar point. Following \citet{Hammond_Guimond_Lichtenberg_Nicholls_Fisher_Luque_Meier_Taylor_Changeat_Dang_et_al._2024} and \citet{Paragas_Knutson_Hu_Ehlmann_Alemanno_Helbert_Maturilli_Zhang_Iyer_Rossman_2025}, the predicted temperature of each section is then calculated such that the total energy emitted from each annular section equals the total energy absorbed from the incident flux:

\begin{equation}
A_{\mathrm{pj}}\left(\frac{R_{\mathrm{*}}}{a_{\mathrm{p}}}\right)^{2} \int f_{\mathrm{*}}(\lambda)\epsilon_{\mathrm{d}}(\lambda) d\lambda = 
A \int \pi B_{\lambda}(T) \epsilon_{\mathrm{h}}(\lambda) d\lambda
\label{eq:energybalance}
\end{equation}

where $f_{\mathrm{*}}(\lambda)$ is the spectral flux density of the host star, and $B_{\lambda}(T)$ the spectral radiance from the planet as a blackbody function (Equation \ref{eq:blackbody_flux}). $A$ and $A_{\mathrm{pj}}$ represent the angular area of each annular section and the corresponding projected angular area above each section respectively. For an annular region on a planet encapsulating the zenith angles $\theta_{1} \leq \theta \leq \theta_{2}$:

\begin{equation}
    A(\theta)  = 2\pi (\cos(\theta_{1})-\cos(\theta_{2}))
    \label{eq:area}
\end{equation}

\begin{equation}
    A_{\mathrm{pj}}(\theta)  = \pi (\sin^{2}(\theta_{2})-\sin^{2}(\theta_{1}))
    \label{eq:areaproj}
\end{equation}

Increasing the total number of regions on the planet's surface improves the precision of the temperature gradient, and produces a temperature balance equivalent to the one originally derived by \citet{Hu_Ehlmann_Seager_2012}.

For each annular section, \texttt{JESTER} calculates the surface temperature $T(\theta)$ numerically using Equation \ref{eq:energybalance}. Using $r_{\mathrm{dh}}(\lambda,\theta)$ and $T(\theta)$, the total spectral flux density of the planet's dayside can be calculated from the projected-area weighted sum of the reflected flux density ($f_{\mathrm{p,r}}$) and emitted flux density ($f_{\mathrm{p,e}}$):

\begin{equation}
    f_{\mathrm{p,r}}(\lambda, \theta) =  \left(\frac{R_{\mathrm{*}}}{a} \right)^{2}f_{\mathrm{*}}(\lambda) r_{\mathrm{dh}}(\lambda, \theta)
    \label{eq:ref_flux}
\end{equation}

\begin{equation}
    f_{\mathrm{p,e}}(\lambda, \theta) = \pi B_{\lambda}(T( \theta)) \epsilon_{\mathrm{d}}(\lambda)
    \label{eq:ref_em}
\end{equation}

\begin{equation}
    \pi f_{\mathrm{p}}(\lambda) = \int_{0}^{\frac{\pi}{2}} A_{\mathrm{pj}}(\theta) \left(f_{\mathrm{p,r}}(\lambda, \theta) + f_{\mathrm{p,e}}(\lambda, \theta) \right) d\theta
    \label{eq:spec_tot}
\end{equation}

From $f_{\mathrm{p}}$, $f_{\mathrm{*}}$, and the planet's parameters, the eclipse depth of the planet may be calculated following Equation \ref{eq:eclipsedepth}. Additionally, we may account for the instrument response when modeling the eclipse depth measured over a specific filter by considering the instrument throughput $W(\lambda)$ \citep{Xue_Bean_Zhang_Mahajan_Ih_Eastman_Lunine_Mansfield_Coy_Kempton_et_al._2024, Mansfield_Xue_Zhang_Mahajan_Ih_Koll_Bean_Coy_Eastman_Kempton_et_al._2024, Coy_Ih_Kite_Koll_Tenthoff_Bean_WeinerMansfield_Zhang_Xue_Kempton_etal._2025}:

\begin{equation}
    (F_{\mathrm{p}}/F_{\mathrm{*}})_{\mathrm{W(\lambda)}} = \left(\frac{R_{\mathrm{p}}}{R_{\mathrm{*}}}\right)^{2} \frac{\int\frac{f_{\mathrm{p}}(\lambda)}{hc/\lambda} W(\lambda) \,d\lambda} {\int\frac{f_{\mathrm{*}}(\lambda)}{hc/\lambda} W(\lambda) \,d\lambda}
    \label{eq:instreclipsedepth}
\end{equation}

\begin{figure*}[t!]
\centering
\includegraphics[width=0.9\linewidth]{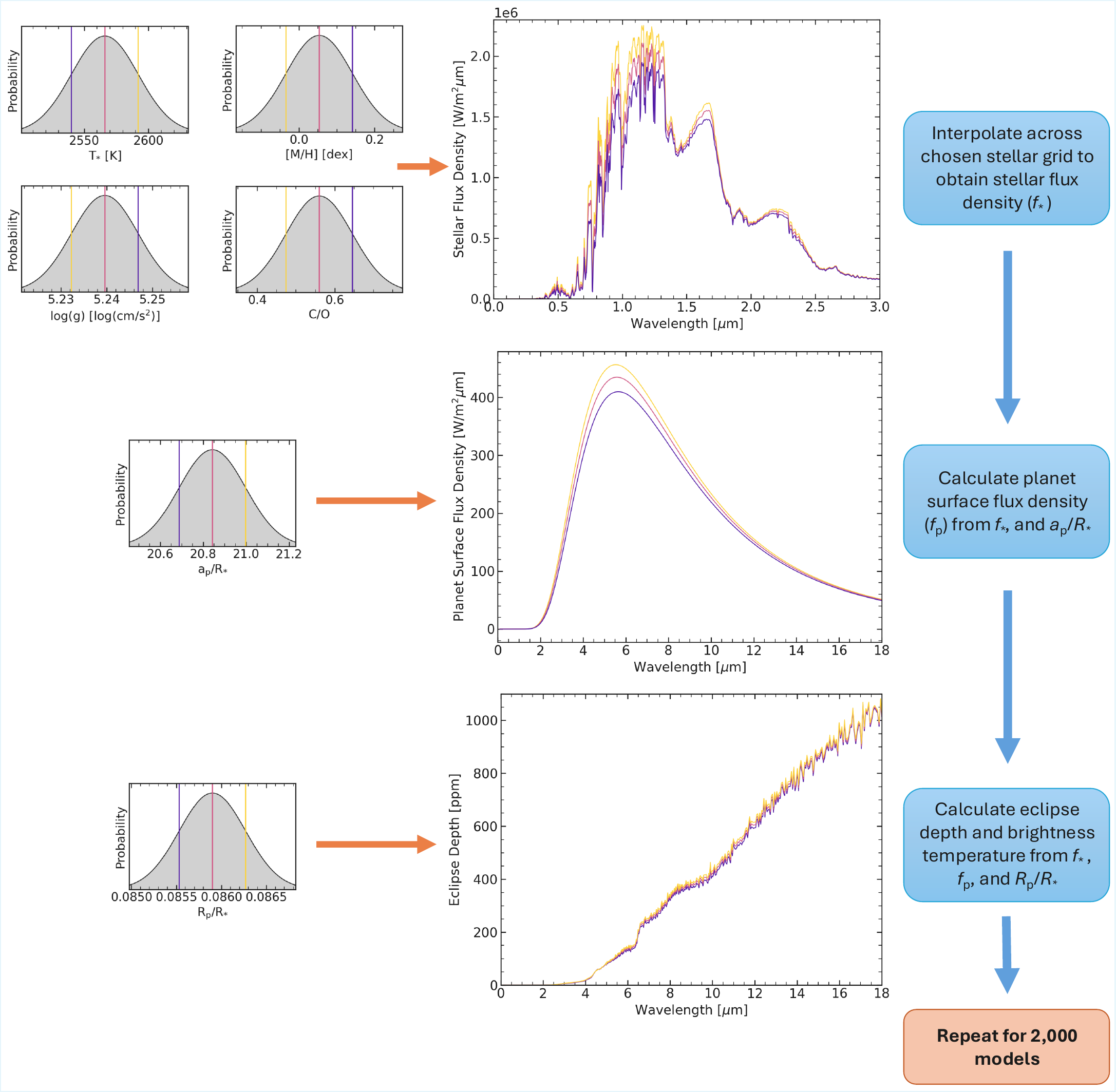}
\caption{Flowchart describing the steps for performing the bare rock analysis of each planet, as outlined in Section \ref{sec:reanalysis}. Three individual model results are shown for TRAPPIST-1\,b, with parameters sampled at the mean ($\mu$) and the mean plus or minus the standard deviation ($\mu\pm \sigma$) as shown on the left side of the chart. Provided models correspond to a dark bare rock with $A_{\mathrm{B}} = 0$, $\epsilon(\lambda) = 1$ and $r_{\mathrm{dh}}(\lambda) = 0$.}
\label{fig:flowchart}
\end{figure*}

\subsection{Modeling Schematic \& Accounting for Model Uncertainty}\label{sec:reanalysis}


We employ \texttt{JESTER} to perform our uniform reanalysis of the planets listed in Table \ref{tab:planets_in_study}. Given the complexity of non-gray atmospheric and surface models, we focus on comparing the published results for each planet to forward models as a dark bare rock with $A_{\mathrm{B}} = 0$, $\epsilon_{\mathrm{h}}(\lambda) = \epsilon_{\mathrm{d}}(\lambda) = 1$ and $r_{\mathrm{dh}}(\lambda) = 0$. We calculate the planet's surface flux by separating the dayside hemisphere into distinct annular regions and calculating the temperature and surface flux density of each section following \ref{sec:jester}. Although increasing the total number of regions results in a more-accurate hemispherical model, the differences between the results become negligible with enough precision. Thus, we choose to simulate 20 distinct annuli for each planet to improve the overall computational efficiency. Using the modeled surface flux for each planet, we calculate the model White Light Curve (WLC) eclipse depth using Equation \ref{eq:instreclipsedepth} and the throughput of the instrument used in the previous study using the SVO Filter Profile Service \citep{2012ivoa.rept.1015R, 2020sea..confE.182R, 2024A&A...689A..93R}, thus obtaining the eclipse depth for a dark bare rock $(F_{\mathrm{p}}/F_{\mathrm{*}})_{\mathrm{dbr}}$. 


Generating a single bare rock model for each planet will fail to account for the problems outlined in Sections  \ref{sec:stellarspectrum} and \ref{sec:astroparams}. To investigate the uncertainties propagated by the intrinsic uncertainty in astrophysical parameters and stellar flux, we generate 2,000 forward models of a dark bare rock and its host star with \texttt{JESTER} by sampling relevant astrophysical parameters from the published values and uncertainties for each planet. These generate 2,000 plausible conditions for the planet and its host star, each with a unique combination of $f_{\mathrm{*}}$ and $f_{\mathrm{p}}$, enabling us to investigate the distribution of potential eclipse depths and brightness temperatures associated with the planet as a dark bare rock. 


For calculating the eclipse depth of all 2,000 models for each planet, we sample the relevant astrophysical parameters from a normal distribution $N(\mu, \sigma^{2})$ using the reported value as the mean $\mu$ and the reported error as the standard deviation $\sigma^{2}$. We choose the largest error bar available when the reported errors are asymmetric to avoid underestimating the uncertainty. The sampled parameters in each model are $R_\mathrm{p}/R_{\mathrm{*}}$, constrained from the planet's transit, $a_{\mathrm{p}}/R_{\mathrm{*}}$, measured from Kepler's 3rd law, the effective stellar temperature $T_{\mathrm{eff}}$, the stellar surface gravity log(g), the stellar metallicity [M/H], and the stellar C/O ratio. Given that the C/O ratio for the majority of the M dwarfs in our survey have not yet been reported, we sample the C/O ratio from $N(0.56, 0.086)$ using the distribution of existing C/O ratios previously reported in other M dwarfs \citep{2014PASJ...66...98T, 2016PASJ...68...13T, 2015PASJ...67...26T, Nakajima_Sorahana_2016}. Furthermore, for M dwarfs without a known metallicity (e.g., LHS 3844), we sample the stellar metallicity from $N(-0.29, 0.27)$ using the distribution of M-dwarf metallicities reported by \citet{SARMENTO2021A&A...649A.147S}. It should be noted that the distribution of C/O and [M/H] in M-dwarf stars is still under study, and different priors on either parameter may result in varied model results \citep[e.g.,][]{Jahandar_Doyon_Artigau_Cook_Cadieux_Donati_Cowan_Cloutier_Pelletier_Alves-Brito_etal._2025}.

We calculate the stellar flux of each model by linearly interpolating across the SPHINX stellar grid using the sampled values of the star's parameters \citep{Iyer_Line_Muirhead_Fortney_Gharib-Nezhad_2023, Iyer_Line_Muirhead_Fortney_Gharib-Nezhad_2024}. For systems beyond the parameter space of the SPHINX grid ($T_{\mathrm{eff}} > 4000$K), we alternatively employ the use of linearly interpolated PHOENIX models \citep{2013A&A...553A...6H}. To investigate the differences between either grid, we generate sets of models using both stellar grids for systems with $T_{\mathrm{eff}} < 4000$K. The results were largely equivalent as a result of the parameter sampling, and we therefore choose to use the SPHINX models for such systems. The differences between using either grid in eclipse forward modeling have previously been elucidated by \citet{fauchez2025stellarmodelslimitexoplanet} and \citet{Coy_Ih_Kite_Koll_Tenthoff_Bean_WeinerMansfield_Zhang_Xue_Kempton_etal._2025}.

From the 2,000 iterations of each planet, we measure the median eclipse depth as the dark bare rock eclipse depth and report the standard deviation as the planet's model uncertainty $\Delta (F_{\mathrm{p}}/F_{\mathrm{*}})_{\mathrm{dbr}}$, which we use in conjunction with the observational uncertainty to constrain the planet's composition. In particular, for a planet consistent with a dark bare rock:

\begin{equation}
\mathcal{F} = \frac{(F_{\mathrm{p}}/F_{\mathrm{*}})_{\mathrm{obs}}}{(F_{\mathrm{p}}/F_{\mathrm{*}})_{\mathrm{dbr}}} \approx 1
\label{eq:barerockfraction}
\end{equation}

In addition to $(F_{\mathrm{p}}/F_{\mathrm{*}})_{\mathrm{dbr}}$, we calculate the correspon- ding brightness temperature of the dark bare rock eclipse depth for each model, reporting the median and uncertainty as $T_{\mathrm{d,dbr}}$ and $\Delta T_{\mathrm{d,dbr}}$, respectively. To compare the bare rock brightness temperatures to the published results, we convert the observed eclipse depth to a dayside brightness temperature for each of the 2,000 individual models, rather than using each planet's reported brightness temperature. While $(F_{\mathrm{p}}/F_{\mathrm{*}})_{\mathrm{obs}}$ is an observational quantity, the brightness temperature of the planet's dayside is modeled from the observed eclipse depth assuming a known stellar flux and set of physical parameters. Thus, comparing the reported brightness temperature $T_{\mathrm{d,obs}}$ to $T_{\mathrm{d,dbr}}$ from our models would be insufficient in representing the distribution of the system's parameters. Therefore, to calculate the derived brightness temperature $T_{\mathrm{d,der}}$ corresponding to the observed eclipse depth, we start by sampling from $N((F_{\mathrm{p}}/F_{\mathrm{*}})_{\mathrm{obs}}, \Delta (F_{\mathrm{p}}/F_{\mathrm{*}})_{\mathrm{obs}})$ to generate a sampled eclipse depth for each model $f_{\mathrm{*}}$ and $\frac{R_{\mathrm{p}}}{R_{\mathrm{*}}}$. We then calculate each sampled eclipse depth's corresponding brightness temperature using Equations \ref{eq:instreclipsedepth} and \ref{eq:blackbody_flux} with the equivalent instrument throughput. In similar fashion to measuring $(F_{\mathrm{p}}/F_{\mathrm{*}})_{\mathrm{dbr}}$ and $\Delta(F_{\mathrm{p}}/F_{\mathrm{*}})_{\mathrm{dbr}}$, we calculate $T_{\mathrm{d,der}}$ and $\Delta T_{\mathrm{d,der}}$ from the distribution of brightness temperatures in the 2,000 iterations, allowing us to report the brightness temperature ratio $\mathcal{R} = T_{\mathrm{d,der}}/T_{\mathrm{d,dbr}}$ and corresponding error for each planet in our reanalysis. For a planet consistent with a dark bare rock, we anticipate that $\mathcal{R} \approx 1$.



\subsection{Target Selection} \label{sec:targets}

We generate suites of dark bare rock models for all 18 rocky exoplanets ($R_{\mathrm{p}} \leq$ 2R$_{\oplus}$) with published secondary eclipse observations (Table \ref{tab:planets_in_study}). These planets span a large range of masses, radii, and surface temperatures, and will enable us to probe for the presence of trends across emission data \citep[e.g.,][]{Coy_Ih_Kite_Koll_Tenthoff_Bean_WeinerMansfield_Zhang_Xue_Kempton_etal._2025, Lin_Daylan_2026}. The majority of these planets have more than a single reported value of $(F_{\mathrm{p}}/F_{\mathrm{*}})_{\mathrm {obs}}$ across multiple publications as a result of both subsequent reanalyses and further observations. Though we only generate a single set of model iterations for each planet, we calculate separate values for $T_{\mathrm{d,der}}, \mathcal{F}$ and $\mathcal{R}$ corresponding to the reported eclipse depth in each publication for completeness.

Additionally, we performed a reanalysis of the TRAPPIST-1\,b MIRI F1280W measurement from \cite{Ducrot_Lagage_Min_Gillon_Bell_Tremblin_Greene_Dyrek_Bouwman_Waters_etal._2025} and the LTT 3780\,b F1500W eclipse depth measurement from \cite{Allen_Espinoza_Diamond-Lowe_Mendonca_Demory_Gressier_Ih_Fortune_August_Holmberg_etal._2025} using the \texttt{Erebus} pipeline following the procedure outlined in \cite{Connors_Monaghan_Benneke_Dang_2025}, shown in Table \ref{tab:eclipsedepthresutls!}. For LTT 3780\,b we start directly with the calibrated \verb|calints| files on MAST. For TRAPPIST-1\,b we first run the \verb|Eureka!| pipeline \citep{Bell2022} on the uncalibrated data before inputting the stage 2 outputs into \texttt{Erebus}, as the default calibrated files had large unhandled systematics. For TRAPPIST-1\,b we get an eclipse depth of $457\pm80$ ppm which is in agreement with the value derived in \cite{Ducrot_Lagage_Min_Gillon_Bell_Tremblin_Greene_Dyrek_Bouwman_Waters_etal._2025} ($452\pm86$ ppm). For LTT\,3780\,b we get an eclipse depth of $322\pm40$ ppm which is in agreement with the value derived in \cite{Allen_Espinoza_Diamond-Lowe_Mendonca_Demory_Gressier_Ih_Fortune_August_Holmberg_etal._2025} ($312\pm38$ ppm). The presented eclipse depths are the average of the fits for each individual visit. Additionally, we present a measured eclipse depth value for the first two visits of the Rocky Worlds DDT target GJ 3929\,b using \texttt{Erebus}. For this planet, we measure an eclipse depth of $154\pm27$ ppm, in agreement with the value derived in \citet{xue2025jwstrockyworldsddt} ($160\pm26$ ppm). A more robust analysis of the planet's occultation data will be presented in a future paper (Connors et al. in prep.). 

\begin{table*}[]

\centering

\begin{tabular}{l|lcc|cc|l}
\hline
\hline
\textbf{Planet}        & Instrument  & $(F_{\mathrm{p}}/F_{\mathrm{*}})_{\mathrm{obs}}$  & $T_{\mathrm{d,der}}$ & $\frac{(F_{\mathrm{p}}/F_{\mathrm{*}})_{\mathrm{obs}}}{(F_{\mathrm{p}}/F_{\mathrm{*}})_{\mathrm{dbr}}}$ & $T_{\mathrm{d,der}}/T_{\mathrm{d,dbr}}$ & Source\\ 
      &   & (ppm) & (K) &  $\mathcal{F}$ & $\mathcal{R}$ &     

\\

\hline
\textbf{TOI-1468\,b}   & F1500W     & 311 $\pm$ 31                                     & 1063 $\pm$ 85        & 1.41 $\pm$ 0.18                                                                                       & \textbf{1.24 $\pm$ 0.10}              & \citet{Meier_Vald_s_2025}                                                                                    \\ 
& F1500W     & 286 $\pm$ 39                                     & 1002 $\pm$ 103       & 1.30 $\pm$ 0.20                                                                                         & 1.17 $\pm$ 0.12                       & \citet{Connors_Monaghan_Benneke_Dang_2025}                                                                    \\ \hline  
\textbf{GJ 1252\,b}    & IRAC Ch2   & 149 $\pm$ 32                                     & 1481 $\pm$ 155       & 1.23 $\pm$ 0.30                                                                                       & \textbf{1.08 $\pm$ 0.13}               & \citet{Crossfield_Malik_Hill_Kane_Foley_Polanski_Coria_Brande_Zhang_Wienke_et_al._2022}                       \\ \hline  
\textbf{GJ 367\,b}     & LRS        & 79 $\pm$ 4                                       & 1836 $\pm$ 101       & 1.13 $\pm$ 0.09                                                                                       & \textbf{1.08 $\pm$ 0.07}              & \citet{Zhang_Hu_Inglis_Dai_Bean_Knutson_Lam_Goffo_Gandolfi_2024}                                             \\ \hline  
\textbf{GJ 3929\,b}    & F1500W     & 160 $\pm$ 26                                     & 781 $\pm$ 80         & 1.14 $\pm$ 0.20                                                                                       & \textbf{1.08 $\pm$ 0.12}              & \citet{Xue_Zhang_Coy_Brady_Ji_Bean_Radica_Seifahrt_Sturmer_Luque_etal._2025}                                  \\
                       & F1500W     & 154 $\pm$ 27                                     & 764 $\pm$ 83         & 1.10 $\pm$ 0.20                                                                                       & 1.05 $\pm$ 0.12                       & This Work                                                                                                   \\ \hline  
\textbf{LHS 1140\,c}   & F1500W     & 273 $\pm$ 43                                     & 548 $\pm$ 46         & 1.08 $\pm$ 0.18                                                                                       & \textbf{1.04 $\pm$ 0.09}              & \citet{2025hotrocksiii}                                                                                       \\
                       & F1500W     & 242 $\pm$ 35                                     & 517 $\pm$ 36         & 0.96 $\pm$ 0.15                                                                                       & 0.98 $\pm$ 0.07                       & \citet{Connors_Monaghan_Benneke_Dang_2025}                                                                    \\
                       & F1500W     & 271 $\pm$ 31                                     & 545 $\pm$ 34         & 1.07 $\pm$ 0.14                                                                                        & 1.03 $\pm$ 0.07                        & \citet{rochon2025reanalysiseclipseslhs1140}                                                                   \\ \hline  
\textbf{LTT 3780\,b}   & F1500W     & 312 $\pm$ 38                                     & 1166 $\pm$ 122       & 1.04 $\pm$ 0.16                                                                                       & \textbf{1.03 $\pm$ 0.11}              & \citet{Allen_Espinoza_Diamond-Lowe_Mendonca_Demory_Gressier_Ih_Fortune_August_Holmberg_etal._2025}            \\
& F1500W     & 322 $\pm$ 40                                     & 1188 $\pm$ 131       & 1.08 $\pm$ 0.15                                                                                       & 1.05 $\pm$ 0.12                       & This Work                                                                                                   \\ \hline  
\textbf{TOI-1685\,b}   & G395H NRS2 & 122 $\pm$ 7                                  & 1403 $\pm$ 46        & 1.07 $\pm$ 0.10                                                                                       & \textbf{1.02 $\pm$ 0.04}              & \citet{Luque_Coy_Xue_Feinstein_Ahrer_Changeat_Zhang_Moran_Bean_Kite_etal._2024}                               \\
& G395H NRS1 & 103 $\pm$ 5                                  & 1460 $\pm$ 40        & 1.22 $\pm$ 0.13                                                                                       & 1.06 $\pm$ 0.04                       & ---                                                                                                         \\ \hline  
\textbf{LHS 3844\,b}   & IRAC Ch2   & 380 $\pm$ 40                                     & 1020 $\pm$ 38        & 1.01 $\pm$ 0.11                                                                                       & \textbf{1.00 $\pm$ 0.05}              & \citet{Kreidberg_Koll_Morley_Hu_Schaefer_Deming_Stevenson_Dittmann_Vanderburg_Berardo_et_al._2019}            \\ \hline  
\textbf{TRAPPIST-1\,b} & F1500W     & 840 $\pm$ 58                                   & 496 $\pm$ 16         & 0.98 $\pm$ 0.07                                                                                       & \textbf{0.99 $\pm$ 0.04}              & \citet{gillon2025jwstthermalphasecurves}                                                                      \\
& F1500W     & 861 $\pm$ 99                                     & 503 $\pm$ 27         & 1.01 $\pm$ 0.12                                                                                       & 1.01 $\pm$ 0.06                       & \citet{Greene_Bell_Ducrot_Dyrek_Lagage_Fortney_2023}                                                          \\
& F1500W     & 775 $\pm$ 90                                     & 481 $\pm$ 26         & 0.90 $\pm$ 0.11                                                                                       & 0.96 $\pm$ 0.06                       & \citet{Ducrot_Lagage_Min_Gillon_Bell_Tremblin_Greene_Dyrek_Bouwman_Waters_etal._2025}                         \\
& F1280W     & 452 $\pm$ 86                                     & 431 $\pm$ 29         & 0.67 $\pm$ 0.13                                                                                       & 0.86 $\pm$ 0.06                         & ---                                                                                                         \\
& F1500W     & 863 $\pm$ 90                                     & 503 $\pm$ 24         & 1.01 $\pm$ 0.11                                                                                       & 1.01 $\pm$ 0.05                        & \citet{Connors_Monaghan_Benneke_Dang_2025}                                                                    \\
& F1500W     & 748 $\pm$ 65                                     & 472 $\pm$ 18         & 0.87 $\pm$ 0.08                                                                                       & 0.94 $\pm$ 0.04                       & \citet{Xue_Zhang_Coy_Brady_Ji_Bean_Radica_Seifahrt_Sturmer_Luque_etal._2025}                                  \\
& F1280W     & 457 $\pm$ 80                                     & 433 $\pm$ 28         & 0.67 $\pm$ 0.12                                                                                       & 0.86 $\pm$ 0.06                       & This Work                                                                                                   \\ \hline  
\textbf{GJ 486\,b}     & LRS        & 136 $\pm$ 5                                  & 832 $\pm$ 16         & 0.89 $\pm$ 0.04                                                                                       & \textbf{0.95 $\pm$ 0.02}              & \citet{Mansfield_Xue_Zhang_Mahajan_Ih_Koll_Bean_Coy_Eastman_Kempton_et_al._2024}                              \\ \hline  
\textbf{LTT 1445 A\,b} & LRS        & 41 $\pm$ 9                                       & 503 $\pm$ 34         & 0.78 $\pm$ 0.22                                                                                       & \textbf{0.93 $\pm$ 0.08}              & \citet{wachiraphan2024thermalemissionspectrumnearby}                                                          \\ \hline  
\textbf{GJ 1132\,b}    & LRS        & 140 $\pm$ 17                                     & 671 $\pm$ 29         & 0.78 $\pm$ 0.11                                                                                       & \textbf{0.92 $\pm$ 0.05}               & \citet{Xue_Bean_Zhang_Mahajan_Ih_Eastman_Lunine_Mansfield_Coy_Kempton_et_al._2024}                            \\ \hline  
\textbf{TRAPPIST-1\,c} & F1500W     & 397 $\pm$ 69                                 & 373 $\pm$ 23         & 0.70 $\pm$ 0.12                                                                                       & \textbf{0.87 $\pm$ 0.06}              & \citet{gillon2025jwstthermalphasecurves}                                                                      \\
& F1500W     & 421 $\pm$ 94                                     & 383 $\pm$ 31         & 0.74 $\pm$ 0.17                                                                                       & 0.89 $\pm$ 0.07                       & \citet{Zieba_Kreidberg_Ducrot_Gillon_Morley_Schaefer_Tamburo_Koll_Lyu_Acuna_et_al._2023}                      \\
& F1500W     & 312 $\pm$ 128                                    & 343 $\pm$ 49         & 0.55 $\pm$ 0.22                                                                                       & 0.80 $\pm$ 0.12                       & \citet{Connors_Monaghan_Benneke_Dang_2025}                                                                    \\
& F1500W     & 329 $\pm$ 79                                     & 349 $\pm$ 29         & 0.58 $\pm$ 0.14                                                                                       & 0.82 $\pm$ 0.07                        & \citet{Xue_Zhang_Coy_Brady_Ji_Bean_Radica_Seifahrt_Sturmer_Luque_etal._2025}                                  \\ \hline  
\textbf{K2-141\,b*}    & IRAC Ch2   & 143 $\pm$ 39                                   & 2242 $\pm$ 348       & 0.76 $\pm$ 0.22                                                                                       & \textbf{0.85 $\pm$ 0.13}              & \citet{Zieba_Zilinskas_Kreidberg_Nguyen_Miguel_Cowan_Pierrehumbert_Carone_Dang_Hammond_et_al._2022}           \\ \hline  
\textbf{55 Cnc\,e*}    & LRS        & 110 $\pm$ 9                                      & 1925 $\pm$ 111       & 0.73 $\pm$ 0.06                                                                                       & \textbf{0.81 $\pm$ 0.05}              & \citet{Hu_Bello-Arufe_Zhang_Paragas_Zilinskas_Van_Buchem_Bess_Patel_Ito_Damiano_et_al._2024}                  \\
& IRAC Ch2   & 84 $\pm$ 14                                      & 1992 $\pm$ 170       & 0.70 $\pm$ 0.12                                                                                       & 0.83 $\pm$ 0.07                       & \citet{Tamburo_2018}                                                                                          \\ \hline  
\textbf{LHS 1478\,b}   & F1500W     & 138 $\pm$ 53                                     & 508 $\pm$ 92         & 0.47 $\pm$ 0.19                                                                                       & \textbf{0.68 $\pm$ 0.13}               & \citet{august2024hotrockssurveyi}                                                                             \\
& F1500W     & 173 $\pm$ 69                                     & 572 $\pm$ 110        & 0.59 $\pm$ 0.24                                                                                       & 0.77 $\pm$ 0.15                       & \citet{Connors_Monaghan_Benneke_Dang_2025}                                                                    \\ \hline  
\textbf{TOI-561\,b*}   & G395H NRS2 & 47 $\pm$ 4                                 & 1935 $\pm$ 83        & 0.47 $\pm$ 0.04                                                                                        & \textbf{0.67 $\pm$ 0.03}               & \citet{teske2025thickvolatileatmosphereultrahot}                                                              \\
& G395H NRS1 & 24 $\pm$ 3                                   & 1704 $\pm$ 80        & 0.29 $\pm$ 0.04                                                                                       & 0.59 $\pm$ 0.03                       & ---                                                                                                         \\ \hline  
\textbf{TOI-431\,b*}   & IRAC Ch2   & 33 $\pm$ 22                                      & 1490 $\pm$ 374       & 0.37 $\pm$ 0.25                                                                                        & \textbf{0.63 $\pm$ 0.16}               & \citet{Monaghan_Roy_Benneke_Crossfield_Coulombe_Piaulet-Ghorayeb_Kreidberg_Dressing_Kane_Dragomir_etal._2025} \\ \hline
\end{tabular}
\caption{Results of our reanalysis of the current suite of rocky exoplanet eclipse observations alongside the observed eclipse depths in the original study, sorted by the planet's derived brightness temperature ratio. The results for the starred planets (*) use an interpolated PHOENIX stellar model \citep{2013A&A...553A...6H}, while the remainder of planets use an interpolated SPHINX stellar model \citep{Iyer_Line_Muirhead_Fortney_Gharib-Nezhad_2023, Iyer_Line_Muirhead_Fortney_Gharib-Nezhad_2024}. Bold results indicate the 'default' value used for the planet in subsequent plots. For planets with multiple values of $(F_{\mathrm{p}}/F_{\mathrm{*}})_{\mathrm{obs}}$, the default value is chosen as the most recently published value from a non-reanalysis paper, where the longest wavelength result is chosen.}
    \label{tab:moderesults}

\end{table*}

\section{Results}\label{sec:results}

Using the schematic outlined in Section \ref{sec:reanalysis} and in Figure \ref{fig:flowchart}, we reanalyze the suite of rocky exoplanets with emission observations by comparing their measured eclipse depths to the distribution of dark bare rock models. The instrument-dependent values of $(F_{\mathrm{p}}/F_{\mathrm{*}})_{\mathrm{dbr}}$ and $T_{\mathrm{d,dbr}}$ from each of the 2,000 models for every planet can be seen in Figures \ref{fig:auxmodels} and \ref{fig:auxmodels2}. For each planet we simulate each model's WLC eclipse depth using the same instrument(s) the planet was observed with, as listed in Table \ref{tab:moderesults}, matching the exact throughput used for spectroscopic observations. From each planet's set of models, we calculate both $\mathcal{F} = \frac{(F_{\mathrm{p}}/F_{\mathrm{*}})_{\mathrm{obs}}}{(F_{\mathrm{p}}/F_{\mathrm{*}})_{\mathrm{dbr}}}$ and $\mathcal{R} = T_{\mathrm{d,der}}/T_{\mathrm{d,dbr}}$ using the distribution of results from \texttt{JESTER} as outlined in Section \ref{sec:reanalysis}, and investigate whether each planet has an eclipse depth and brightness temperature consistent with that of a dark bare rock. A planet consistent with a dark bare rock should have $\mathcal{R} = 1$, while planets with uniform heat recirculation ($f = \frac{1}{4}, A_{\mathrm{B}} = 0$) should have $\mathcal{R} \approx 0.78$ \citep{2008ApJS..179..484H, 48d96982-3471-3bab-9b6f-f8b21bafd6fb, Crossfield_Malik_Hill_Kane_Foley_Polanski_Coria_Brande_Zhang_Wienke_et_al._2022}.

\begin{figure*}[]
\centering

\includegraphics[width=0.8\linewidth]{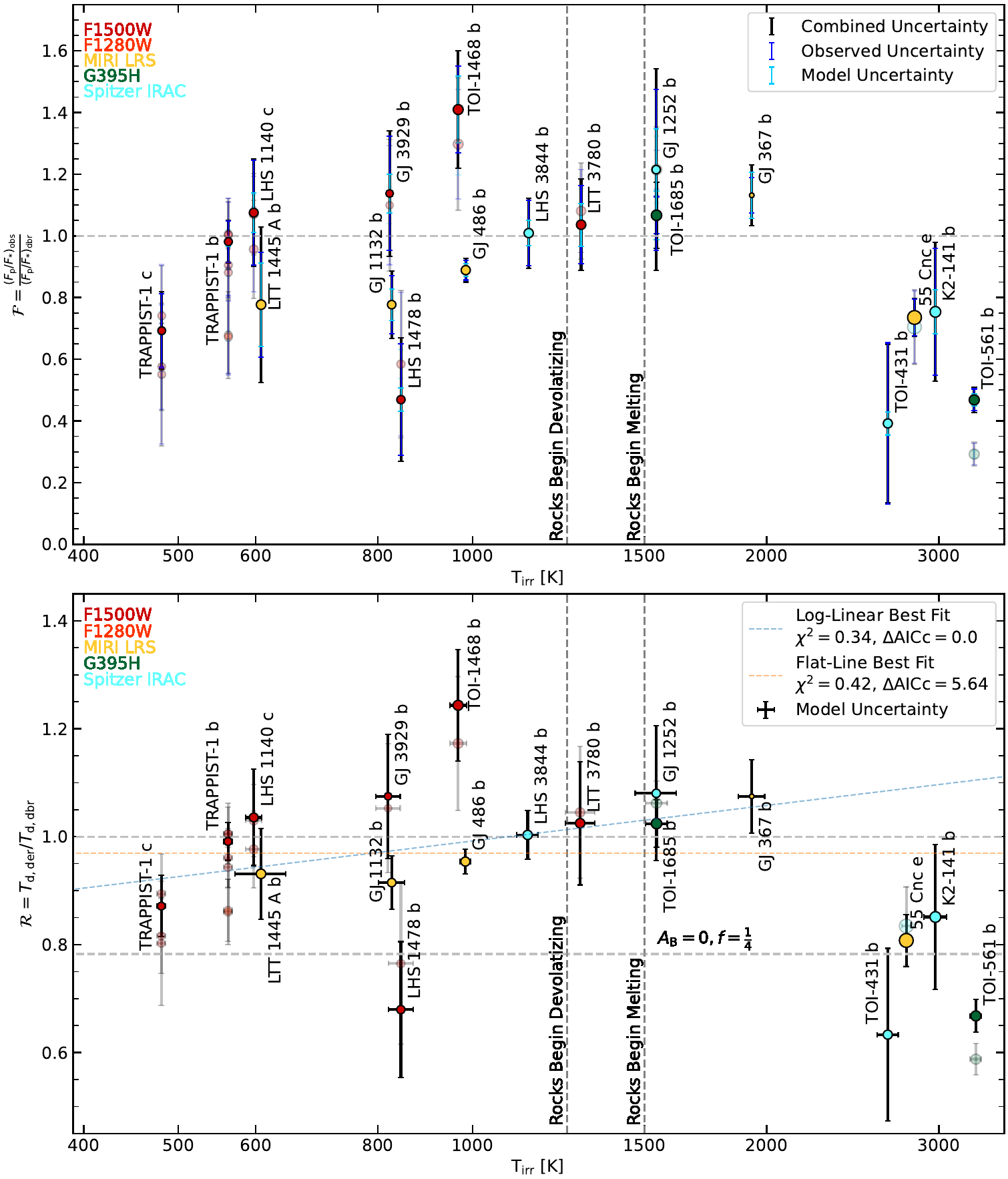}

\caption{$\mathcal{F}$ and $\mathcal{R}$ plotted as a function of irradiation temperature for the current suite of rocky exoplanets with thermal emission observations. Data point radii are proportional to planet radii. Planets where $\mathcal{F} \approx 1$ and $\mathcal{R} \approx 1$ are consistent with a dark bare rock composition. The approximate temperatures at which silicate rock begins to devolatize and melt are shown at $\sim1250$ K and $\sim1500$ K, respectively \citep{lutgens, Mansfield_Kite_Hu_Koll_Malik_Bean_Kempton_2019}. The error bars indicate the source of uncertainty for each point: observed uncertainty refers to the uncertainty in the observed eclipse depth, while model uncertainty refers to the model variation generated following Section \ref{sec:reanalysis}. The color of each point corresponds to the instrument used in the original observation to measure $(F_{\mathrm{p}}/F_{\mathrm{*}})_{\mathrm{obs}}$. Multiple points are plotted at a lower opacity for planets with more than one published eclipse depth. We include the $\chi^{2}$ statistic for two fits to $\mathcal{R}$ as a function of $T_{\mathrm{irr}}$ for all the data points from M-dwarf orbiting planets in Table \ref{tab:moderesults}.}

\label{fig:tirr}
\end{figure*}

Table \ref{tab:moderesults} and Figure \ref{fig:tirr} summarize the results of our reanalysis for all of the planets in our survey. From the $\mathcal{F}$ and $\mathcal{R}$ results, we can see that the majority of planets analyzed are consistent with a dark bare rock to $1\sigma$. These planets likely have compositions similar to those of the Moon, Mercury, and Ceres in our own solar system. Thin atmospheres ($< 1$ bar) with minimal heat recirculation to the planet's nightside, however, cannot be ruled out. In general, these results are largely consistent with the previously published results for each planet. However, the addition of model uncertainty in our analysis enables some variation in our knowledge of the planet's composition. Notably, planets hosting intermediate values of $\mathcal{R}$ between 1.0 and 0.78 may be consistent to $\sim1\sigma$ with both a dark bare rock and a completely redistributive atmosphere; for K2-141\,b in particular, this result deviates from past literature \citep{Zieba_Zilinskas_Kreidberg_Nguyen_Miguel_Cowan_Pierrehumbert_Carone_Dang_Hammond_et_al._2022, Crossfield_Malik_Hill_Kane_Foley_Polanski_Coria_Brande_Zhang_Wienke_et_al._2022}.


\begin{figure}[]
\centering

\includegraphics[width=0.9\linewidth]{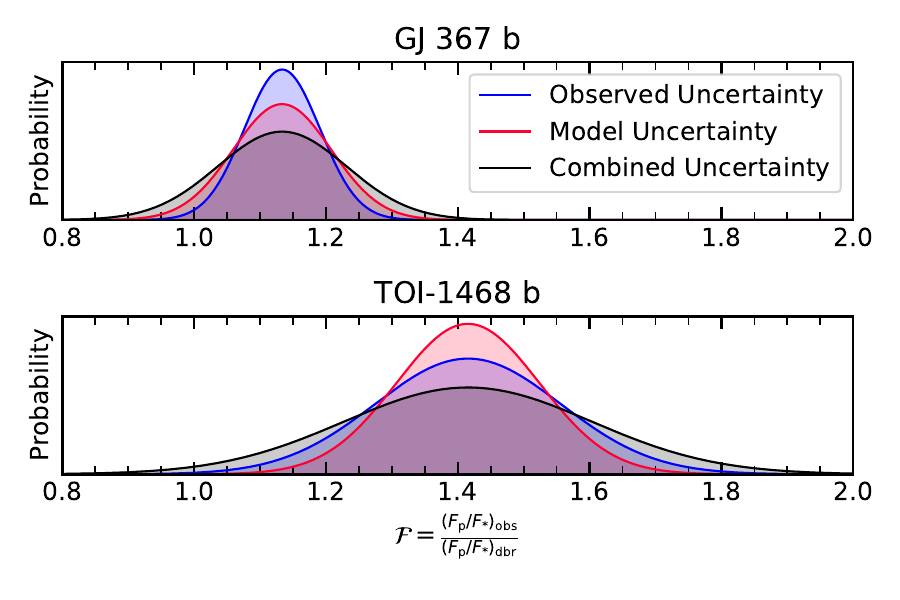}

\caption{Demonstration of how the observed and model uncertainty convolve when deriving the combined uncertainty in $\mathcal{F}$, using the MIRI LRS white light eclipse depths of GJ 367\,b \citep{Zhang_Hu_Inglis_Dai_Bean_Knutson_Lam_Goffo_Gandolfi_2024} and TOI-1468\,b \citep{Meier_Vald_s_2025}. The observed uncertainty accounts the reported uncertainty in $(F_{\mathrm{p}}/F_{\mathrm{*}})_{\mathrm{obs}}$, while the model uncertainty derived from our modeling schematic (Section \ref{sec:reanalysis}) accounts for the measured uncertainty in $(F_{\mathrm{p}}/F_{\mathrm{*}})_{\mathrm{dbr}}$. When combining the two sources of error, it's clear that the model uncertainty provides a fundamental limit to our confidence in constraining both $\mathcal{F}$ and $\mathcal{R}$ when comparing observations to the null hypothesis of a dark bare rock.}
\label{fig:errorprop}
\end{figure}

A number of planets have observations that disfavor the null hypothesis of a dark bare rock when compared to $(F_{\mathrm{p}}/F_{\mathrm{*}})_{\mathrm{dbr}}$ and $T_{\mathrm{d,dbr}}$. For example, F1280W observations of TRAPPIST-1\,b's eclipse depth results in a measured brightness temperature that is $\sim2\sigma$ away from the dark bare rock model \citep{gillon2025jwstthermalphasecurves, Ducrot_Lagage_Min_Gillon_Bell_Tremblin_Greene_Dyrek_Bouwman_Waters_etal._2025}. Additionally, the F1500W brightness temperature ratio of TRAPPIST-1\,c has been found to deviate from $\mathcal{R}=1$ in a number of analyses \citep{Zieba_Kreidberg_Ducrot_Gillon_Morley_Schaefer_Tamburo_Koll_Lyu_Acuna_et_al._2023, Connors_Monaghan_Benneke_Dang_2025, xue2025jwstrockyworldsddt}. Similarly, LHS 1478\,b was found to have $\mathcal{R} \sim2.5\sigma$ away from unity for the two existing data reductions on a single eclipse visit \citep{august2024hotrockssurveyi, Connors_Monaghan_Benneke_Dang_2025}. Furthermore, the lava worlds 55 Cnc\,e, TOI-431\,b, and TOI-561\,b all host particularly low brightness temperatures $\geq2\sigma$ below the dark bare rock models \citep{Hu_Bello-Arufe_Zhang_Paragas_Zilinskas_Van_Buchem_Bess_Patel_Ito_Damiano_et_al._2024, Monaghan_Roy_Benneke_Crossfield_Coulombe_Piaulet-Ghorayeb_Kreidberg_Dressing_Kane_Dragomir_etal._2025, teske2025thickvolatileatmosphereultrahot}. 

It should be noted that a low eclipse depth or brightness temperature ratio is degenerate between a number of surface compositions and cannot be considered definitive evidence for an atmosphere by itself. High albedo surfaces formed from fresh regolith and partially devolatized rock may produce similarly low eclipse depths at infrared wavelengths, and as such produce brightness temperatures consistent with strong heat recirculation \citep{Kite_Fegley_Schaefer_Gaidos_2016, Mansfield_Kite_Hu_Koll_Malik_Bean_Kempton_2019, Hammond_Guimond_Lichtenberg_Nicholls_Fisher_Luque_Meier_Taylor_Changeat_Dang_et_al._2024, Paragas_Knutson_Hu_Ehlmann_Alemanno_Helbert_Maturilli_Zhang_Iyer_Rossman_2025}. Other potential degeneracies are outlined in Section \ref{sec:modelinglimitations}.

The measured eclipse depth and brightness tempera- ture of TOI-1468\,b is much larger than anticipated from a dark bare rock \citep{Meier_Vald_s_2025, Connors_Monaghan_Benneke_Dang_2025}. A number of astrophysical and systematic causes are presented by \citet{Meier_Vald_s_2025} to explain the origin of the increased flux. If TOI-1468\,b hosts an atmosphere with efficient absorption of stellar radiation, a thermal inversion may be generated in the planet's atmosphere, leading to the generation of emission features in the planet's emission spectrum \citep{thermal_inversion_2011ApJ...742L..19M, thermal_inversion_Castan_2011, thermal_inversion_2013cctp.book.....M, thermal_inversion_Zilinskas_2022}. Alternatively, the energy budget of the planet may be inflated through magnetic induction heating \citep{induction_heating_2014MNRAS.441.2361V, induction_heating_2016PhR...663....1S, induction_heating_Kislyakova_2018, induction_heating_2020MNRAS.498.5684D}. Tidal heating driven by eccentric orbits may also increase the planet's energy budget, resulting in significantly hotter dayside temperatures \citep{2010exop.book..239C, 2018exha.book.....P}. Unaccounted for systematics and instrumental artifacts in MIRI may also result in poorly measured eclipse depths \citep{miri_artifacts_Morrison_2023, bell2023_miri_lrs_settling_ramp, miri_artifacts_libralato2024highprecisionastrometryphotometryjwstmiri, Dyrek_2024_miri_lrs_settling}, though follow-up study of the data through frame-normalized principal component analysis was unable to find evidence of linear, detector-level systematic error \citep{Connors_Monaghan_Benneke_Dang_2025}. Additionally, thermal beaming from a non-isotropic bare rock with macroscopic surface roughness may result in eclipse depths deeper than those simulated by \texttt{JESTER} \citep[e.g.,][]{Wohlfarth_Wohler_Hiesinger_Helbert_2023, Tenthoff_Wohlfarth_Wohler_Zieba_Kreidberg_2024, Coy_Ih_Kite_Koll_Tenthoff_Bean_WeinerMansfield_Zhang_Xue_Kempton_etal._2025}. 


Our uniform reanalysis elucidates a notable result: for many of the planets in our survey, the model uncertainty derived from the system parameters and stellar properties is comparable to the reported uncer- tainty in the observed eclipse depths (Figure \ref{fig:errorprop}). Effectively, the model uncertainty presents itself as a fundamental limit to precisely constraining a planet's composition from eclipse depth observations. Although increasing the number of visits may decrease the observational uncertainty, the model uncertainty may only be decreased by improving the precision of reported astrophysical parameters and proper characterization of the host star's SED. Thus, it is critical that the model uncertainty be accounted for in future observations and data reductions for a more refined analysis of the planet's composition.

\begin{figure*}[]
\centering

\includegraphics[width=1\linewidth]{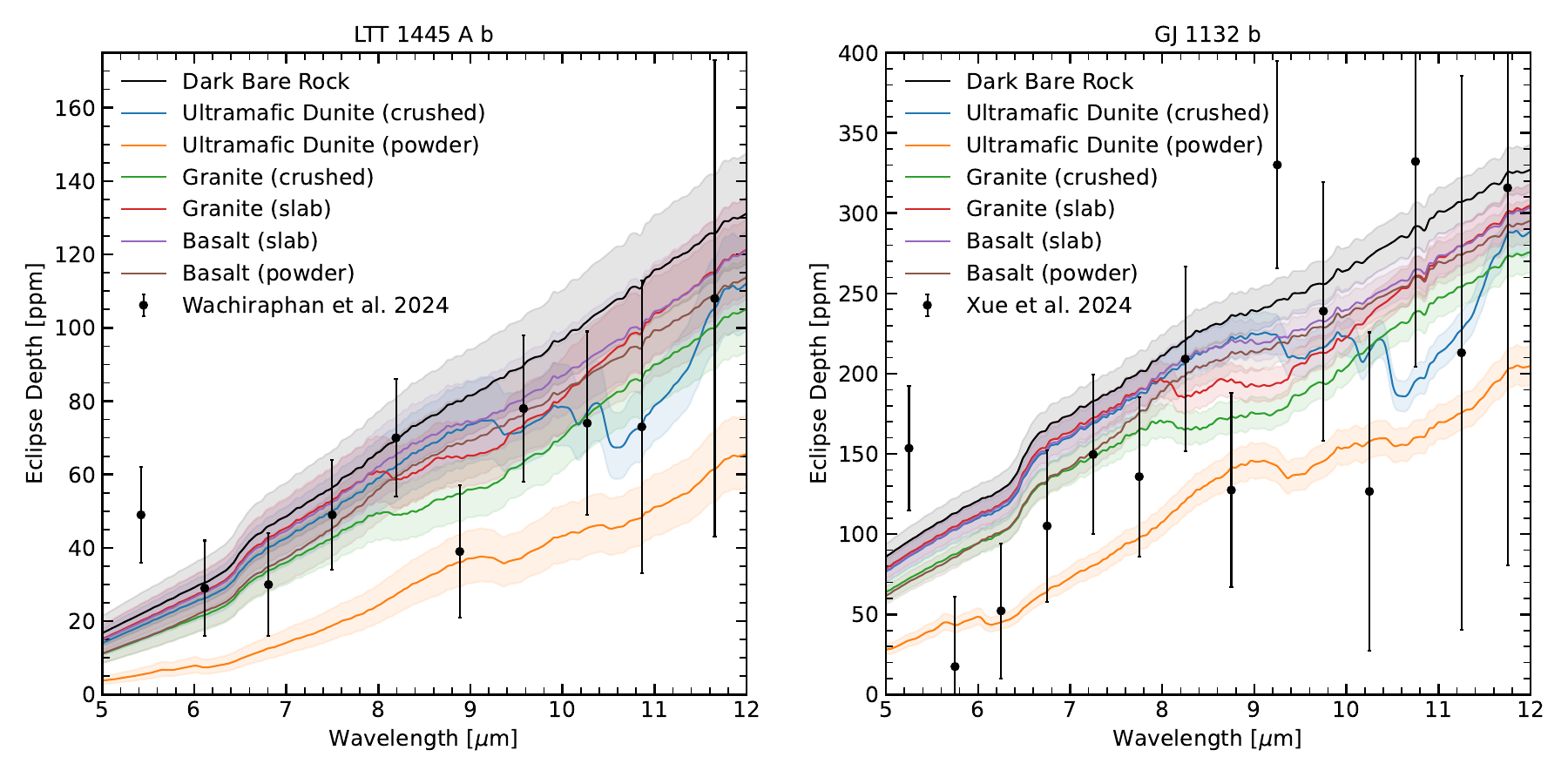}

\caption{Eclipse spectra of various surface models from \citet{Paragas_Knutson_Hu_Ehlmann_Alemanno_Helbert_Maturilli_Zhang_Iyer_Rossman_2025} as simulated by \texttt{JESTER}, compared to the observed eclipse spectra LTT 1445 A\,b \citep{wachiraphan2024thermalemissionspectrumnearby} and GJ 1132\,b \citep{Xue_Bean_Zhang_Mahajan_Ih_Eastman_Lunine_Mansfield_Coy_Kempton_et_al._2024}. Model uncertainty is calculated following Section \ref{sec:reanalysis}. Significant model overlap renders photometric observations incapable of producing strong constraints on the planet's composition, though exogeological study may remain possible through spectroscopy as the shape and location of wavelength dependent spectral features are preserved.}

\label{fig:surfacecom}
\end{figure*}

\subsection{Population-level Trends}

A number of studies in the past have investigated trends in the rocky exoplanet population by plotting the brightness temperature ratio $\mathcal{R}$ as a function of the planet's irradiation temperature \citep{Crossfield_Malik_Hill_Kane_Foley_Polanski_Coria_Brande_Zhang_Wienke_et_al._2022, Luque_Coy_Xue_Feinstein_Ahrer_Changeat_Zhang_Moran_Bean_Kite_etal._2024, Coy_Ih_Kite_Koll_Tenthoff_Bean_WeinerMansfield_Zhang_Xue_Kempton_etal._2025, kreidberg2025lookrockyexoplanetsjwst, Lin_Daylan_2026}. The irradiation temperature corresponds to the planet's surface temperature at the substellar point, and can be calculated following:

\begin{equation}
    T_{\mathrm{irr}} = T_{\mathrm{*}}\sqrt{\frac{R_{\mathrm{*}}}{a_{\mathrm{p}}}}
    \label{eq:irradiation temperature}
\end{equation}

To further search for trends in the current suite of rocky exoplanets with published emission observations, we plot both $\mathcal{F}$ and $\mathcal{R}$ as a function of $T_{\mathrm{irr}}$ (Figure \ref{fig:tirr}). In particular, we search for trends between $\mathcal{R}$ and $T_{\mathrm{irr}}$ as an extension of \citet{Coy_Ih_Kite_Koll_Tenthoff_Bean_WeinerMansfield_Zhang_Xue_Kempton_etal._2025} and \citet{Lin_Daylan_2026}. The results from planets orbiting M-dwarf stars ($T_{\mathrm{irr}} < 2000$K) indicate a tenuous trend in brightness temperature as a function of irradiation temperature, wherein $\mathcal{R}$ tends to decrease with colder surface temperatures. These results are in strong agreement with those presented by both \citet{Coy_Ih_Kite_Koll_Tenthoff_Bean_WeinerMansfield_Zhang_Xue_Kempton_etal._2025} and \citet{Lin_Daylan_2026}, who noted similar trends in the results for planets orbiting M-dwarf stars. Following the analysis of \citet{Coy_Ih_Kite_Koll_Tenthoff_Bean_WeinerMansfield_Zhang_Xue_Kempton_etal._2025}, we similarly fit the our results using a log-linear regression:
\begin{equation}
    \mathcal{R} = \alpha \mathrm{ln(}T_{\mathrm{irr}}\mathrm{)} + \beta 
\end{equation}

and compare our result to the best fit model for the best-fitting flat line where $\mathcal{R} = 0.97\pm0.01$ for every planet.

When including every M-dwarf measurement listed in Table \ref{tab:moderesults}, our regression returns $\alpha = 0.095\pm0.026$ and $\beta = 0.334\pm0.201$. We derive the $\chi^{2}$ statistic for our log-linear fit and calculate $\chi^{2} = 0.34$, while the flat line model returns $\chi^{2} = 0.42$ and a relative corrected
Akaike Information Criterion (AIC) of $\Delta\mathrm{AICc} = 5.29$. We additionally perform a regression on the M-dwarf planets when removing the notable outliers TOI-1468\,b and LHS 1478\,b from our dataset, resulting in a value of $\chi^{2} = 0.10$ for the log-linear fit and values of $\chi^{2} = 0.17$ and $\Delta\mathrm{AICc} = 14.24$ for the flat line. These results indicate that the log-linear model is slightly preferred when fitting the data, indicating a potential log-linear trend between $\mathcal{R}$ and $T_{\mathrm{irr}}$, in agreement with the results from \citet{Coy_Ih_Kite_Koll_Tenthoff_Bean_WeinerMansfield_Zhang_Xue_Kempton_etal._2025}. \citet{Lin_Daylan_2026} similarly determine that hotter M-dwarf planets produce larger $\mathcal{R}$ values, following a broken power law fit to the data.

\citet{Coy_Ih_Kite_Koll_Tenthoff_Bean_WeinerMansfield_Zhang_Xue_Kempton_etal._2025} and \citet{Lin_Daylan_2026} present a number of options to explain the potential trend in $\mathcal{R}$ as a function of $T_{\mathrm{irr}}$ for M-dwarf orbiting planets. For example, hotter planets may maintain regolith surfaces with macroscopic grains through lava resurfacing and high temperature sintering, which decreases the net reflectivity of the planet's surface \citep{Zaini_van, Zhuang_Zhang_Ma_Jiang_Yang_Milliken_Hsu_2023}. These trends are supported by the models provided by \citet{Paragas_Knutson_Hu_Ehlmann_Alemanno_Helbert_Maturilli_Zhang_Iyer_Rossman_2025}, where fine-grain materials tend to be more reflective than their crushed or smooth counterparts. Space weathering from stellar winds and micro-meteor impacts may also contribute to the trend in $\mathcal{R}$ by forming thin layers of non-reflective metallic iron and/or graphite on the planet's surface \citep{2015aste.book..597B, 2019A&A...627A..43D}. The trend may also indicate that colder planets are more likely to retain atmospheres as a result of decreased instellation or increased outgassing rates.

It should be noted that the trends noted above in \citet{Coy_Ih_Kite_Koll_Tenthoff_Bean_WeinerMansfield_Zhang_Xue_Kempton_etal._2025} and \citet{Lin_Daylan_2026} are tenuous as best. In particular, \citet{Coy_Ih_Kite_Koll_Tenthoff_Bean_WeinerMansfield_Zhang_Xue_Kempton_etal._2025} find that a number of trendlines suitably fit the data when using stellar models from the PHOENIX stellar grid as the values of $\mathcal{R}$ tend towards unity. Further research is required in order to confirm the existence of this trend, fill in gaps in the parameter space and to explain remaining outliers. 

Our analysis furthermore reveals that the most highly irradiated of planets tend to host particularly low $\mathcal{R}$ values that may be indicative of atmospheric heat recirculation. \citet{Lin_Daylan_2026} suggest that either thick atmospheres or high albedos are required to explain the measured values of $\mathcal{R}$ for the hotter, non M-dwarf orbiting planets ($T_{\mathrm{irr}} > 2000$K). At such temperatures, the daysides of such planets are anticipated to be partially or entirely molten. Generally, such surfaces are expected to host relatively low albedos, and therefore maintain temperatures similar to that of a dark bare rock \citep{Essack_Seager_Pajusalu_2020}. Nevertheless, the particularly low $\mathcal{R}$ values of these planets may be indicative of atmospheric heat recirculation despite the strong stellar insolation. Different atmospheric scenarios have previously been investigated for these planets, including stochastic volatile rich atmospheres and tenuous mineral vapor envelopes \citep[e.g.,][]{Zieba_Zilinskas_Kreidberg_Nguyen_Miguel_Cowan_Pierrehumbert_Carone_Dang_Hammond_et_al._2022, rasmussen2023nondetectionironhighresolutionemission, Hu_Bello-Arufe_Zhang_Paragas_Zilinskas_Van_Buchem_Bess_Patel_Ito_Damiano_et_al._2024, Patel_Brandeker_Kitzmann_de_la_Roche_Bello-Arufe_Heng_Valdes_Persson_Zhang_Demory_et_al._2024, Monaghan_Roy_Benneke_Crossfield_Coulombe_Piaulet-Ghorayeb_Kreidberg_Dressing_Kane_Dragomir_etal._2025}. Although a rapidly outgassing magma ocean may exceed the rate of photoevaporative escape from XUV irradiation, the outgassing rate on lava worlds is difficult to constrain \citep[e.g.,][]{Dorn_2018, 2020A&A...643A..44S, 2023A&A...675A.122B}, and thus further study is required to determine whether lava worlds may host redistributive atmospheres.

Planets with more than one set of published eclipses tend to have some variation in the derived values of $\mathcal{R}$, indicating the presence of varying systematics inherent in different data pipelines. Although these values tend to agree with one another to $\sim1\sigma$, determining the cause of such variation will require further observation and study. Additionally, the trends noted in $\mathcal{R}$ may be impacted by a number of other astrophysical parameters beyond $T_{\mathrm{irr}}$. Our calculated results for $\mathcal{R}$ are directly compared to those from \citet{Coy_Ih_Kite_Koll_Tenthoff_Bean_WeinerMansfield_Zhang_Xue_Kempton_etal._2025} and \citet{Lin_Daylan_2026} in Figure \ref{fig:comparisontopastresults}.

\begin{figure*}[t]
  \centering
      \includegraphics[width=\linewidth]{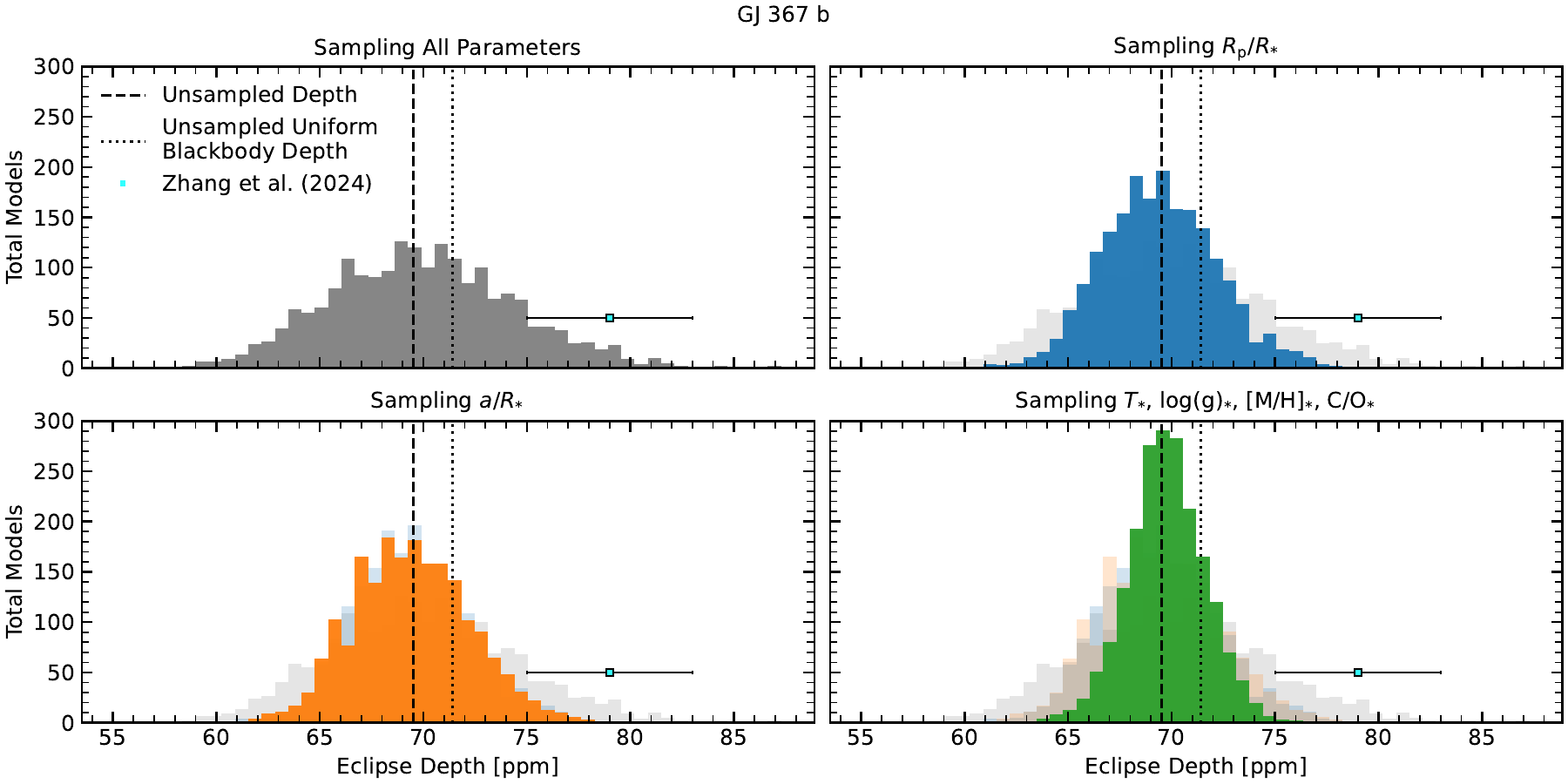} 
  \caption{Distribution of modeled instrument eclipse depths for GJ 367\,b using the MIRI LRS filter when sampling different parameters. Published eclipse depths and error bars are shown for each planet, alongside the unsampled eclipse depth for both a dark bare rock and a uniform temperature blackbody. The distribution of modeled eclipse depths when sampling every relevant parameter is shown at the top left in gray, and in light gray for all other subplots.}
  \label{fig:hists}
  
\end{figure*}

\section{Discussion}\label{sec:Discussion}

\subsection{Simulating different Surface Compositions}

The prospect of performing exogeological study on planets without an atmosphere has grown significantly alongside the increased interest in emission spectroscopy of rocky exoplanets \citep{Hu_Ehlmann_Seager_2012, Hammond_Guimond_Lichtenberg_Nicholls_Fisher_Luque_Meier_Taylor_Changeat_Dang_et_al._2024, Paragas_Knutson_Hu_Ehlmann_Alemanno_Helbert_Maturilli_Zhang_Iyer_Rossman_2025, First_Mishra_Gazel_Lewis_Letai_Hanssen_2025}. In addition to generating dark bare rock models for each planet, we generate bare rock models of varying surface composition following Section \ref{sec:reanalysis} and Figure \ref{fig:flowchart} for a select group of planets from our survey. This enables us to investigate how the uncertainties propagate for compositions with a nonuniform emissivity and reflectivity. We generate a range of models to simulate the emission spectra of a basalt, granite, and ultramafic surface using the $r_{\mathrm{dh}}(\lambda)$ data provided by \citet{Paragas_Knutson_Hu_Ehlmann_Alemanno_Helbert_Maturilli_Zhang_Iyer_Rossman_2025} for the K1919 basalt, dalmatian granite, and dunite xenolith samples, respectively (Figure \ref{fig:surfacecom}). The variation in eclipse depths shown for these bare rock models may also be used as a probe for forward models of different atmospheres, which should be affected by the same uncertainties.

There is significant overlap between the different surface models simulated across all wavelengths. When combined with the existing uncertainty from observa- tions, distinguishing between different models becomes incredibly challenging. This further illustrates the difficulty in constraining a planet's surface composition from eclipse observations, particularly from photometric measurements alone. However, in depth analysis of a planet's composition may nonetheless be possible from spectroscopic observations, as wavelength dependent spectral features from a surface or atmosphere can be probed regardless of our knowledge of the planet's true parameters. The shape and location of absorption and emission features are preserved across the 2,000 iterations of each model, allowing us to make inferences regarding a planet's composition despite the variation in the system's physical parameters. Thus, both atmospheric and exogeological study of rocky exoplanets remain possible in ongoing and future spectroscopic JWST observations.

\begin{figure*}[t]
\centering
\includegraphics[width=\linewidth]{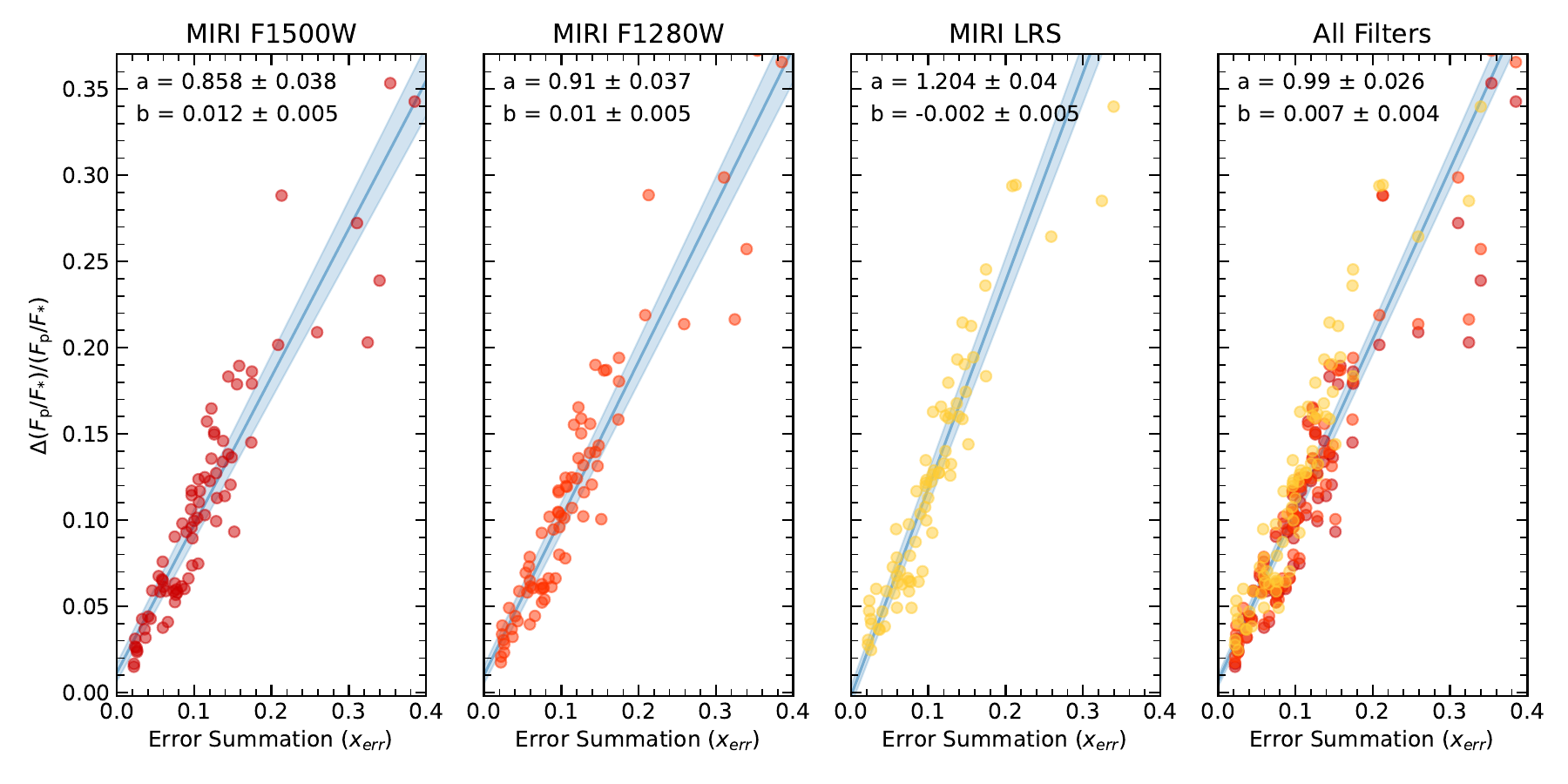}
\caption{$\frac{\Delta (F_{\mathrm{p}}/F_{\mathrm{*}})_{\mathrm{dbr}}}{(F_{\mathrm{p}}/F_{\mathrm{*}})_{\mathrm{dbr}}}$ for all of the candidate targets of the Rocky Worlds DDT plotted as a function of $x_{\mathrm{err}}$ (Equations \ref{eq:eclipsedepthunceratainty} and \ref{eq:xerr}), revealing an approximately linear correlation between the two variables. The best fit line is shown for each axis.}

\label{fig:Fpdepallfilters}
\end{figure*}

\subsection{Parameter Dependence}
For a selection of planets from Table \ref{tab:moderesults}, we repeat the model generation outlined in Section \ref{sec:reanalysis} for a dark bare rock when sampling a single parameter at a time, enabling us to determine how different parameter uncertainties are injected into our bare rock forward models (Figure \ref{fig:hists}). From the observed distributions, it's clear that all of the parameters can individually generate a wide range of simulated instrument eclipse depths if they are not well constrained, and that no single parameter dominates the overall uncertainty. Thus, to properly estimate the uncertainty inherent in all forward models, the imprecision in each parameter must be weighed equally.

Assuming that the modeling schematic presented in Section \ref{sec:reanalysis} is applicable to all secondary eclipse forward models, it should be possible to approximate the uncertainty in a forward model's eclipse depth without generating 2,000 samples of each planet. We find that $\frac{\Delta (F_{\mathrm{p}}/F_{\mathrm{*}})_{\mathrm{dbr}}}{(F_{\mathrm{p}}/F_{\mathrm{*}})_{\mathrm{dbr}}}$ can be well modeled by a linear relationship with the uncertainty in the sampled astrophysical parameters. For a dark bare rock with an instrument eclipse depth $(F_{\mathrm{p}}/F_{\mathrm{*}})$ and model uncertainty $\Delta (F_{\mathrm{p}}/F_{\mathrm{*}})$:

\begin{equation}
    \frac{\Delta(F_{\mathrm{p}}/F_{\mathrm{*}})}{F_{\mathrm{p}}/F_{\mathrm{*}}} \approx ax_{\mathrm{err}} +b
    \label{eq:eclipsedepthunceratainty}
\end{equation}

where $x_{\mathrm{err}}$ represents a linear sum of how each parameter uncertainty modifies the model eclipse depth:

\begin{equation}
    x_{\mathrm{err}} = \frac{\Delta T_{\mathrm{*}}}{T_{\mathrm{*}}} + \frac{\Delta (a/R_\mathrm{*})}{a/R_\mathrm{*}} + \frac{\Delta (R_{\mathrm{p}}/{R_{\mathrm{*}})}}{R_{\mathrm{p}}/{R_{\mathrm{*}}}}
    \label{eq:xerr}
\end{equation}

This enables us to approximate the error bars associated with a modeled instrumental eclipse depth, as shown in Figure \ref{fig:Fpdepallfilters} for the candidate targets listed in the Rocky Worlds DDT \citep{Redfield_Batalha_Benneke_Biller_Espinoza_France_Konopacky_Kreidberg_Rauscher_Sing_2024}. Although the exact slope and intercept of the best-fit linear trend varies depending on the filter and surface model, the linear correlation remains consistent. Thus, the trend elucidated by Equation \ref{eq:eclipsedepthunceratainty} may be particularly useful in comparing candidate targets and for estimating the instrumental model uncertainty for different forward models without necessitating the use of the framework outlined in Section \ref{sec:reanalysis}.

\subsection{Disagreements between Published Parameters}\label{sec:disagreements}
Many planets in our survey have a number of published solutions to the system's astrophysical parameters. Although the most precise set of parameters are most often chosen to represent the system as a whole, there is no guarantee that this set of parameters is the most accurate. The solutions may disagree with one another despite using similar datasets and methods of analysis, and as such may produce surface forward models that differ significantly.

One notable example of this phenomenon is LTT 1445 A\,b,  which has two sets of astrophysical parameters published from the same year. \citet{PassLTTparams} used joint HST/TESS transit observations and radial velocity measurements to derive physical parameters of the system, while a prior study by \citet{OddoLTTparams} using TESS and CHEOPS transit observations returned different estimations for $R_{\mathrm{p}}/R_{\mathrm{*}}$, $a_{\mathrm{p}}/R_{\mathrm{*}}$, and $T_{\mathrm{*}}$.

\begin{figure}[]
\centering
\includegraphics[width=0.9\linewidth]{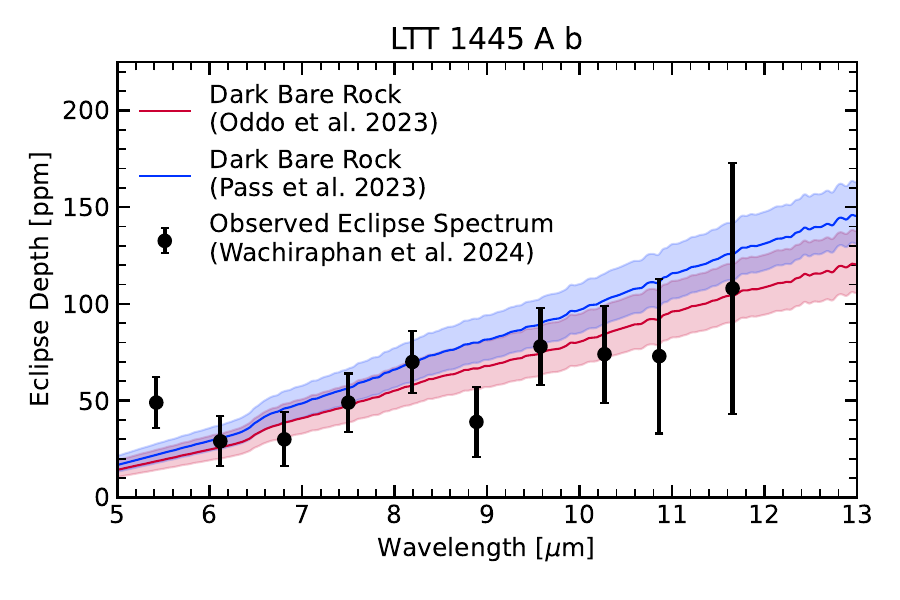}
\caption{Comparison of the dark bare rock emission spectra for LTT 1445 A\,b using the parameters from both \citet{PassLTTparams} and \citet{OddoLTTparams} as simulated by \texttt{JESTER} following the modeling schematic outlined in Section \ref{sec:reanalysis}. The observed eclipse spectrum and WLC depth from \citet{wachiraphan2024thermalemissionspectrumnearby} are overplotted.}

\label{fig:LTT1445}
\end{figure}

Following our parameter sampling method from before, we generate 2,000 bare rock models of LTT 1445 A\,b following the parameters from \citet{PassLTTparams} and \citet{OddoLTTparams} separately. We then compare these forward models to the eclipse spectrum observed by \citet{wachiraphan2024thermalemissionspectrumnearby} (Figure \ref{fig:LTT1445}). Although both sets of models agree with the reported eclipse spectrum, they notably deviate from one another by $\sim1\sigma$. This disagreement complicates our ability to accurately characterize the planet's composition from emission observations, and is not unique to LTT 1445 A\,b.

Inherently, the reliability of a planetary emission model depends on the accuracy of the reported astrophysical parameters. Although our ability to estimate the value of such parameters has significantly improved with the development of increasingly powerful observatories and robust modeling schematics \citep[e.g.,][]{Mahajan_2024}, photon noise and instrumental systematics may hinder our capability to precisely measure such values. Grazing transits and stellar variation may also hinder the accuracy of fitting astrophysical parameters from transit observations \citep{Heller_2019, Alexoudi_Mallonn_Keles_Poppenhager_Essen_Strassmeier_2020, Morris}. Systematics from reduction pipelines may also bias astrophysical parameters, potentially leading to inconsistencies across entire planet populations. For example, \citet{Han_Robertson} predict that the radii of hundreds of TESS exoplanets may be underestimated by $\sim6\%$ due to background stars blending with the light from a host star. Underestimating a planet's radius will result in shallower eclipse depth models that may lead to incorrect constraints on the planet's composition. 



\subsection{Modeling Limitations}\label{sec:modelinglimitations}

There are a number of caveats that remain in our modeling schematic. Many of these issues stem from the use of stellar models when simulating the eclipse depth of each planetary iteration (see Section \ref{sec:stellarspectrum}). Interpolating across the chosen stellar grid enables a large degree of variation in the SED of the simulated host star, allowing us to estimate the model uncertainty for systems with a poorly characterized host. However, there is no physical justification in assuming that the true SED is proportional to any of these stellar models. These interpolated models also assume that the host star’s flux is constant in time. Variations in the stellar flux will result in a nonconstant surface flux for a bare rock planet, though these are difficult to quantify mathematically.

\begin{figure}[]
\centering
\includegraphics[width=0.9\linewidth]{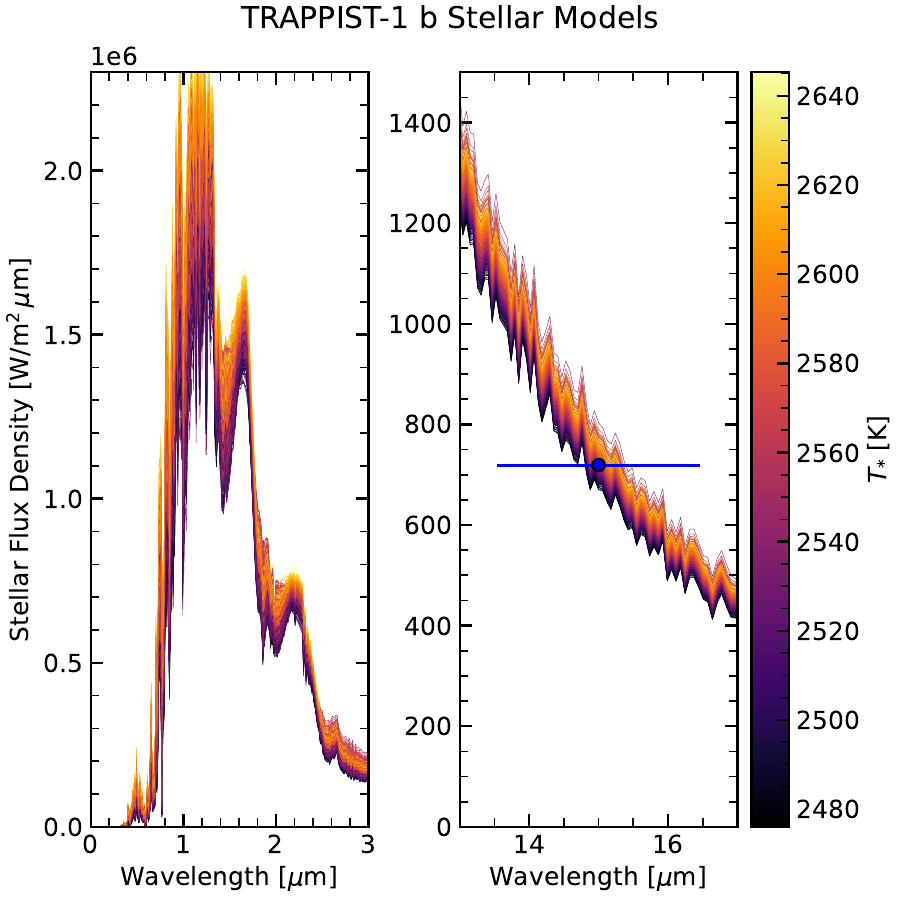}
\caption{SED generated across all 2,000 models of TRAPPIST-1 when sampling parameters that modify the stellar flux density ($T_{\mathrm{*}}$, log(g), [M/H], and C/O). The colorbar indicates the model's corresponding $T_{\mathrm{*}}$ value. The blue marker indicates the uniform stellar flux density most consistent with the absolute calibrated flux of 2.528 mJy of TRAPPIST-1 as measured by \citet{Ducrot_Lagage_Min_Gillon_Bell_Tremblin_Greene_Dyrek_Bouwman_Waters_etal._2025} in the F1500W filter.}

\label{fig:T1bSEDvariation}
\end{figure}

Sampling a new SED for every planet's iteration will introduce uncertainty for systems with a well characterized host star, which may offset the results significantly. For example, \citet{Hu_Bello-Arufe_Zhang_Paragas_Zilinskas_Van_Buchem_Bess_Patel_Ito_Damiano_et_al._2024} generate a stellar flux model for 55 Cnc using measurements from MIRI LRS and a calibrated flux model of the host star from \citet{Crossfield_2012}. This stellar model differs from our sampled range of PHOENIX models generated for 55 Cnc, leading to differences in the interpreted dayside brightness temperature and calculated value of $\mathcal{R}$ \citep{Lin_Daylan_2026}. In similar fashion, the SED of TRAPPIST-1 is well constrained in the NIR and across two broadband MIR filters from different JWST observations, though we choose to sample the SED for all iterations of TRAPPIST-1 b and TRAPPIST-1 c to maintain consistency across our reanalysis \citep[][Figure \ref{fig:T1bSEDvariation}]{fauchez2025stellarmodelslimitexoplanet}. Future work will further investigate the importance of a well-characterized SED in the generation of forward emission models.

\begin{figure}[t!]
\centering
\includegraphics[width=0.9\linewidth]{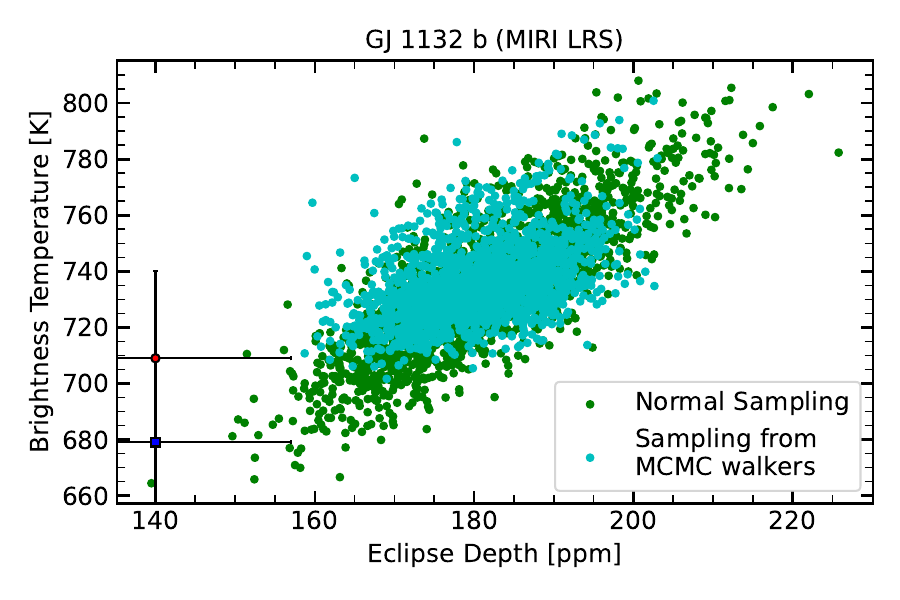}
\caption{Simulated eclipse and brightness temperature of GJ 1132\,b with MIRI LRS using two distinct parameter sampling methods. Green samples were generated by sampling each physical and stellar parameter independently from a gaussian distribution, while cyan samples were generated by sampling from the \texttt{ExoFASTv2} chains provided in \citet{Xue_Bean_Zhang_Mahajan_Ih_Eastman_Lunine_Mansfield_Coy_Kempton_et_al._2024}. The red dot indicate the observed eclipse depth and observed brightness temperature from \citet{Xue_Bean_Zhang_Mahajan_Ih_Eastman_Lunine_Mansfield_Coy_Kempton_et_al._2024}, while the blue dot indicates the observed eclipse depth and derived brightness temperature as calculated using the distribution of relevant astrophysical parameters.}

\label{fig:GJ1132bsamplingtime}
\end{figure}

Our choice of assuming an independent, normal distribution for each relevant parameter in our modeling schematic may also fail to account for parameter covariances and asymmetric error bars. Ideally, our framework would account for existing correlations by directly sampling the chain/walkers from which the parameter bounds are sourced from, rather than assuming each follows an uncorrelated normal distribu- tion. We investigate the impact of such covariances using the parameter samples provided by \citet{Xue_Bean_Zhang_Mahajan_Ih_Eastman_Lunine_Mansfield_Coy_Kempton_et_al._2024} for GJ 1132\,b from their \texttt{ExoFASTv2} analysis of the physical and stellar conditions. We repeat our modeling analysis of GJ 1132\,b as described in Section \ref{sec:reanalysis} with these samples. 

Comparing these model results to our those when sampling each parameter from a normal distribution reveals that, while sampling directly from the \texttt{ExoFASTv2} samples results in less model variation, the two schematics generate similar bare rock eclipse depths and brightness temperatures (Figure \ref{fig:GJ1132bsamplingtime}). Using these results instead of those derived from out independly-sampled parameters thus generates a nearly identical estimation of $\mathcal{R}$. Although it's clear that the covariances in the parameters of GJ 1132\,b do not significantly impact the validity of our model results, further analysis of other targets in our survey would be necessary to confirm whether this is consistent for all planets.

\begin{figure}[t!]
\centering
\includegraphics[width=0.9\linewidth]{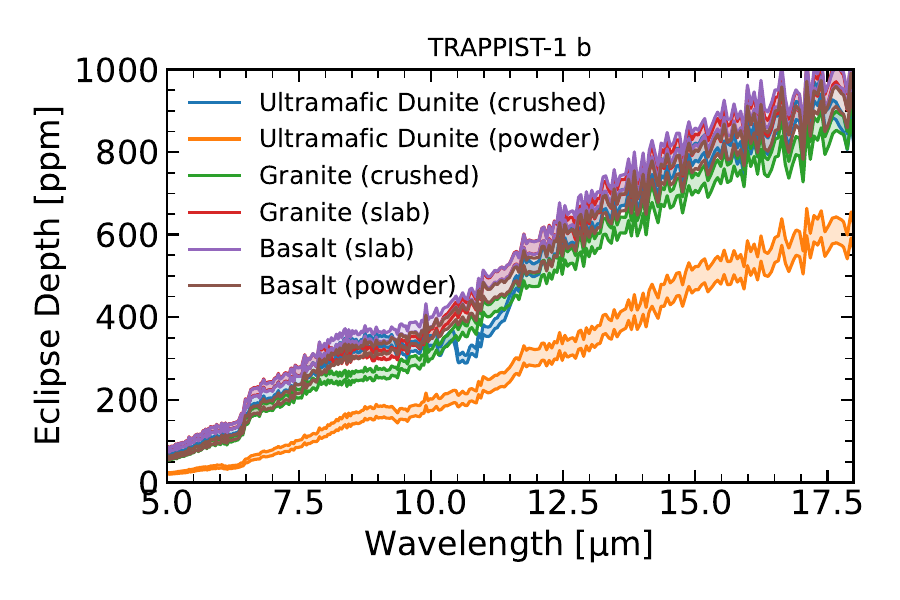}
\caption{Eclipse spectra of various surface models from \citet{Paragas_Knutson_Hu_Ehlmann_Alemanno_Helbert_Maturilli_Zhang_Iyer_Rossman_2025} as simulated by \texttt{JESTER} for TRAPPIST-1\,b, assuming uniform single-scattering albedos of 0 and 1 for $\lambda < 0.4\mu$m and $\lambda>20\mu$m. The range of potential models between the two extremes is shown in the light colors.}

\label{fig:Albedos}
\end{figure}

Beyond the treatment of the stellar flux and parameter sampling, there are a number of approximations and simplifications required in order to generate bare rock models, even with our improved treatment of the surface temperature. The conversions presented in \citet{Hapke_2012} for calculating $r_{\mathrm{dh}}(\lambda)$, $\epsilon_{\mathrm{h}}(\lambda)$ and $\epsilon_{\mathrm{d}}(\lambda)$ from the single-scattering albedo assume that the particles scatter isotropically and do not account for surface roughness, thermal beaming, or space weathering. Furthermore, the impact of the opposition effect and surface porosity is assumed to be negligible \citep{2001JGR...10610039H, Paragas_Knutson_Hu_Ehlmann_Alemanno_Helbert_Maturilli_Zhang_Iyer_Rossman_2025}. Additionally, the presence of a solid-state greenhouse or anti-greenhouse effect formed by subsurface temperature gradients within the planet's regolith is neglected \citep{lyu2025impactsubsurfacetemperaturegradients}. Finally, these models also assume that the emissivity and reflectivity of the surface is independent of the planet's temperature. In reality, the emissivity of various substances are known to vary depending on a number of parameters, including the surface temperature. \citep{2013E&PSL.371..252H, 2020E&PSL.53416089F,  2021PSJ.....2...43T, 2021Icar..35414040P, First_Mishra_Gazel_Lewis_Letai_Hanssen_2025, Paragas_Knutson_Hu_Ehlmann_Alemanno_Helbert_Maturilli_Zhang_Iyer_Rossman_2025}. The inclusion of such effects would provide further uncertainty to the validity of existing bare rock models.

Our surface models also rely on a consistent energy balance between the incident light from the host star and the sum of the planet's reflected and re-radiated light. Balancing the energy budget requires accounting for the entirety of the bolometric wavelength space even when simulating the emissions within a single filter. Thus, simulating the emissions from planets with a non-uniform emissivity becomes difficult. As described in Section \ref{sec:jester}, we assume a uniform value of $\omega(\lambda)$ for values of $\lambda$ outside of the included model. Depending on the true values of $w(\lambda)$ outside of the model boundaries, the true temperature gradient for each surface composition may vary, resulting in a range of potential eclipse depths for each model (Figure \ref{fig:Albedos}).



Finally, although we focus on the use of energy- balanced bare rock models to investigate the propagation of model uncertainty, additional sources of uncertainty may arise when generating different kinds of forward models. The inclusion of atmospheres, clouds, hazes, molten surfaces, and interior heating will all introduce further complexity into the generation of forward models. Nevertheless, model uncertainty will still be injected into these models, affecting the accuracy of our conclusions.

\subsection{Future Observations \& Recommendations}\label{sec:future}

Future observations will continue to fill in the unexplored parameter space in Figure \ref{fig:tirr}, which may further clarify the observed trends and explore the transitional region between an outgassing lava world and a partially melted bare rock. In particular, GO 4818 (PI: Mansfield) will perform MIRI LRS emission spectroscopy on 10 candidate lava worlds spanning multiple stellar classes and may further investigate the nature of highly irradiated rocky worlds. Furthermore, the tenuous trend observed in M-dwarf rocky exoplanets may be further elucidated by the ongoing Rocky Worlds DDT \citep{Redfield_Batalha_Benneke_Biller_Espinoza_France_Konopacky_Kreidberg_Rauscher_Sing_2024} and Hot Rocks Survey \citep{2023jwst.prop.3730D}.

As the JWST continues to observe rocky exoplanets in emission, it is critical that we begin properly accounting for the model uncertainty when characterizing exoplanets using emission spectroscopy. The impact of model uncertainty can be limited by precise characterization of the system's astrophysical parameters and the host star's bolometric SED \citep[e.g.,][]{Mahajan_2024}. In particular, model uncertainty introduced by the use of an imprecise stellar model can be limited by collecting the host star's bolometric SED, which is possible using a number of instruments on the JWST \citep{fauchez2025stellarmodelslimitexoplanet}. For planets without a precisely characterized stellar spectrum, equations \ref{eq:eclipsedepthunceratainty} and \ref{eq:xerr} can be used to estimate the model uncertainty in forward models for all exoplanets, including those with an atmosphere, using the best-fitting values for $a$ and $b$ derived in Figure \ref{fig:Fpdepallfilters}. Furthermore, the model uncertainty can still be estimated in systems with a well-defined stellar spectrum by assuming that $\Delta T_{\mathrm{*}}$ is negligible in Equation \ref{eq:xerr}. 


\section{Conclusion} \label{sec:con}
In this work, we presented a framework that enables us to account for the propagation of stellar and orbital uncertainties when producing forward models of emission spectra, and use this framework to reanalyze the results from published thermal emission observations of rocky exoplanets. In particular, we compare the published eclipse depths of each planet to dark bare rock models as simulated by \texttt{JESTER}, which accounts for the temperature gradient across the planet's dayside by solving for the localized energy balance. For each planet, we generated 2,000 models of the planet and host star by sampling relevant astrophysical parameters and generating a dark bare-rock model for each planet-star iteration. We then calculated the expected eclipse depth of the planet as a dark bare rock alongside the model uncertainty from the distribution of simulated depths. Our analysis presents the model uncertainty as a strongly limiting factor when constraining the composition of exoplanets from emission-based studies, and impacts the validity of both bare rock and atmospheric model spectra. In particular, the presence of model uncertainty causes a number of potential surface compositions to become degenerate with one another, further complicating our capabilities to constrain a planet's composition through eclipse observations. 

Repeating our modeling schematic for the candidate targets from the Rocky Worlds DDT reveals a linear correlation between the model uncertainty and the error in $R_{\mathrm{p}}/R_{\mathrm{*}}$, $ a_{\mathrm{p}}/R_{\mathrm{*}}$, and $T_{\mathrm{*}}$. The relationship between these variables may be utilized in future eclipse analyses to provide a rough estimate of uncertainty for both atmospheric and surface toy models, therefore enabling a more honest characterization of the planet's composition. 

Model uncertainty can be mitigated by providing more precise constraints on the system's astrophysical parameters and through proper characterization of the host star's SED. Furthermore, the impact of model uncertainty is largely uniform across wavelength space and maintains the shape and location of spectral features. Therefore, compositional constraints may still be defined through the use of emission spectroscopy.

Our uniform reinterpretation of existing results further elucidates a tentative trend seen in rocky planets orbiting M-dwarfs between the planet's irradiation temperature and the brightness temperature ratio, where $\mathcal{R}$ tends to decrease with colder temperatures \citep{Coy_Ih_Kite_Koll_Tenthoff_Bean_WeinerMansfield_Zhang_Xue_Kempton_etal._2025, Lin_Daylan_2026}. This trend, however, remains tenuous, and necessitates further observation and study. Upcoming eclipse observations by the JWST may further clarify the nature of rocky exoplanets by probing regions of the parameter space that remain unexplored. Future emission-based studies must account for the model uncertainty to avoid the development of erroneous conclusions when implementing the use of forward modeling. By accounting for such uncertainty, we will ultimately advance our understanding of the compositions of rocky exoplanets.

\section{Acknowledgments}

This work made use of the NASA Exoplanet Archive, which is operated by the California Institute of Technology, under contract with NASA under the Exoplanet Exploration Program. C.M., N.C. and P.-A.R. acknowledge financial support from the University of Montreal, and C.M. further acknowledges financial support from Jean-Marc Lauzon. C.M., B.B., and P.-A.R. acknowledge financial support from the Natural Sciences and Engineering Research Council (NSERC) of Canada. C.M., N.C., B.B., and P.-A.R. acknowledge financial support from the Canadian Space Agency (CSA). This work was made with the support of the Institut Trottier de Recherche sur les Exoplanetes (iREx). C.M. thanks Kimberly Paragas for an insightful discussion regarding the energy balance of bare rock exoplanets, and Charles Cadieux, Hannah Diamond-Lowe, Brandon Park-Coy, and Michiel Min for providing valuable suggestions that greatly enhanced the quality of our modeling schematic.

We used the following code resources in our data analysis: \verb!astropy! \citep{astropy:2013, astropy:2018, astropy:2022}, \verb!NumPy! \citep{harris2020array_numpy}, \verb!matplotlib! \citep{matplotlib}, and \verb!SciPy! \citep{2020SciPy-NMeth}. This research has made use of the SVO Filter Profile Service by Carlos Rodrigo, funded by MCIN/AEI/10.13039/ 501100011033/ through grant PID2023-146210NB-I00.

\appendix

\begin{figure*}[h!]
  \centering
      \includegraphics[width=1.0\linewidth]{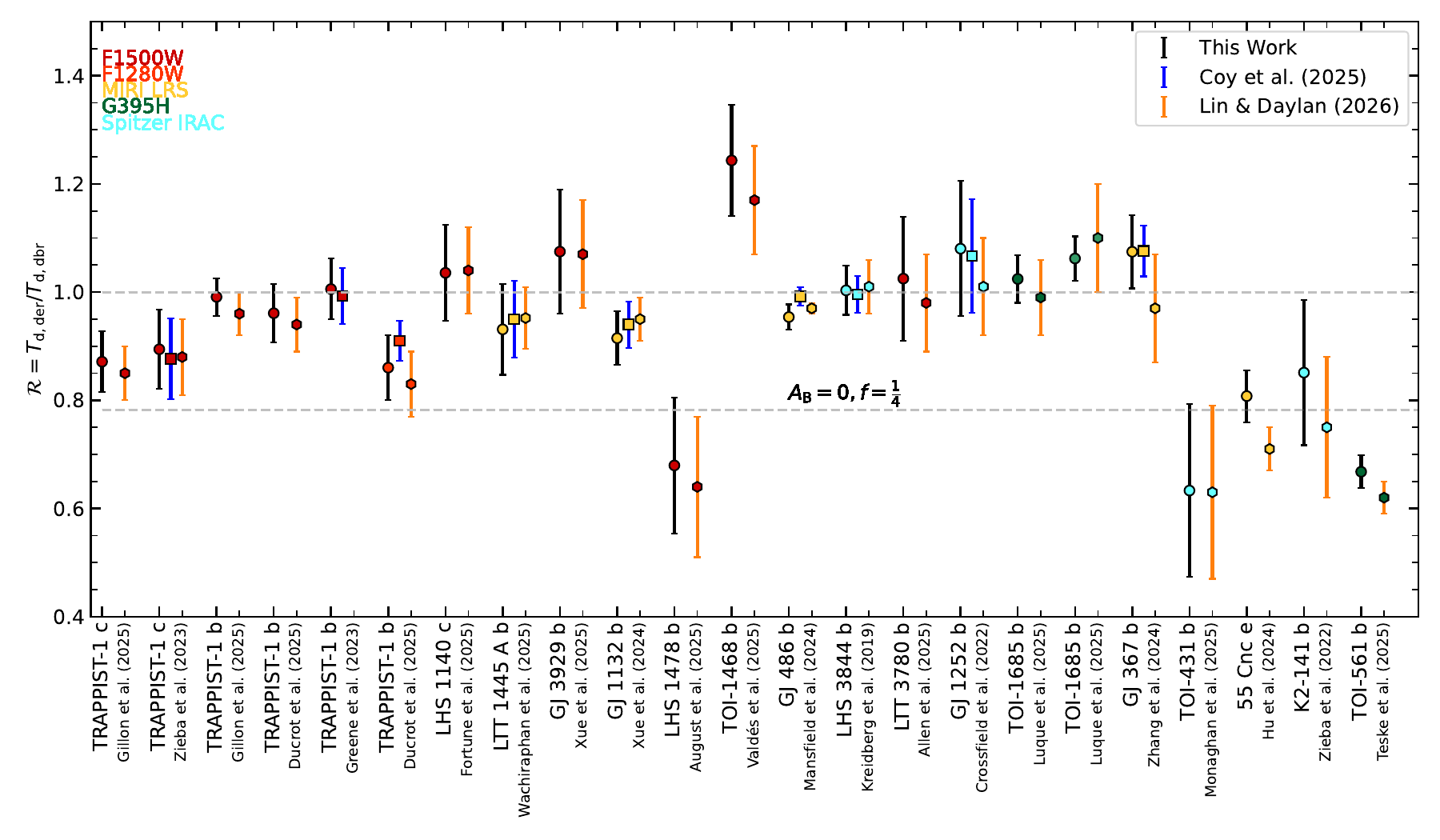} 
  \caption{Comparison of our $\mathcal{R}$ values with those calculated in \citet{Coy_Ih_Kite_Koll_Tenthoff_Bean_WeinerMansfield_Zhang_Xue_Kempton_etal._2025} and \citet{Lin_Daylan_2026}. Most results agree to $\sim 1 \sigma$, but our error bars are often larger due to the inclusion of the model uncertainty in our analysis.}
  \label{fig:comparisontopastresults}
  
\end{figure*}


\begin{figure}[]
\centering
\begin{subfigure}{}
\includegraphics[width=0.45\linewidth]{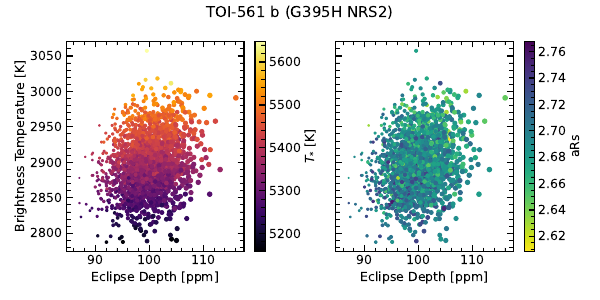}
\end{subfigure}
\begin{subfigure}{}
\includegraphics[width=0.45\linewidth]{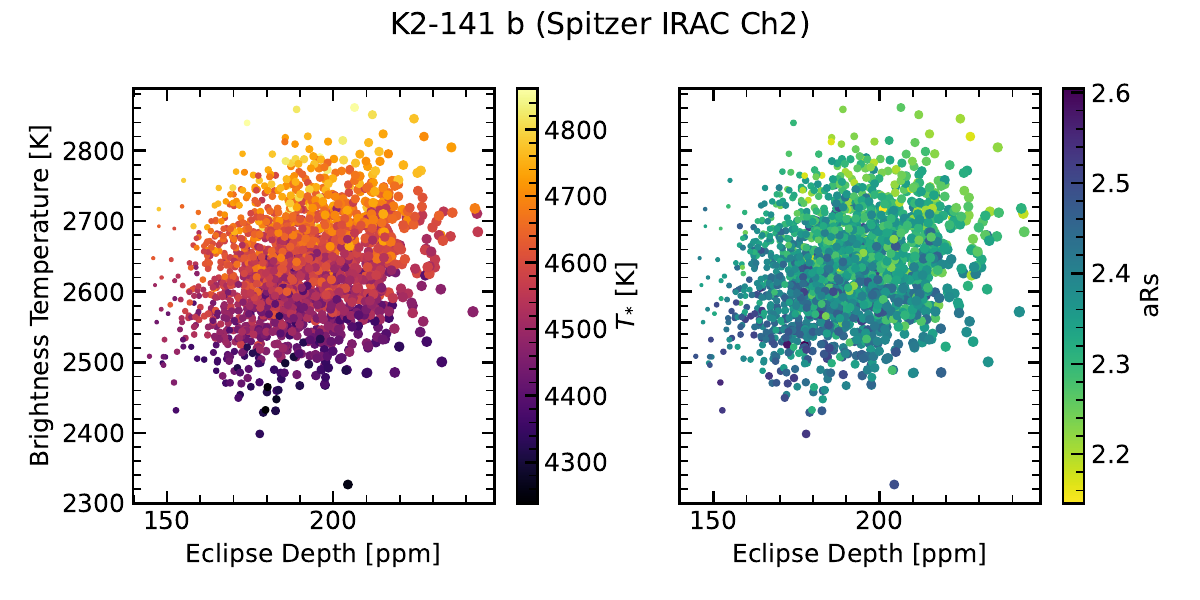}
\end{subfigure}
\begin{subfigure}{}
\includegraphics[width=0.45\linewidth]{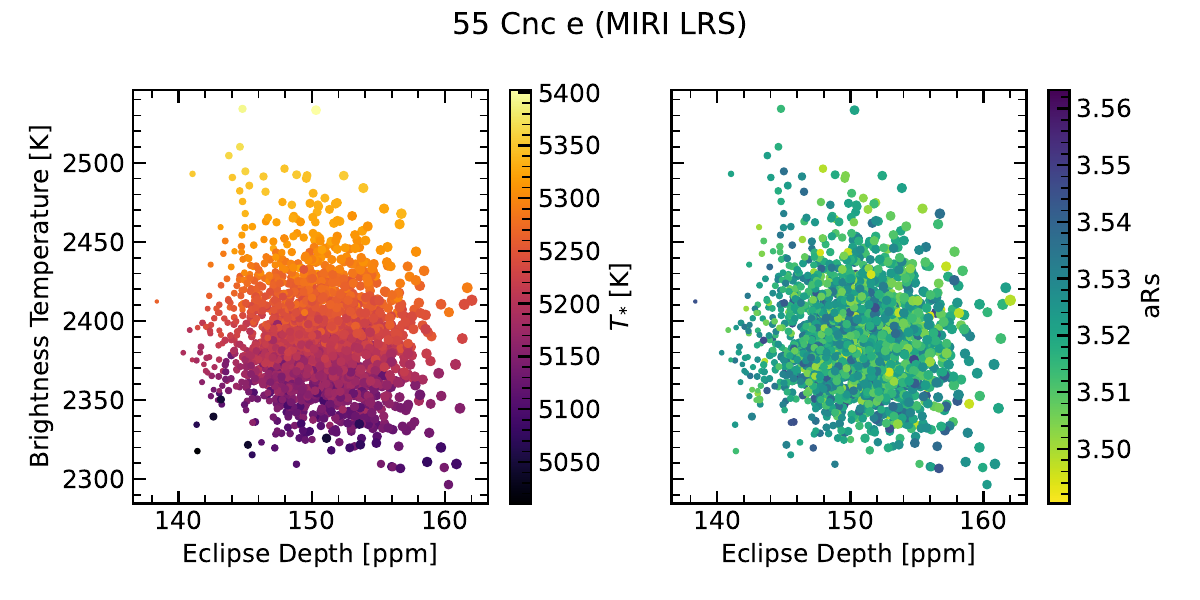}
\end{subfigure}
\begin{subfigure}{}
\includegraphics[width=0.45\linewidth]{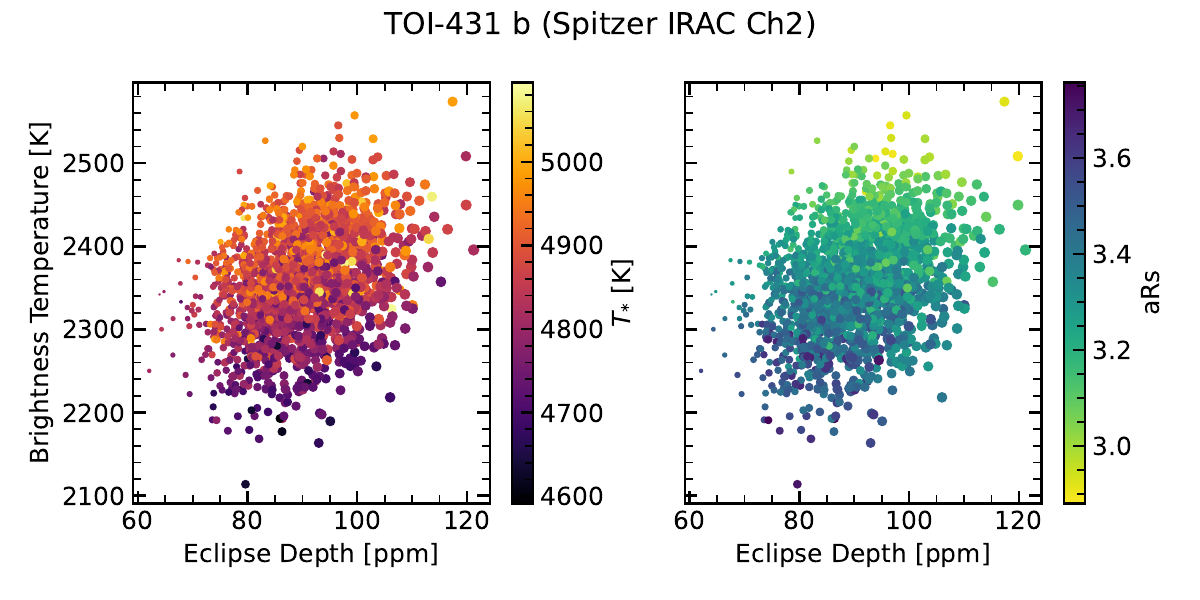}
\end{subfigure}
\begin{subfigure}{}
\includegraphics[width=0.45\linewidth]{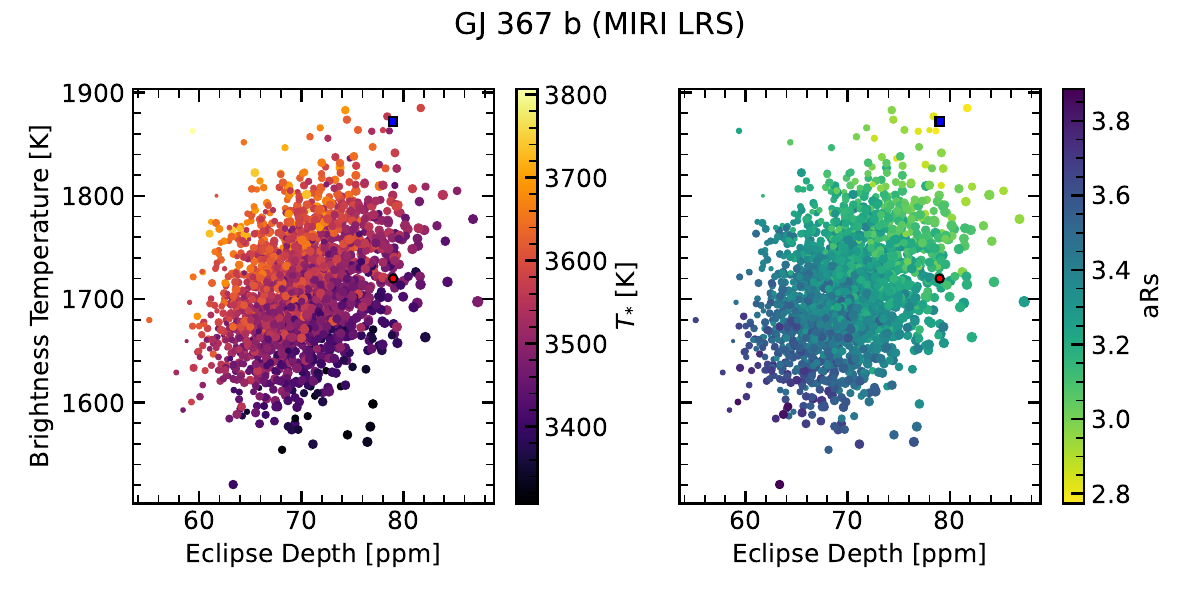}
\end{subfigure}
\begin{subfigure}{}
\includegraphics[width=0.45\linewidth]{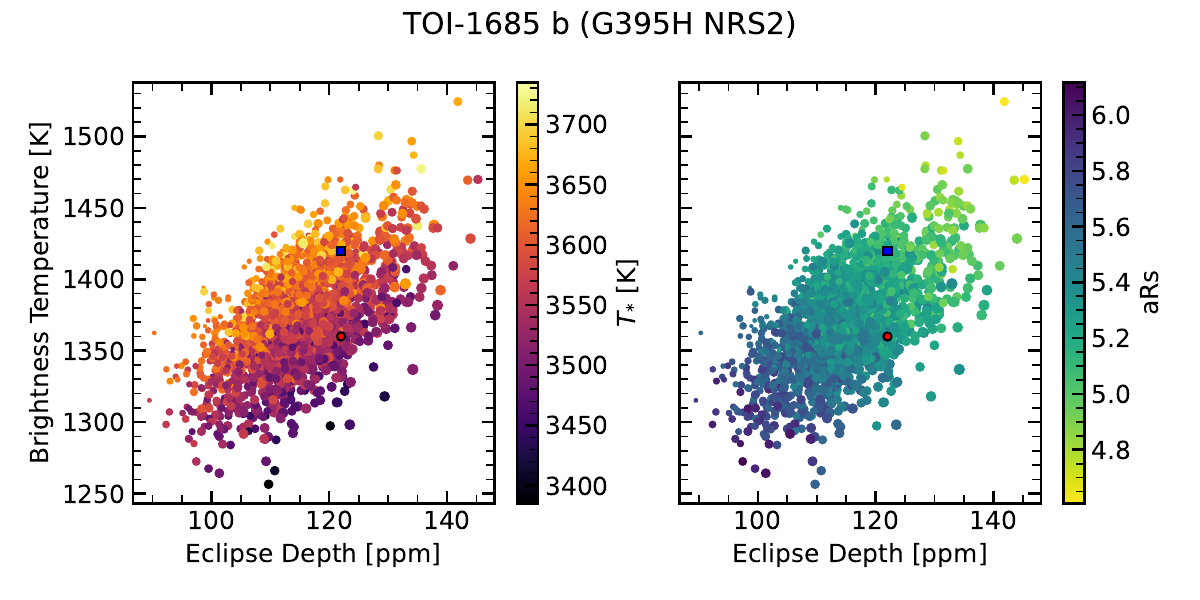}
\end{subfigure}
\begin{subfigure}{}
\includegraphics[width=0.45\linewidth]{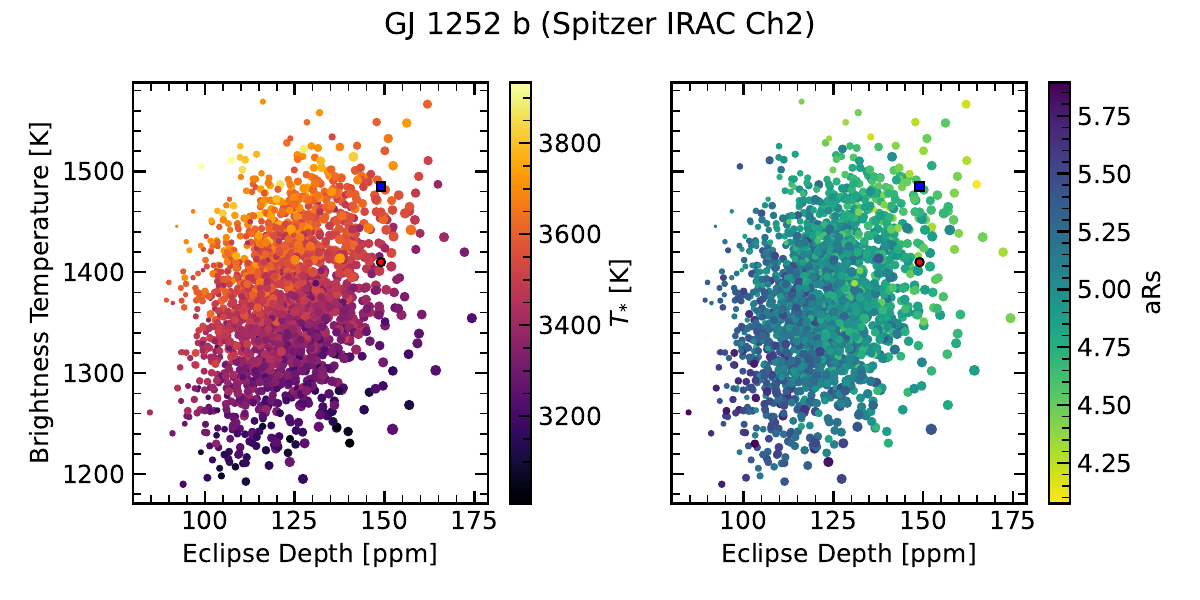}
\end{subfigure}
\begin{subfigure}{}
\includegraphics[width=0.45\linewidth]{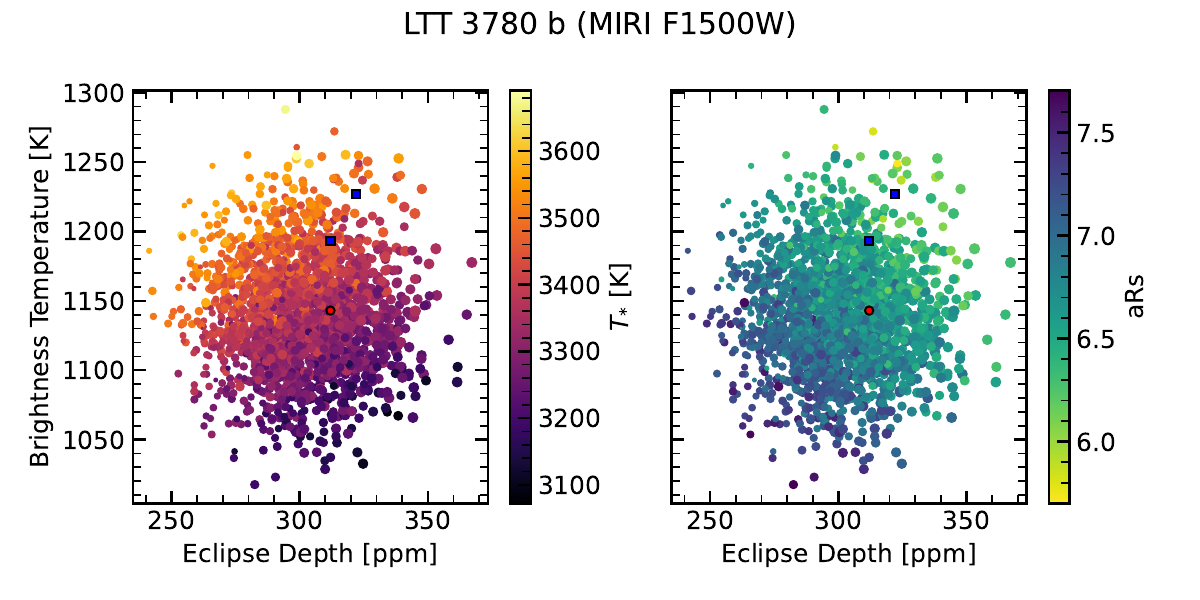}
\end{subfigure}

\caption{Simulated eclipse and brightness temperature of all 2,000 models simulated for each planet, using the filter associated with the bolded observations in Table \ref{tab:moderesults}. Marker size scales with the model's $R_{\mathrm{*}}$ on the leftmost axis, and $R_{\mathrm{p}}/R_{\mathrm{*}}$ on the rightmost axis. Observed results are overplotted if they are within the regime of $F_{\mathrm{p}} / F_{\mathrm{*}}$ and $T_{\mathrm{d}}$ explored by the 2,000 models. Red dots indicate the observed eclipse depth and observed brightness temperature from existing publications, while blue squares indicate the observed eclipse depth and derived brightness temperature as calculated using the distribution of relevant astrophysical parameters. Error bars are ignored for clarity.}

\label{fig:auxmodels}
\end{figure}

\clearpage

\begin{figure}[] 
\centering
\begin{subfigure}{}
\includegraphics[width=0.45\linewidth]{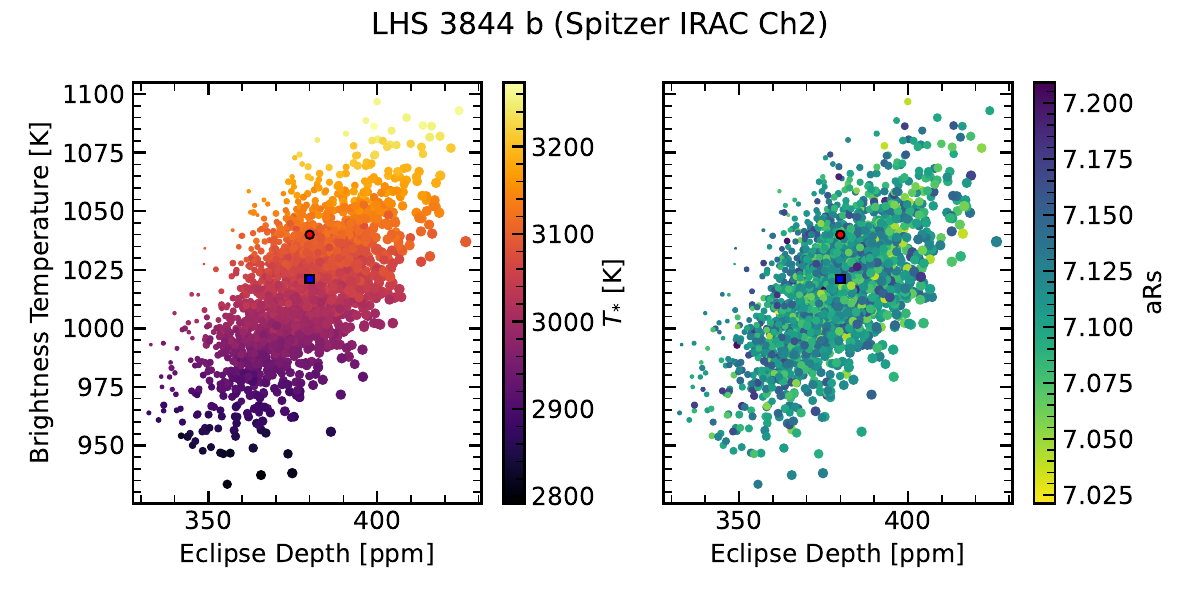}
\end{subfigure}
\begin{subfigure}{}
\includegraphics[width=0.45\linewidth]{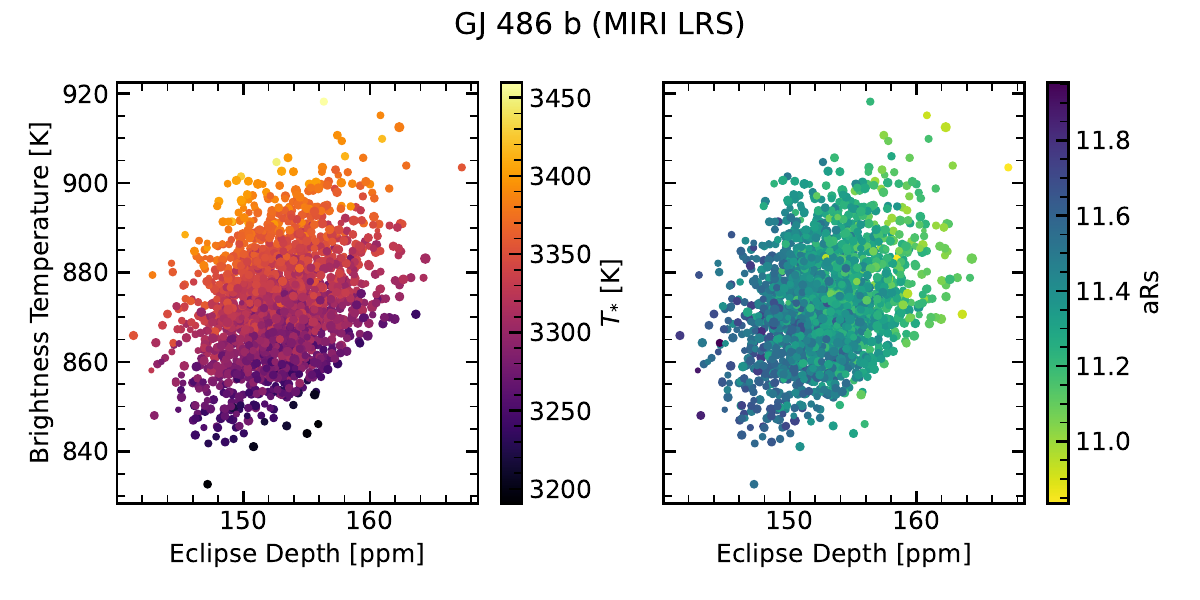}
\end{subfigure}
\begin{subfigure}{}
\includegraphics[width=0.45\linewidth]{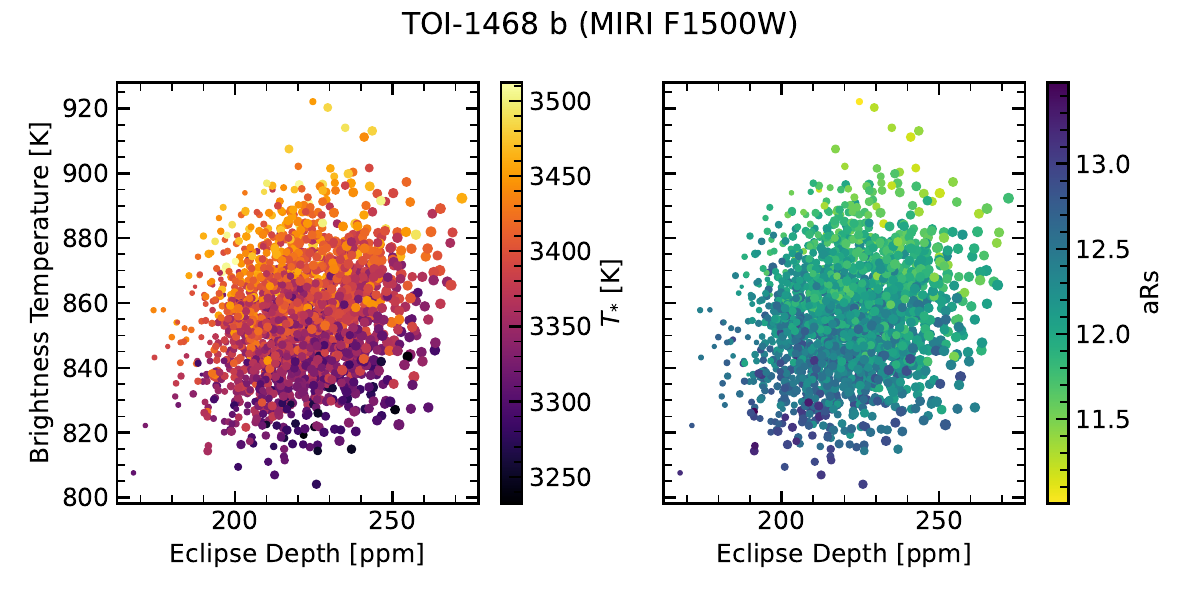}
\end{subfigure}
\begin{subfigure}{}
\includegraphics[width=0.45\linewidth]{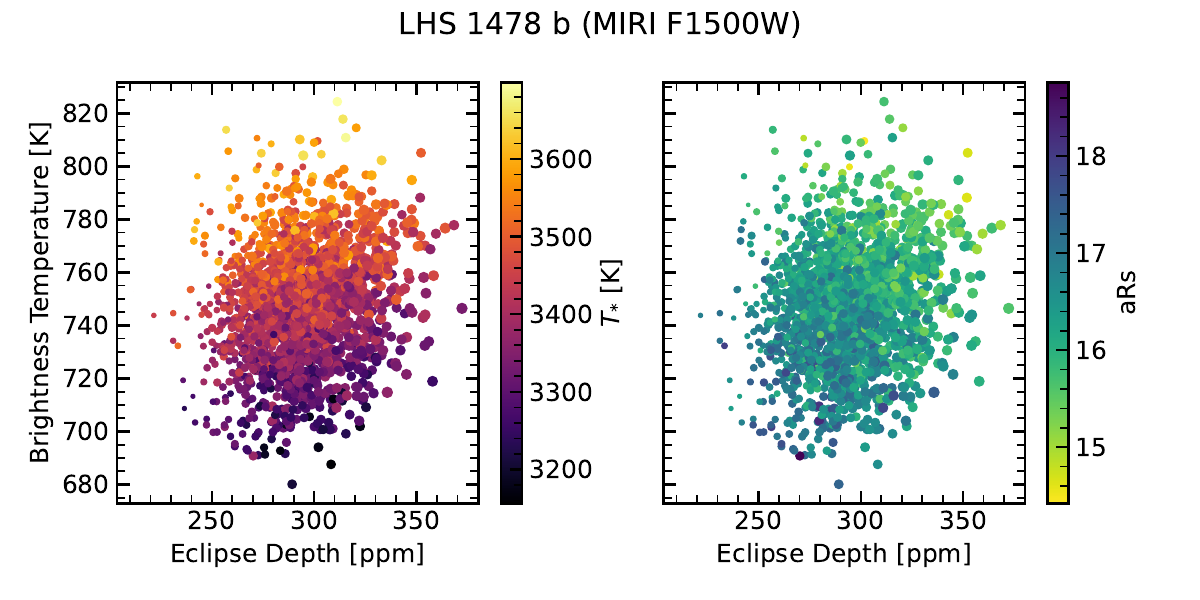}
\end{subfigure}
\begin{subfigure}{}
\includegraphics[width=0.45\linewidth]{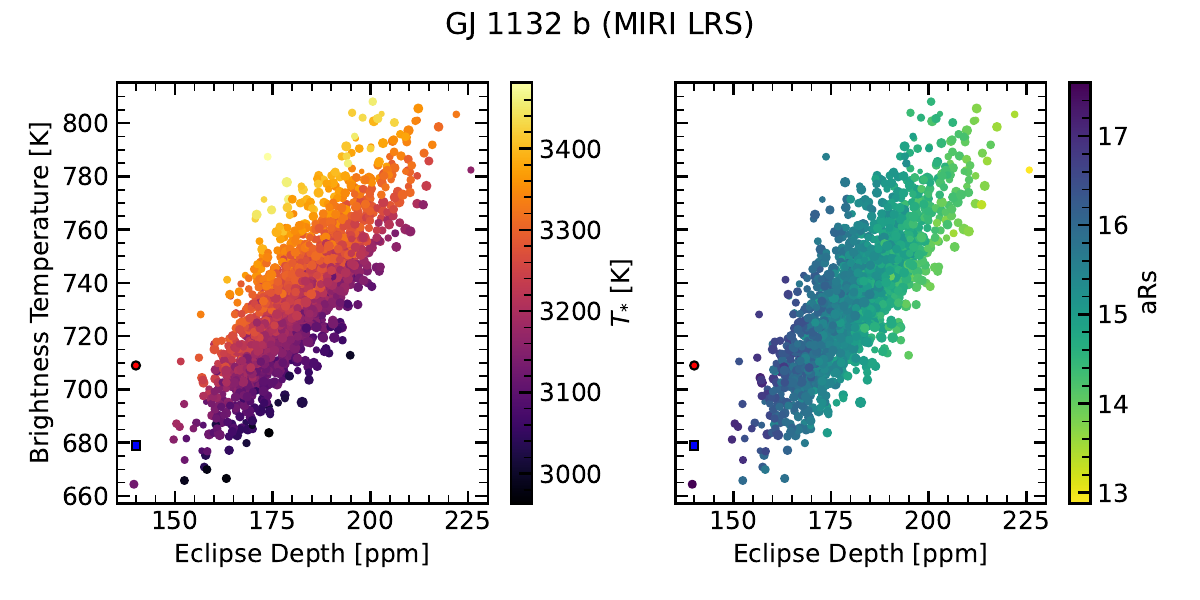}
\end{subfigure}
\begin{subfigure}{}
\includegraphics[width=0.45\linewidth]{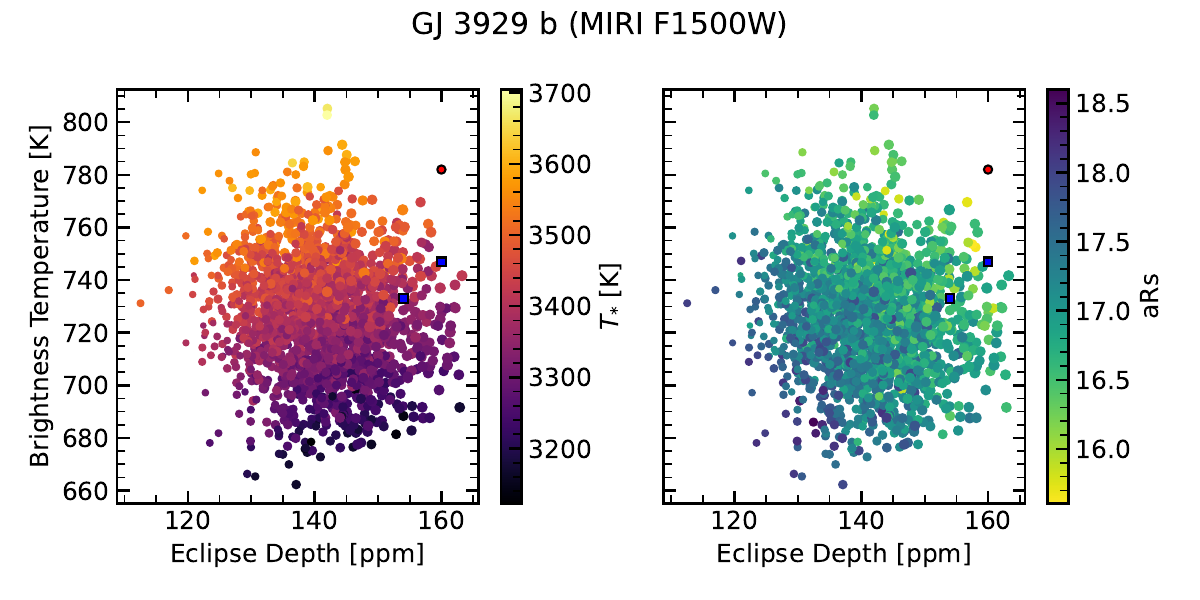}
\end{subfigure}
\begin{subfigure}{}
\includegraphics[width=0.45\linewidth]{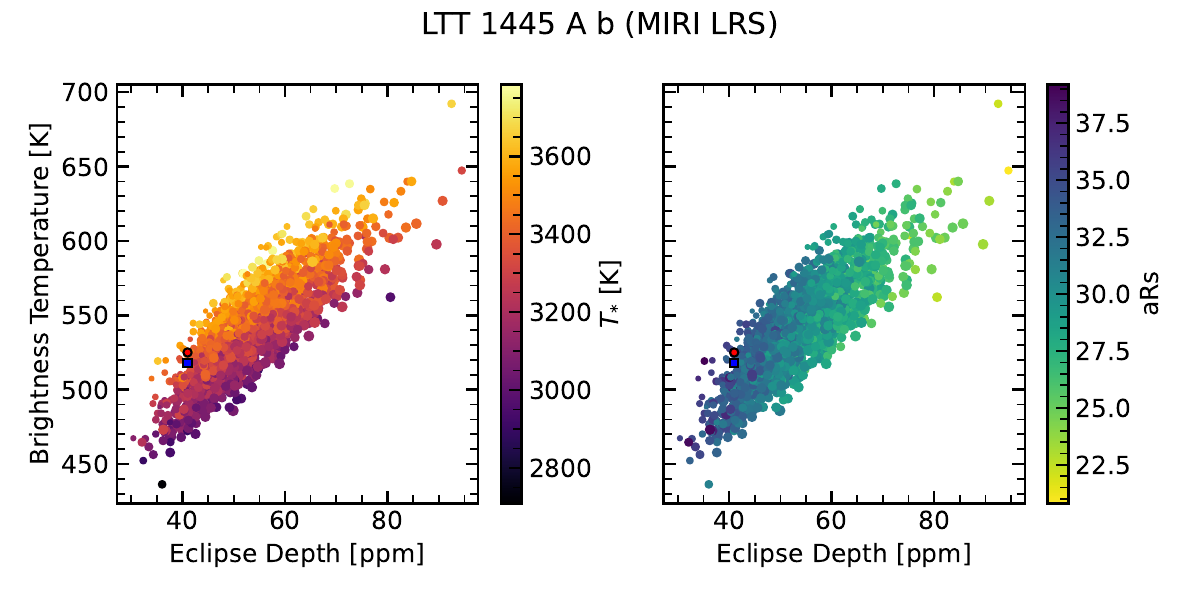}
\end{subfigure}
\begin{subfigure}{}
\includegraphics[width=0.45\linewidth]{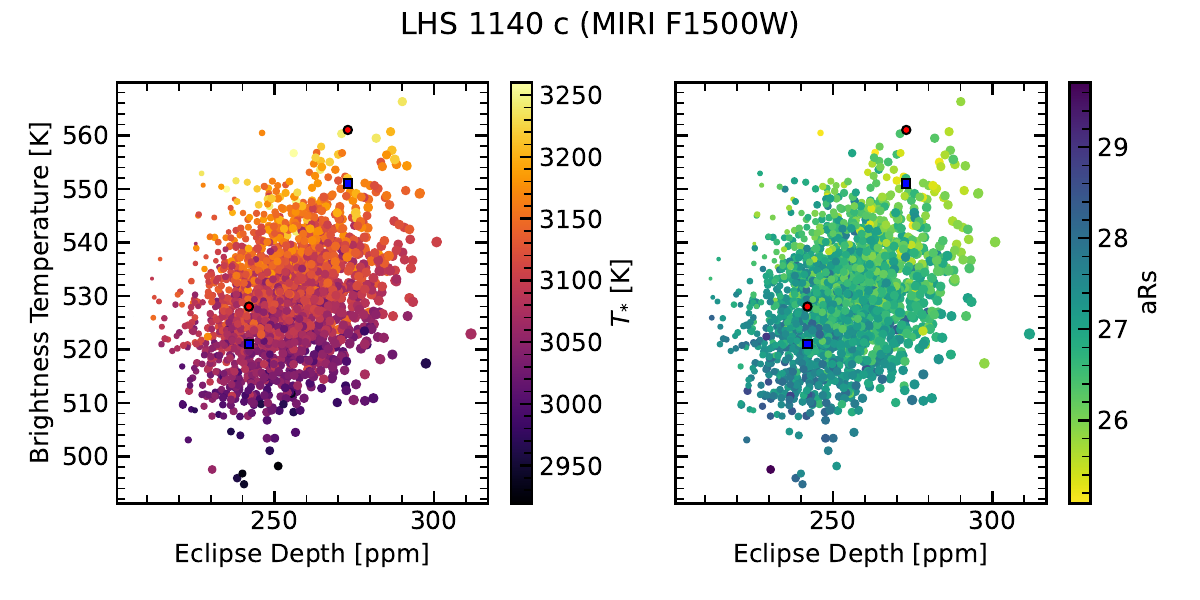}
\end{subfigure}
\begin{subfigure}{}
\includegraphics[width=0.45\linewidth]{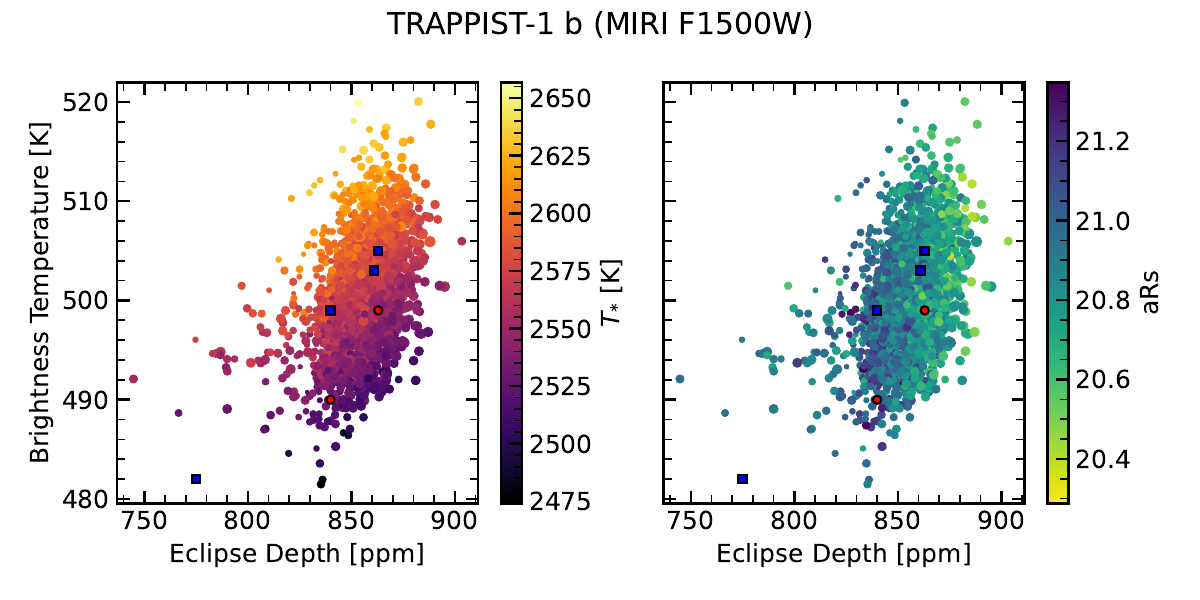}
\end{subfigure}
\begin{subfigure}{}
\includegraphics[width=0.45\linewidth]{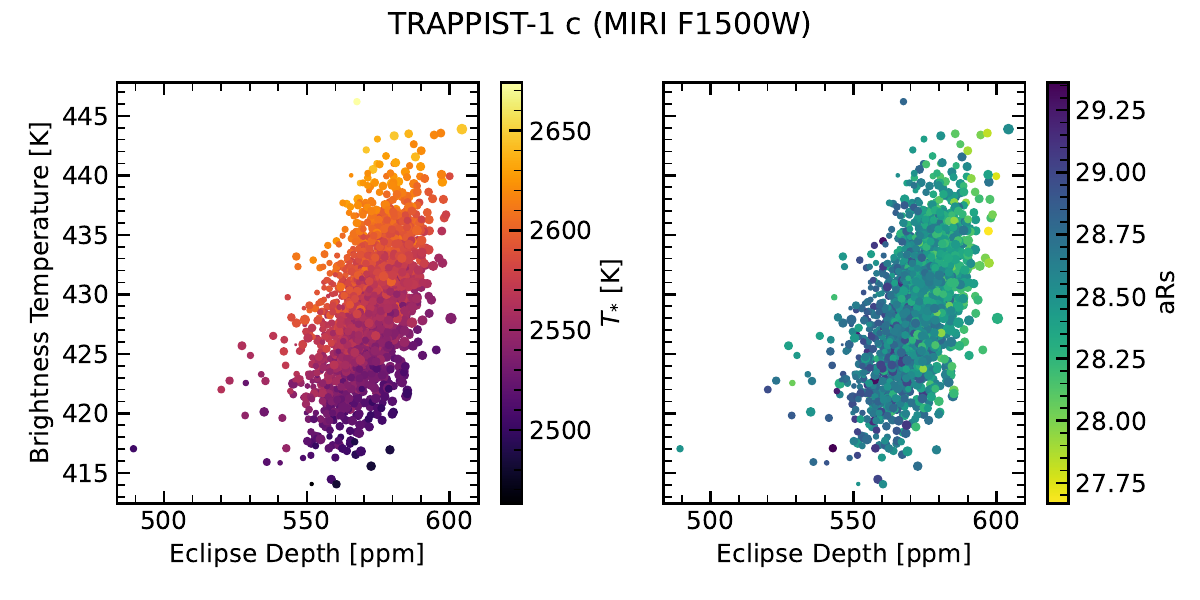}
\end{subfigure}
\caption{Simulated eclipse and brightness temperature of all 2,000 models simulated for each planet (cont.)}
\label{fig:auxmodels2}
\end{figure}

\clearpage


\begin{table*}[]
    \centering
    {
    \footnotesize
    \begin{tabular}{>{\bfseries}l|c c c | c | c c | c}
    \hline
    \hline

         & \multicolumn{6}{c|}{\textbf{Dark Bare Rock}}&   \\             & \multicolumn{6}{c|}{$F_{\mathrm{p}}/F_{\mathrm{s}}$ (ppm)} 
        
        \\ \hline \textbf{Planet} & \multicolumn{3}{c|}{\textbf{MIRI}} & \textbf{Spitzer} & \multicolumn{2}{c|}{\textbf{NIRSpec G395H}} & $T_{\mathrm{irr}}$\\ 
         & F1500W & F1280W & LRS & IRAC Ch2 & NRS1 & NRS2 & K\\ 

    \hline
    \hline
TOI-561\,b    & 139 $\pm$ 5   & 135 $\pm$ 5   & 119 $\pm$ 4  & 101 $\pm$ 4  & 99 $\pm$ 3   & 82 $\pm$ 3   & 3261 $\pm$ 50 \\
K2-141\,b     & 261 $\pm$ 23  & 254 $\pm$ 22  & 221 $\pm$ 20 & 188 $\pm$ 17 & 182 $\pm$ 17 & 144 $\pm$ 13 & 2976 $\pm$ 78 \\
55 Cnc\,e     & 174 $\pm$ 4   & 169 $\pm$ 4   & 150 $\pm$ 3  & 119 $\pm$ 2  & 115 $\pm$ 2  & 89 $\pm$ 2   & 2833 $\pm$ 41 \\
TOI-431\,b    & 133 $\pm$ 12  & 129 $\pm$ 11  & 110 $\pm$ 10 & 89 $\pm$ 8   & 86 $\pm$ 8   & 66 $\pm$ 6   & 2660 $\pm$ 62 \\
GJ 367\,b     & 89 $\pm$ 5    & 86 $\pm$ 5    & 69 $\pm$ 4   & 51 $\pm$ 3   & 48 $\pm$ 3   & 40 $\pm$ 3   & 1930 $\pm$ 59 \\
TOI-1685\,b   & 265 $\pm$ 15  & 253 $\pm$ 15  & 191 $\pm$ 12 & 121 $\pm$ 8  & 114 $\pm$ 8  & 84 $\pm$ 7   & 1542 $\pm$ 39 \\
GJ 1252\,b    & 266 $\pm$ 27  & 254 $\pm$ 25  & 192 $\pm$ 19 & 121 $\pm$ 13 & 114 $\pm$ 12 & 84 $\pm$ 10  & 1541 $\pm$ 74 \\
LTT 3780\,b   & 299 $\pm$ 20  & 281 $\pm$ 18  & 197 $\pm$ 13 & 108 $\pm$ 8  & 101 $\pm$ 8  & 67 $\pm$ 6   & 1291 $\pm$ 43 \\
LHS 3844\,b   & 1226 $\pm$ 42 & 1127 $\pm$ 37 & 755 $\pm$ 26 & 376 $\pm$ 15 & 345 $\pm$ 15 & 210 $\pm$ 12 & 1141 $\pm$ 37 \\
GJ 486\,b     & 281 $\pm$ 6   & 256 $\pm$ 5   & 152 $\pm$ 3  & 63 $\pm$ 1   & 58 $\pm$ 1   & 29 $\pm$ 1   & 984 $\pm$ 13  \\
TOI-1468\,b   & 219 $\pm$ 16  & 199 $\pm$ 15  & 118 $\pm$ 9  & 48 $\pm$ 4   & 44 $\pm$ 4   & 22 $\pm$ 2   & 966 $\pm$ 18  \\
LHS 1478\,b   & 292 $\pm$ 23  & 259 $\pm$ 21  & 138 $\pm$ 12 & 46 $\pm$ 5   & 42 $\pm$ 5   & 17 $\pm$ 2   & 844 $\pm$ 24  \\
GJ 1132\,b    & 390 $\pm$ 16  & 344 $\pm$ 15  & 180 $\pm$ 11 & 59 $\pm$ 6   & 54 $\pm$ 5   & 22 $\pm$ 3   & 826 $\pm$ 25  \\
GJ 3929\,b    & 140 $\pm$ 7   & 124 $\pm$ 6   & 65 $\pm$ 3   & 20 $\pm$ 1   & 19 $\pm$ 1   & 7 $\pm$ 0    & 822 $\pm$ 26  \\
LTT 1445 A\,b & 170 $\pm$ 18  & 141 $\pm$ 17  & 52 $\pm$ 9   & 10 $\pm$ 3   & 8 $\pm$ 2    & 2 $\pm$ 1    & 607 $\pm$ 35  \\
LHS 1140\,c   & 253 $\pm$ 14  & 209 $\pm$ 12  & 84 $\pm$ 5   & 14 $\pm$ 1   & 12 $\pm$ 1   & 3 $\pm$ 0    & 597 $\pm$ 11  \\
TRAPPIST-1\,b & 855 $\pm$ 12  & 677 $\pm$ 12  & 277 $\pm$ 7  & 44 $\pm$ 1   & 37 $\pm$ 1   & 9 $\pm$ 0    & 562 $\pm$ 6   \\
TRAPPIST-1\,c & 571 $\pm$ 10  & 432 $\pm$ 9   & 148 $\pm$ 4  & 15 $\pm$ 0   & 12 $\pm$ 0   & 2 $\pm$ 0    & 480 $\pm$ 5 \\

\hline
\hline
    \end{tabular} }
    \caption{Simulated eclipse depth of each planet as a dark bare rock as derived with \texttt{JESTER} following Section \ref{sec:reanalysis}, sorted by the planet's irradiation temperature (Equation \ref{eq:irradiation temperature}). Values are rounded to the nearest integer. The depths shown here are used to calculate $\mathcal{F}$ and $\mathcal{R}$ in Table \ref{tab:moderesults}.}
    \label{tab:eclipsedepthresutls!}
\end{table*}


\begin{thebibliography}{}
\expandafter\ifx\csname natexlab\endcsname\relax\def\natexlab#1{#1}\fi
\providecommand{\url}[1]{\href{#1}{#1}}
\providecommand{\dodoi}[1]{doi:~\href{http://doi.org/#1}{\nolinkurl{#1}}}
\providecommand{\doeprint}[1]{\href{http://ascl.net/#1}{\nolinkurl{http://ascl.net/#1}}}
\providecommand{\doarXiv}[1]{\href{https://arxiv.org/abs/#1}{\nolinkurl{https://arxiv.org/abs/#1}}}

\bibitem[{Agol {et~al.}(2021)Agol, Dorn, Grimm, Turbet, Ducrot, Delrez, Gillon, Demory, Burdanov, Barkaoui, Benkhaldoun, Bolmont, Burgasser, Carey, de~Wit, Fabrycky, Foreman-Mackey, Haldemann, Hernandez, Ingalls, Jehin, Langford, Leconte, Lederer, Luger, Malhotra, Meadows, Morris, Pozuelos, Queloz, Raymond, Selsis, Sestovic, Triaud, \& Grootel}]{agol2021refiningtransittimingphotometric}
Agol, E., Dorn, C., Grimm, S.~L., {et~al.} 2021, Refining the transit timing and photometric analysis of TRAPPIST-1: Masses, radii, densities, dynamics, and ephemerides.
\newblock \doarXiv{2010.01074}

\bibitem[{Alexoudi {et~al.}(2020)Alexoudi, Mallonn, Keles, Poppenhäger, Essen, \& Strassmeier}]{Alexoudi_Mallonn_Keles_Poppenhager_Essen_Strassmeier_2020}
Alexoudi, X., Mallonn, M., Keles, E., {et~al.} 2020, Astronomy \& Astrophysics, 640, A134, \dodoi{10.1051/0004-6361/202038080}

\bibitem[{Allen {et~al.}(2025)Allen, Espinoza, Diamond-Lowe, Mendonça, Demory, Gressier, Ih, Fortune, August, Holmberg, Valdés, Zgraggen, Buchhave, Burgasser, Fisher, Gibson, Heng, Hoeijmakers, Kitzmann, Prinoth, Rathcke, \& Morris}]{Allen_Espinoza_Diamond-Lowe_Mendonca_Demory_Gressier_Ih_Fortune_August_Holmberg_etal._2025}
Allen, N.~H., Espinoza, N., Diamond-Lowe, H., {et~al.} 2025, \dodoi{10.48550/arXiv.2508.14210}

\bibitem[{{Alonso}(2018)}]{2018haex.bookE..40A}
{Alonso}, R. 2018, in Handbook of Exoplanets, ed. H.~J. {Deeg} \& J.~A. {Belmonte}, 40, \dodoi{10.1007/978-3-319-55333-7_40}

\bibitem[{{Astropy Collaboration} {et~al.}(2013){Astropy Collaboration}, {Robitaille}, {Tollerud}, {Greenfield}, {Droettboom}, {Bray}, {Aldcroft}, {Davis}, {Ginsburg}, {Price-Whelan}, {Kerzendorf}, {Conley}, {Crighton}, {Barbary}, {Muna}, {Ferguson}, {Grollier}, {Parikh}, {Nair}, {Unther}, {Deil}, {Woillez}, {Conseil}, {Kramer}, {Turner}, {Singer}, {Fox}, {Weaver}, {Zabalza}, {Edwards}, {Azalee Bostroem}, {Burke}, {Casey}, {Crawford}, {Dencheva}, {Ely}, {Jenness}, {Labrie}, {Lim}, {Pierfederici}, {Pontzen}, {Ptak}, {Refsdal}, {Servillat}, \& {Streicher}}]{astropy:2013}
{Astropy Collaboration}, {Robitaille}, T.~P., {Tollerud}, E.~J., {et~al.} 2013, \aap, 558, A33, \dodoi{10.1051/0004-6361/201322068}

\bibitem[{{Astropy Collaboration} {et~al.}(2018){Astropy Collaboration}, {Price-Whelan}, {Sip{\H{o}}cz}, {G{\"u}nther}, {Lim}, {Crawford}, {Conseil}, {Shupe}, {Craig}, {Dencheva}, {Ginsburg}, {Vand erPlas}, {Bradley}, {P{\'e}rez-Su{\'a}rez}, {de Val-Borro}, {Aldcroft}, {Cruz}, {Robitaille}, {Tollerud}, {Ardelean}, {Babej}, {Bach}, {Bachetti}, {Bakanov}, {Bamford}, {Barentsen}, {Barmby}, {Baumbach}, {Berry}, {Biscani}, {Boquien}, {Bostroem}, {Bouma}, {Brammer}, {Bray}, {Breytenbach}, {Buddelmeijer}, {Burke}, {Calderone}, {Cano Rodr{\'\i}guez}, {Cara}, {Cardoso}, {Cheedella}, {Copin}, {Corrales}, {Crichton}, {D'Avella}, {Deil}, {Depagne}, {Dietrich}, {Donath}, {Droettboom}, {Earl}, {Erben}, {Fabbro}, {Ferreira}, {Finethy}, {Fox}, {Garrison}, {Gibbons}, {Goldstein}, {Gommers}, {Greco}, {Greenfield}, {Groener}, {Grollier}, {Hagen}, {Hirst}, {Homeier}, {Horton}, {Hosseinzadeh}, {Hu}, {Hunkeler}, {Ivezi{\'c}}, {Jain}, {Jenness}, {Kanarek}, {Kendrew}, {Kern}, {Kerzendorf}, {Khvalko}, {King}, {Kirkby}, {Kulkarni},
  {Kumar}, {Lee}, {Lenz}, {Littlefair}, {Ma}, {Macleod}, {Mastropietro}, {McCully}, {Montagnac}, {Morris}, {Mueller}, {Mumford}, {Muna}, {Murphy}, {Nelson}, {Nguyen}, {Ninan}, {N{\"o}the}, {Ogaz}, {Oh}, {Parejko}, {Parley}, {Pascual}, {Patil}, {Patil}, {Plunkett}, {Prochaska}, {Rastogi}, {Reddy Janga}, {Sabater}, {Sakurikar}, {Seifert}, {Sherbert}, {Sherwood-Taylor}, {Shih}, {Sick}, {Silbiger}, {Singanamalla}, {Singer}, {Sladen}, {Sooley}, {Sornarajah}, {Streicher}, {Teuben}, {Thomas}, {Tremblay}, {Turner}, {Terr{\'o}n}, {van Kerkwijk}, {de la Vega}, {Watkins}, {Weaver}, {Whitmore}, {Woillez}, {Zabalza}, \& {Astropy Contributors}}]{astropy:2018}
{Astropy Collaboration}, {Price-Whelan}, A.~M., {Sip{\H{o}}cz}, B.~M., {et~al.} 2018, \aj, 156, 123, \dodoi{10.3847/1538-3881/aabc4f}

\bibitem[{{Astropy Collaboration} {et~al.}(2022){Astropy Collaboration}, {Price-Whelan}, {Lim}, {Earl}, {Starkman}, {Bradley}, {Shupe}, {Patil}, {Corrales}, {Brasseur}, {N{"o}the}, {Donath}, {Tollerud}, {Morris}, {Ginsburg}, {Vaher}, {Weaver}, {Tocknell}, {Jamieson}, {van Kerkwijk}, {Robitaille}, {Merry}, {Bachetti}, {G{"u}nther}, {Aldcroft}, {Alvarado-Montes}, {Archibald}, {B{'o}di}, {Bapat}, {Barentsen}, {Baz{'a}n}, {Biswas}, {Boquien}, {Burke}, {Cara}, {Cara}, {Conroy}, {Conseil}, {Craig}, {Cross}, {Cruz}, {D'Eugenio}, {Dencheva}, {Devillepoix}, {Dietrich}, {Eigenbrot}, {Erben}, {Ferreira}, {Foreman-Mackey}, {Fox}, {Freij}, {Garg}, {Geda}, {Glattly}, {Gondhalekar}, {Gordon}, {Grant}, {Greenfield}, {Groener}, {Guest}, {Gurovich}, {Handberg}, {Hart}, {Hatfield-Dodds}, {Homeier}, {Hosseinzadeh}, {Jenness}, {Jones}, {Joseph}, {Kalmbach}, {Karamehmetoglu}, {Ka{l}uszy{'n}ski}, {Kelley}, {Kern}, {Kerzendorf}, {Koch}, {Kulumani}, {Lee}, {Ly}, {Ma}, {MacBride}, {Maljaars}, {Muna}, {Murphy}, {Norman}, {O'Steen},
  {Oman}, {Pacifici}, {Pascual}, {Pascual-Granado}, {Patil}, {Perren}, {Pickering}, {Rastogi}, {Roulston}, {Ryan}, {Rykoff}, {Sabater}, {Sakurikar}, {Salgado}, {Sanghi}, {Saunders}, {Savchenko}, {Schwardt}, {Seifert-Eckert}, {Shih}, {Jain}, {Shukla}, {Sick}, {Simpson}, {Singanamalla}, {Singer}, {Singhal}, {Sinha}, {Sip{H{o}}cz}, {Spitler}, {Stansby}, {Streicher}, {{{S}}umak}, {Swinbank}, {Taranu}, {Tewary}, {Tremblay}, {Val-Borro}, {Van Kooten}, {Vasovi{'c}}, {Verma}, {de Miranda Cardoso}, {Williams}, {Wilson}, {Winkel}, {Wood-Vasey}, {Xue}, {Yoachim}, {Zhang}, {Zonca}, \& {Astropy Project Contributors}}]{astropy:2022}
{Astropy Collaboration}, {Price-Whelan}, A.~M., {Lim}, P.~L., {et~al.} 2022, \apj, 935, 167, \dodoi{10.3847/1538-4357/ac7c74}

\bibitem[{August {et~al.}(2024)August, Buchhave, Diamond-Lowe, Mendonça, Gressier, Rathcke, Allen, Fortune, Jones, Meier-Valdés, Demory, Espinoza, Fisher, Gibson, Heng, Hoeijmakers, Hooton, Kitzmann, \& Prinoth}]{august2024hotrockssurveyi}
August, P.~C., Buchhave, L.~A., Diamond-Lowe, H., {et~al.} 2024, Hot Rocks Survey I : A shallow eclipse for LHS 1478 b.
\newblock \doarXiv{2410.11048}

\bibitem[{{Baumeister} {et~al.}(2023){Baumeister}, {Tosi}, {Brachmann}, {Grenfell}, \& {Noack}}]{2023A&A...675A.122B}
{Baumeister}, P., {Tosi}, N., {Brachmann}, C., {Grenfell}, J.~L., \& {Noack}, L. 2023, \aap, 675, A122, \dodoi{10.1051/0004-6361/202245791}

\bibitem[{Bell {et~al.}(2022)Bell, Ahrer, Brande, Carter, Feinstein, {Guzman Caloca}, Mansfield, Zieba, Piaulet, Benneke, Filippazzo, May, Roy, Kreidberg, \& Stevenson}]{Bell2022}
Bell, T.~J., Ahrer, E.-M., Brande, J., {et~al.} 2022, Journal of Open Source Software, 7, 4503, \dodoi{10.21105/joss.04503}

\bibitem[{Bell {et~al.}(2023)Bell, Kreidberg, Kendrew, Bean, Crouzet, Ducrot, Dyrek, Gao, Lagage, \& Moses}]{bell2023_miri_lrs_settling_ramp}
Bell, T.~J., Kreidberg, L., Kendrew, S., {et~al.} 2023, A First Look at the JWST MIRI/LRS Phase Curve of WASP-43b.
\newblock \doarXiv{2301.06350}

\bibitem[{Berta-Thompson {et~al.}(2025)Berta-Thompson, Wachiraphan, \& Murray}]{bertathompson20253dcosmicshorelinenurturing}
Berta-Thompson, Z.~K., Wachiraphan, P., \& Murray, C. 2025, The 3D Cosmic Shoreline for Nurturing Planetary Atmospheres.
\newblock \doarXiv{2507.02136}

\bibitem[{{Bonfanti} {et~al.}(2024){Bonfanti}, {Brady}, {Wilson}, {Venturini}, {Egger}, {Brandeker}, {Sousa}, {Lendl}, {Simon}, {Queloz}, {Olofsson}, {Adibekyan}, {Alibert}, {Fossati}, {Hooton}, {Kubyshkina}, {Luque}, {Murgas}, {Mustill}, {Santos}, {Van Grootel}, {Alonso}, {Asquier}, {Bandy}, {B{\'a}rczy}, {Barrado Navascues}, {Barros}, {Baumjohann}, {Bean}, {Beck}, {Beck}, {Benz}, {Bergomi}, {Billot}, {Borsato}, {Broeg}, {Collier Cameron}, {Csizmadia}, {Cubillos}, {Davies}, {Deleuil}, {Deline}, {Delrez}, {Demangeon}, {Demory}, {Ehrenreich}, {Erikson}, {Fortier}, {Fridlund}, {Gandolfi}, {Gillon}, {G{\"u}del}, {G{\"u}nther}, {Heitzmann}, {Helling}, {Hoyer}, {Isaak}, {Kasper}, {Kiss}, {Lam}, {Laskar}, {Lecavelier des Etangs}, {Magrin}, {Maxted}, {Mordasini}, {Nascimbeni}, {Ottensamer}, {Pagano}, {Pall{\'e}}, {Peter}, {Piotto}, {Pollacco}, {Ragazzoni}, {Rando}, {Rauer}, {Ribas}, {Scandariato}, {S{\'e}gransan}, {Seifahrt}, {Smith}, {Stalport}, {Stef{\'a}nsson}, {Steinberger}, {St{\"u}rmer}, {Szab{\'o}}, {Thomas},
  {Udry}, {Villaver}, {Walton}, {Westerdorff}, \& {Zingales}}]{2024A&A...682A..66B}
{Bonfanti}, A., {Brady}, M., {Wilson}, T.~G., {et~al.} 2024, \aap, 682, A66, \dodoi{10.1051/0004-6361/202348180}

\bibitem[{Bonomo {et~al.}(2023)Bonomo, Dumusque, Massa, Mortier, Bongiolatti, Malavolta, Sozzetti, Buchhave, Damasso, Haywood, Morbidelli, Latham, Molinari, Pepe, Poretti, Udry, Affer, Boschin, Charbonneau, Cosentino, Cretignier, Ghedina, Lega, López-Morales, Margini, Fiorenzano, Mayor, Micela, Pedani, Pinamonti, Rice, Sasselov, Tronsgaard, \& Vanderburg}]{Bonomo_Dumusque_Massa_Mortier_Bongiolatti_Malavolta_Sozzetti_Buchhave_Damasso_Haywood_etal._2023}
Bonomo, A.~S., Dumusque, X., Massa, A., {et~al.} 2023, Astronomy \& Astrophysics, 677, A33, \dodoi{10.1051/0004-6361/202346211}

\bibitem[{{Brunetto} {et~al.}(2015){Brunetto}, {Loeffler}, {Nesvorn{\'y}}, {Sasaki}, \& {Strazzulla}}]{2015aste.book..597B}
{Brunetto}, R., {Loeffler}, M.~J., {Nesvorn{\'y}}, D., {Sasaki}, S., \& {Strazzulla}, G. 2015, in Asteroids IV, ed. P.~{Michel}, F.~E. {DeMeo}, \& W.~F. {Bottke}, 597--616, \dodoi{10.2458/azu_uapress_9780816532131-ch031}

\bibitem[{{Burt} {et~al.}(2024){Burt}, {Hooton}, {Mamajek}, {Barrag{\'a}n}, {Millholland}, {Fairnington}, {Fisher}, {Halverson}, {Huang}, {Brady}, {Seifahrt}, {Gaidos}, {Luque}, {Kasper}, \& {Bean}}]{2024ApJ...971L..12B}
{Burt}, J.~A., {Hooton}, M.~J., {Mamajek}, E.~E., {et~al.} 2024, \apjl, 971, L12, \dodoi{10.3847/2041-8213/ad5b52}

\bibitem[{{Cadieux} {et~al.}(2024){Cadieux}, {Plotnykov}, {Doyon}, {Valencia}, {Jahandar}, {Dang}, {Turbet}, {Fauchez}, {Cloutier}, {Cherubim}, {Artigau}, {Cook}, {Edwards}, {Hallatt}, {Charnay}, {Bouchy}, {Allart}, {Mignon}, {Baron}, {Barros}, {Benneke}, {Canto Martins}, {Cowan}, {De Medeiros}, {Delfosse}, {Delgado-Mena}, {Dumusque}, {Ehrenreich}, {Frensch}, {Gonz{\'a}lez Hern{\'a}ndez}, {Hara}, {Lafreni{\`e}re}, {Lo Curto}, {Malo}, {Melo}, {Mounzer}, {Passeger}, {Pepe}, {Poulin-Girard}, {Santos}, {Sosnowska}, {Su{\'a}rez Mascare{\~n}o}, {Thibault}, {Vaulato}, {Wade}, \& {Wildi}}]{2024ApJ...960L...3C}
{Cadieux}, C., {Plotnykov}, M., {Doyon}, R., {et~al.} 2024, \apjl, 960, L3, \dodoi{10.3847/2041-8213/ad1691}

\bibitem[{Castan \& Menou(2011)}]{thermal_inversion_Castan_2011}
Castan, T., \& Menou, K. 2011, The Astrophysical Journal, 743, L36, \dodoi{10.1088/2041-8205/743/2/l36}

\bibitem[{Connors {et~al.}(2025)Connors, Monaghan, Benneke, \& Dang}]{Connors_Monaghan_Benneke_Dang_2025}
Connors, N.~J., Monaghan, C., Benneke, B., \& Dang, L. 2025, The Astrophysical Journal Letters, 989, L11, \dodoi{10.3847/2041-8213/adee0d}

\bibitem[{{Correia} \& {Laskar}(2010)}]{2010exop.book..239C}
{Correia}, A.~C.~M., \& {Laskar}, J. 2010, in Exoplanets, ed. S.~{Seager}, 239--266, \dodoi{10.48550/arXiv.1009.1352}

\bibitem[{{Cowan} \& {Agol}(2011)}]{COWAN2011ApJ...726...82C}
{Cowan}, N.~B., \& {Agol}, E. 2011, \apj, 726, 82, \dodoi{10.1088/0004-637X/726/2/82}

\bibitem[{Coy {et~al.}(2025)Coy, Ih, Kite, Koll, Tenthoff, Bean, Weiner~Mansfield, Zhang, Xue, Kempton, Wohlfarth, Hu, Lyu, \& Wöhler}]{Coy_Ih_Kite_Koll_Tenthoff_Bean_WeinerMansfield_Zhang_Xue_Kempton_etal._2025}
Coy, B.~P., Ih, J., Kite, E.~S., {et~al.} 2025, The Astrophysical Journal, 987, 22, \dodoi{10.3847/1538-4357/add3f7}

\bibitem[{Crossfield(2012)}]{Crossfield_2012}
Crossfield, I. J.~M. 2012, Astronomy and Astrophysics, 545, A97, \dodoi{10.1051/0004-6361/201219826}

\bibitem[{Crossfield {et~al.}(2022)Crossfield, Malik, Hill, Kane, Foley, Polanski, Coria, Brande, Zhang, Wienke, Kreidberg, Cowan, Dragomir, Gorjian, Mikal-Evans, Benneke, Christiansen, Deming, \& Morales}]{Crossfield_Malik_Hill_Kane_Foley_Polanski_Coria_Brande_Zhang_Wienke_et_al._2022}
Crossfield, I. J.~M., Malik, M., Hill, M.~L., {et~al.} 2022, The Astrophysical Journal Letters, 937, L17, \dodoi{10.3847/2041-8213/ac886b}

\bibitem[{Deming {et~al.}(2018)Deming, Louie, \& Sheets}]{Deming_Louie_Sheets_2018}
Deming, D., Louie, D., \& Sheets, H. 2018, Publications of the Astronomical Society of the Pacific, 131, 013001, \dodoi{10.1088/1538-3873/aae5c5}

\bibitem[{{Diamond-Lowe} {et~al.}(2023){Diamond-Lowe}, {Mendonca}, {Akin}, {Allen}, {Baungaard}, {Borsato}, {Buchhave}, {Burgasser}, {Demory}, {Espinoza}, {Fisher}, {Fortune}, {Gibson}, {Gressier}, {Guzman Mesa}, {Heng}, {Hoeijmakers}, {Hooton}, {Jones}, {Kitzmann}, {Lueber}, {Meier Valdes}, {Prinoth}, {Rathcke}, \& {Tian}}]{2023jwst.prop.3730D}
{Diamond-Lowe}, H., {Mendonca}, J.~M., {Akin}, C.~J., {et~al.} 2023, {The Hot Rocks Survey: Testing 9 Irradiated Terrestrial Exoplanets for Atmospheres}, JWST Proposal. Cycle 2, ID. \#3730

\bibitem[{{Donati} {et~al.}(2020){Donati}, {Kouach}, {Moutou}, {Doyon}, {Delfosse}, {Artigau}, {Baratchart}, {Lacombe}, {Barrick}, {H{\'e}brard}, {Bouchy}, {Saddlemyer}, {Par{\`e}s}, {Rabou}, {Micheau}, {Dolon}, {Reshetov}, {Challita}, {Carmona}, {Striebig}, {Thibault}, {Martioli}, {Cook}, {Fouqu{\'e}}, {Vermeulen}, {Wang}, {Arnold}, {Pepe}, {Boisse}, {Figueira}, {Bouvier}, {Ray}, {Feugeade}, {Morin}, {Alencar}, {Hobson}, {Castilho}, {Udry}, {Santos}, {Hernandez}, {Benedict}, {Vall{\'e}e}, {Gallou}, {Dupieux}, {Larrieu}, {Perruchot}, {Sottile}, {Moreau}, {Usher}, {Baril}, {Wildi}, {Chazelas}, {Malo}, {Bonfils}, {Loop}, {Kerley}, {Wevers}, {Dunn}, {Pazder}, {Macdonald}, {Dubois}, {Carri{\'e}}, {Valentin}, {Henault}, {Yan}, \& {Steinmetz}}]{induction_heating_2020MNRAS.498.5684D}
{Donati}, J.~F., {Kouach}, D., {Moutou}, C., {et~al.} 2020, \mnras, 498, 5684, \dodoi{10.1093/mnras/staa2569}

\bibitem[{Dorn {et~al.}(2018)Dorn, Noack, \& Rozel}]{Dorn_2018}
Dorn, C., Noack, L., \& Rozel, A.~B. 2018, Astronomy \& Astrophysics, 614, A18, \dodoi{10.1051/0004-6361/201731513}

\bibitem[{Ducrot {et~al.}(2025)Ducrot, Lagage, Min, Gillon, Bell, Tremblin, Greene, Dyrek, Bouwman, Waters, Güdel, Henning, Vandenbussche, Absil, Barrado, Boccaletti, Coulais, Decin, Edwards, Gastaud, Glasse, Kendrew, Olofsson, Patapis, Pye, Rouan, Whiteford, Argyriou, Cossou, Glauser, Krause, Lahuis, Royer, Scheithauer, Colina, van Dishoeck, Ostlin, Ray, \& Wright}]{Ducrot_Lagage_Min_Gillon_Bell_Tremblin_Greene_Dyrek_Bouwman_Waters_etal._2025}
Ducrot, E., Lagage, P.-O., Min, M., {et~al.} 2025, Nature Astronomy, 9, 358–369, \dodoi{10.1038/s41550-024-02428-z}

\bibitem[{{Dumusque} {et~al.}(2019){Dumusque}, {Turner}, {Dorn}, {Eastman}, {Allart}, {Adibekyan}, {Sousa}, {Santos}, {Mordasini}, {Bourrier}, {Bouchy}, {Coffinet}, {Davies}, {D{\'\i}az}, {Fausnaugh}, {Glidden}, {Guerrero}, {Henze}, {Jenkins}, {Latham}, {Lovis}, {Mayor}, {Pepe}, {Quintana}, {Ricker}, {Rowden}, {Segransan}, {Su{\'a}rez Mascare{\~n}o}, {Seager}, {Twicken}, {Udry}, {Vanderspek}, \& {Winn}}]{2019A&A...627A..43D}
{Dumusque}, X., {Turner}, O., {Dorn}, C., {et~al.} 2019, \aap, 627, A43, \dodoi{10.1051/0004-6361/201935457}

\bibitem[{Dyrek {et~al.}(2024)Dyrek, Ducrot, Lagage, Tremblin, Kendrew, Bouwman, \& Bouffet}]{Dyrek_2024_miri_lrs_settling}
Dyrek, A., Ducrot, E., Lagage, P.-O., {et~al.} 2024, Astronomy \& Astrophysics, 683, A212, \dodoi{10.1051/0004-6361/202347127}

\bibitem[{Essack {et~al.}(2020)Essack, Seager, \& Pajusalu}]{Essack_Seager_Pajusalu_2020}
Essack, Z., Seager, S., \& Pajusalu, M. 2020, The Astrophysical Journal, 898, 160, \dodoi{10.3847/1538-4357/ab9cba}

\bibitem[{Fauchez {et~al.}(2025)Fauchez, Ducrot, Rackham, Stevenson, Mayorga, \& de~Wit}]{fauchez2025stellarmodelslimitexoplanet}
Fauchez, T.~J., Ducrot, E., Rackham, B.~V., {et~al.} 2025, The Astrophysical Journal, 989, 170, \dodoi{10.3847/1538-4357/adf068}

\bibitem[{{Ferrari} {et~al.}(2020){Ferrari}, {Maturilli}, {Carli}, {D'Amore}, {Helbert}, {Nestola}, \& {Hiesinger}}]{2020E&PSL.53416089F}
{Ferrari}, S., {Maturilli}, A., {Carli}, C., {et~al.} 2020, Earth and Planetary Science Letters, 534, 116089, \dodoi{10.1016/j.epsl.2020.116089}

\bibitem[{First {et~al.}(2025)First, Mishra, Gazel, Lewis, Letai, \& Hanssen}]{First_Mishra_Gazel_Lewis_Letai_Hanssen_2025}
First, E.~C., Mishra, I., Gazel, E., {et~al.} 2025, Nature Astronomy, 9, 370–379, \dodoi{10.1038/s41550-024-02412-7}

\bibitem[{{Fortune} {et~al.}(2025){Fortune}, {Gibson}, {Diamond-Lowe}, {Mendon{\c{c}}a}, {Gressier}, {Kitzmann}, {Allen}, {August}, {Ih}, {Meier Vald{\'e}s}, {Zgraggen}, {Buchhave}, {Demory}, {Espinoza}, {Heng}, {Jones}, \& {Rathcke}}]{2025hotrocksiii}
{Fortune}, M., {Gibson}, N.~P., {Diamond-Lowe}, H., {et~al.} 2025, arXiv e-prints, arXiv:2505.22186, \dodoi{10.48550/arXiv.2505.22186}

\bibitem[{Gillon {et~al.}(2025)Gillon, Ducrot, Bell, Huang, Lincowski, Lyu, Maurel, Revol, Agol, Bolmont, Dong, Fauchez, Koll, Leconte, Meadows, Selsis, Turbet, Charnay, Delre, Demory, Householder, Zieba, Berardo, Dyrek, Edwards, de~Wit, Greene, Hu, Iro, Kreidberg, Lagage, Lustig-Yaeger, \& Iyer}]{gillon2025jwstthermalphasecurves}
Gillon, M., Ducrot, E., Bell, T.~J., {et~al.} 2025, First JWST thermal phase curves of temperate terrestrial exoplanets reveal no thick atmosphere around TRAPPIST-1 b and c.
\newblock \doarXiv{2509.02128}

\bibitem[{{Goffo} {et~al.}(2023){Goffo}, {Gandolfi}, {Egger}, {Mustill}, {Albrecht}, {Hirano}, {Kochukhov}, {Astudillo-Defru}, {Barragan}, {Serrano}, {Hatzes}, {Alibert}, {Guenther}, {Dai}, {Lam}, {Csizmadia}, {Smith}, {Fossati}, {Luque}, {Rodler}, {Winther}, {R{\o}rsted}, {Alarcon}, {Bonfils}, {Cochran}, {Deeg}, {Jenkins}, {Korth}, {Livingston}, {Meech}, {Murgas}, {Orell-Miquel}, {Osborne}, {Palle}, {Persson}, {Redfield}, {Ricker}, {Seager}, {Vanderspek}, {Van Eylen}, \& {Winn}}]{GoffoGJ367}
{Goffo}, E., {Gandolfi}, D., {Egger}, J.~A., {et~al.} 2023, \apjl, 955, L3, \dodoi{10.3847/2041-8213/ace0c7}

\bibitem[{Greene {et~al.}(2023)Greene, Bell, Ducrot, Dyrek, Lagage, \& Fortney}]{Greene_Bell_Ducrot_Dyrek_Lagage_Fortney_2023}
Greene, T.~P., Bell, T.~J., Ducrot, E., {et~al.} 2023, Nature, 618, 39–42, \dodoi{10.1038/s41586-023-05951-7}

\bibitem[{Hammond {et~al.}(2024)Hammond, Guimond, Lichtenberg, Nicholls, Fisher, Luque, Meier, Taylor, Changeat, Dang, Herbort, \& Teske}]{Hammond_Guimond_Lichtenberg_Nicholls_Fisher_Luque_Meier_Taylor_Changeat_Dang_et_al._2024}
Hammond, M., Guimond, C.~M., Lichtenberg, T., {et~al.} 2024.
\newblock \url{http://arxiv.org/abs/2409.04386}

\bibitem[{Han {et~al.}(2025)Han, Robertson, Brandt, Kanodia, Cañas, Shporer, Ricker, \& Beard}]{Han_Robertson}
Han, T., Robertson, P., Brandt, T.~D., {et~al.} 2025, \dodoi{10.48550/arXiv.2506.19985}

\bibitem[{{Hansen}(2008)}]{2008ApJS..179..484H}
{Hansen}, B. M.~S. 2008, \apjs, 179, 484, \dodoi{10.1086/591964}

\bibitem[{{Hapke}(2001)}]{2001JGR...10610039H}
{Hapke}, B. 2001, \jgr, 106, 10039, \dodoi{10.1029/2000JE001338}

\bibitem[{Hapke(2012)}]{Hapke_2012}
Hapke, B. 2012, Theory of Reflectance and Emittance Spectroscopy, 2nd edn. (Cambridge University Press)

\bibitem[{Harris {et~al.}(2020)Harris, Millman, van~der Walt, Gommers, Virtanen, Cournapeau, Wieser, Taylor, Berg, Smith, Kern, Picus, Hoyer, van Kerkwijk, Brett, Haldane, del R{\'{i}}o, Wiebe, Peterson, G{\'{e}}rard-Marchant, Sheppard, Reddy, Weckesser, Abbasi, Gohlke, \& Oliphant}]{harris2020array_numpy}
Harris, C.~R., Millman, K.~J., van~der Walt, S.~J., {et~al.} 2020, Nature, 585, 357, \dodoi{10.1038/s41586-020-2649-2}

\bibitem[{{Helbert} {et~al.}(2013){Helbert}, {Nestola}, {Ferrari}, {Maturilli}, {Massironi}, {Redhammer}, {Capria}, {Carli}, {Capaccioni}, \& {Bruno}}]{2013E&PSL.371..252H}
{Helbert}, J., {Nestola}, F., {Ferrari}, S., {et~al.} 2013, Earth and Planetary Science Letters, 371, 252, \dodoi{10.1016/j.epsl.2013.03.038}

\bibitem[{Heller(2019)}]{Heller_2019}
Heller, R. 2019, Astronomy \& Astrophysics, 623, A137, \dodoi{10.1051/0004-6361/201834620}

\bibitem[{Hu {et~al.}(2012)Hu, Ehlmann, \& Seager}]{Hu_Ehlmann_Seager_2012}
Hu, R., Ehlmann, B.~L., \& Seager, S. 2012, The Astrophysical Journal, 752, 7, \dodoi{10.1088/0004-637X/752/1/7}

\bibitem[{Hu {et~al.}(2024)Hu, Bello-Arufe, Zhang, Paragas, Zilinskas, Van~Buchem, Bess, Patel, Ito, Damiano, Scheucher, Oza, Knutson, Miguel, Dragomir, Brandeker, \& Demory}]{Hu_Bello-Arufe_Zhang_Paragas_Zilinskas_Van_Buchem_Bess_Patel_Ito_Damiano_et_al._2024}
Hu, R., Bello-Arufe, A., Zhang, M., {et~al.} 2024, Nature, \dodoi{10.1038/s41586-024-07432-x}

\bibitem[{Hunter(2007)}]{matplotlib}
Hunter, J.~D. 2007, Computing in Science \& Engineering, 9, 90, \dodoi{10.1109/MCSE.2007.55}

\bibitem[{{Husser} {et~al.}(2013){Husser}, {Wende-von Berg}, {Dreizler}, {Homeier}, {Reiners}, {Barman}, \& {Hauschildt}}]{2013A&A...553A...6H}
{Husser}, T.~O., {Wende-von Berg}, S., {Dreizler}, S., {et~al.} 2013, \aap, 553, A6, \dodoi{10.1051/0004-6361/201219058}

\bibitem[{Iyer {et~al.}(2024)Iyer, Line, Muirhead, Fortney, \& Gharib-Nezhad}]{Iyer_Line_Muirhead_Fortney_Gharib-Nezhad_2024}
Iyer, A., Line, M., Muirhead, P., Fortney, J., \& Gharib-Nezhad, E. 2024, in , 631.03.
\newblock \url{https://ui.adsabs.harvard.edu/abs/2024ESS.....563103I}

\bibitem[{Iyer {et~al.}(2023)Iyer, Line, Muirhead, Fortney, \& Gharib-Nezhad}]{Iyer_Line_Muirhead_Fortney_Gharib-Nezhad_2023}
Iyer, A.~R., Line, M.~R., Muirhead, P.~S., Fortney, J.~J., \& Gharib-Nezhad, E. 2023, The Astrophysical Journal, 944, 41, \dodoi{10.3847/1538-4357/acabc2}

\bibitem[{Jahandar {et~al.}(2025)Jahandar, Doyon, Artigau, Cook, Cadieux, Donati, Cowan, Cloutier, Pelletier, Alves-Brito, Martins, Shang, \& Carmona}]{Jahandar_Doyon_Artigau_Cook_Cadieux_Donati_Cowan_Cloutier_Pelletier_Alves-Brito_etal._2025}
Jahandar, F., Doyon, R., Artigau, E., {et~al.} 2025, The Astrophysical Journal, 978, 154, \dodoi{10.3847/1538-4357/ad91a0}

\bibitem[{Ji {et~al.}(2025)Ji, Chatterjee, Coy, \& Kite}]{Ji_Chatterjee_Coy_Kite_2025}
Ji, X., Chatterjee, R.~D., Coy, B.~P., \& Kite, E.~S. 2025, \dodoi{10.48550/arXiv.2504.19872}

\bibitem[{Kislyakova {et~al.}(2018)Kislyakova, Fossati, Johnstone, Noack, Lüftinger, Zaitsev, \& Lammer}]{induction_heating_Kislyakova_2018}
Kislyakova, K.~G., Fossati, L., Johnstone, C.~P., {et~al.} 2018, The Astrophysical Journal, 858, 105, \dodoi{10.3847/1538-4357/aabae4}

\bibitem[{Kite {et~al.}(2016)Kite, Fegley, Schaefer, \& Gaidos}]{Kite_Fegley_Schaefer_Gaidos_2016}
Kite, E.~S., Fegley, Bruce, J., Schaefer, L., \& Gaidos, E. 2016, The Astrophysical Journal, 828, 80, \dodoi{10.3847/0004-637X/828/2/80}

\bibitem[{Koll {et~al.}(2019)Koll, Malik, Mansfield, Kempton, Kite, Abbot, \& Bean}]{Koll_2019}
Koll, D. D.~B., Malik, M., Mansfield, M., {et~al.} 2019, The Astrophysical Journal, 886, 140, \dodoi{10.3847/1538-4357/ab4c91}

\bibitem[{Kreidberg(2018)}]{Kreidberg_2018}
Kreidberg, L. 2018, Exoplanet Atmosphere Measurements from Transmission Spectroscopy and Other Planet Star Combined Light Observations (Springer, Cham), 2083–2105, \dodoi{10.1007/978-3-319-55333-7_100}

\bibitem[{Kreidberg \& Stevenson(2025)}]{kreidberg2025lookrockyexoplanetsjwst}
Kreidberg, L., \& Stevenson, K.~B. 2025, A first look at rocky exoplanets with JWST.
\newblock \doarXiv{2507.00933}

\bibitem[{Kreidberg {et~al.}(2019)Kreidberg, Koll, Morley, Hu, Schaefer, Deming, Stevenson, Dittmann, Vanderburg, Berardo, Guo, Stassun, Crossfield, Charbonneau, Latham, Loeb, Ricker, Seager, \& Vanderspek}]{Kreidberg_Koll_Morley_Hu_Schaefer_Deming_Stevenson_Dittmann_Vanderburg_Berardo_et_al._2019}
Kreidberg, L., Koll, D. D.~B., Morley, C., {et~al.} 2019, Nature, 573, 87–90, \dodoi{10.1038/s41586-019-1497-4}

\bibitem[{Libralato {et~al.}(2024)Libralato, Argyriou, Dicken, Marín, Guillard, Hines, Kavanagh, Kendrew, Law, Noriega-Crespo, \& Álvarez Márquez}]{miri_artifacts_libralato2024highprecisionastrometryphotometryjwstmiri}
Libralato, M., Argyriou, I., Dicken, D., {et~al.} 2024, High-precision astrometry and photometry with the JWST/MIRI imager.
\newblock \doarXiv{2311.12145}

\bibitem[{Lin \& Daylan(2026)}]{Lin_Daylan_2026}
Lin, Z., \& Daylan, T. 2026, \dodoi{10.48550/arXiv.2601.00412}

\bibitem[{Luque {et~al.}(2024)Luque, Coy, Xue, Feinstein, Ahrer, Changeat, Zhang, Moran, Bean, Kite, Mansfield, \& Pallé}]{Luque_Coy_Xue_Feinstein_Ahrer_Changeat_Zhang_Moran_Bean_Kite_etal._2024}
Luque, R., Coy, B.~P., Xue, Q., {et~al.} 2024, \dodoi{10.48550/arXiv.2412.03411}

\bibitem[{Lutgens {et~al.}(2017)Lutgens, Tarbuck, \& Tasa}]{lutgens}
Lutgens, F.~K., Tarbuck, E.~J., \& Tasa, D. 2017, Essentials of Geology, 13th edition (Pearson)

\bibitem[{Lyu \& Koll(2025)}]{lyu2025impactsubsurfacetemperaturegradients}
Lyu, X., \& Koll, D. D.~B. 2025, Impact of Subsurface Temperature Gradients on Emission Spectra of Airless Exoplanets: the Solid-state Greenhouse and Anti-Greenhouse.
\newblock \doarXiv{2510.22932}

\bibitem[{Lyu {et~al.}(2024)Lyu, Koll, Cowan, Hu, Kreidberg, \& Rose}]{lyu2024superearthlhs3844btidallylocked}
Lyu, X., Koll, D. D.~B., Cowan, N.~B., {et~al.} 2024, Super-Earth LHS3844b is tidally locked.
\newblock \doarXiv{2310.01725}

\bibitem[{{Mackwell} {et~al.}(2013){Mackwell}, {Simon-Miller}, {Harder}, \& {Bullock}}]{thermal_inversion_2013cctp.book.....M}
{Mackwell}, S.~J., {Simon-Miller}, A.~A., {Harder}, J.~W., \& {Bullock}, M.~A. 2013, {Comparative Climatology of Terrestrial Planets} (University of Arizona Press), \dodoi{10.2458/azu_uapress_9780816530595}

\bibitem[{Mahajan {et~al.}(2024)Mahajan, Eastman, \& Kirk}]{Mahajan_2024}
Mahajan, A.~S., Eastman, J.~D., \& Kirk, J. 2024, The Astrophysical Journal Letters, 963, L37, \dodoi{10.3847/2041-8213/ad29f3}

\bibitem[{Mansfield {et~al.}(2019)Mansfield, Kite, Hu, Koll, Malik, Bean, \& Kempton}]{Mansfield_Kite_Hu_Koll_Malik_Bean_Kempton_2019}
Mansfield, M., Kite, E.~S., Hu, R., {et~al.} 2019, The Astrophysical Journal, 886, 141, \dodoi{10.3847/1538-4357/ab4c90}

\bibitem[{Mansfield {et~al.}(2024)Mansfield, Xue, Zhang, Mahajan, Ih, Koll, Bean, Coy, Eastman, Kempton, Kite, \& Lunine}]{Mansfield_Xue_Zhang_Mahajan_Ih_Koll_Bean_Coy_Eastman_Kempton_et_al._2024}
Mansfield, M.~W., Xue, Q., Zhang, M., {et~al.} 2024.
\newblock \url{http://arxiv.org/abs/2408.15123}

\bibitem[{Meier~Valdés {et~al.}(2025)Meier~Valdés, Demory, Diamond-Lowe, Mendonça, August, Fortune, Allen, Kitzmann, Gressier, Hooton, Jones, Buchhave, Espinoza, Fisher, Gibson, Heng, Hoeijmakers, Prinoth, Rathcke, \& Eastman}]{Meier_Vald_s_2025}
Meier~Valdés, E.~A., Demory, B.-O., Diamond-Lowe, H., {et~al.} 2025, Astronomy \& Astrophysics, 698, A68, \dodoi{10.1051/0004-6361/202453449}

\bibitem[{Meni-Gallardo \& Pallé(2025)}]{menigallardo2025empiricaldeterminationcosmicshoreline}
Meni-Gallardo, P., \& Pallé, E. 2025, An empirical determination of the Cosmic Shoreline.
\newblock \doarXiv{2508.12865}

\bibitem[{{Miguel} {et~al.}(2011){Miguel}, {Kaltenegger}, {Fegley}, \& {Schaefer}}]{thermal_inversion_2011ApJ...742L..19M}
{Miguel}, Y., {Kaltenegger}, L., {Fegley}, B., \& {Schaefer}, L. 2011, \apjl, 742, L19, \dodoi{10.1088/2041-8205/742/2/L19}

\bibitem[{Miles {et~al.}(2023)Miles, Biller, Patapis, Worthen, Rickman, Hoch, Skemer, Perrin, Whiteford, Chen, Sargent, Mukherjee, Morley, Moran, Bonnefoy, Petrus, Carter, Choquet, Hinkley, Ward-Duong, Leisenring, Millar-Blanchaer, Pueyo, Ray, Sallum, Stapelfeldt, Stone, Wang, Absil, Balmer, Boccaletti, Bonavita, Booth, Bowler, Chauvin, Christiaens, Currie, Danielski, Fortney, Girard, Grady, Greenbaum, Henning, Hines, Janson, Kalas, Kammerer, Kennedy, Kenworthy, Kervella, Lagage, Lew, Liu, Macintosh, Marino, Marley, Marois, Matthews, Matthews, Mawet, McElwain, Metchev, Meyer, Molliere, Pantin, Quirrenbach, Rebollido, Ren, Schneider, Vasist, Wyatt, Zhou, Briesemeister, Bryan, Calissendorff, Cantalloube, Cugno, De~Furio, Dupuy, Factor, Faherty, Fitzgerald, Franson, Gonzales, Hood, Howe, Kraus, Kuzuhara, Lagrange, Lawson, Lazzoni, Liu, Llop-Sayson, Lloyd, Martinez, Mazoyer, Quanz, Redai, Samland, Schlieder, Tamura, Tan, Uyama, Vigan, Vos, Wagner, Wolff, Ygouf, Zhang, Zhang, \&
  Zhang}]{Miles_Biller_Patapis_Worthen_Rickman_Hoch_Skemer_Perrin_Whiteford_Chen_etal._2023}
Miles, B.~E., Biller, B.~A., Patapis, P., {et~al.} 2023, The Astrophysical Journal Letters, 946, L6, \dodoi{10.3847/2041-8213/acb04a}

\bibitem[{Monaghan {et~al.}(2025)Monaghan, Roy, Benneke, Crossfield, Coulombe, Piaulet-Ghorayeb, Kreidberg, Dressing, Kane, Dragomir, Werner, Parmentier, Christiansen, Morales, Berardo, \& Gorjian}]{Monaghan_Roy_Benneke_Crossfield_Coulombe_Piaulet-Ghorayeb_Kreidberg_Dressing_Kane_Dragomir_etal._2025}
Monaghan, C., Roy, P.-A., Benneke, B., {et~al.} 2025, The Astronomical Journal, 169, 239, \dodoi{10.3847/1538-3881/adbe75}

\bibitem[{Morris {et~al.}(2020)Morris, Bobra, Agol, Lee, \& Hawley}]{Morris}
Morris, B.~M., Bobra, M.~G., Agol, E., Lee, Y.~J., \& Hawley, S.~L. 2020, Monthly Notices of the Royal Astronomical Society, 493, 5489–5498, \dodoi{10.1093/mnras/staa618}

\bibitem[{Morrison {et~al.}(2023)Morrison, Dicken, Argyriou, Ressler, Gordon, Regan, Cracraft, Rieke, Engesser, Alberts, Alvarez-Marquez, Colbert, Fox, Gasman, Law, Garcia~Marin, Gáspár, Guillard, Kendrew, Labiano, Laine, Noriega-Crespo, Shivaei, \& Sloan}]{miri_artifacts_Morrison_2023}
Morrison, J.~E., Dicken, D., Argyriou, I., {et~al.} 2023, Publications of the Astronomical Society of the Pacific, 135, 075004, \dodoi{10.1088/1538-3873/acdea6}

\bibitem[{Nakajima \& Sorahana(2016)}]{Nakajima_Sorahana_2016}
Nakajima, T., \& Sorahana, S. 2016, The Astrophysical Journal, 830, 159, \dodoi{10.3847/0004-637X/830/2/159}

\bibitem[{Nixon {et~al.}(2024)Nixon, Welbanks, McGill, \& Kempton}]{Nixon_Welbanks_McGill_Kempton_2024}
Nixon, M.~C., Welbanks, L., McGill, P., \& Kempton, E. M.-R. 2024, The Astrophysical Journal, 966, 156, \dodoi{10.3847/1538-4357/ad354e}

\bibitem[{{Oddo} {et~al.}(2023){Oddo}, {Dragomir}, {Brandeker}, {Osborn}, {Collins}, {Stassun}, {Astudillo-Defru}, {Bieryla}, {Howell}, {Ciardi}, {Quinn}, {Almenara}, {Brice{\~n}o}, {Collins}, {Col{\'o}n}, {Conti}, {Crouzet}, {Furlan}, {Gan}, {Gnilka}, {Goeke}, {Gonzales}, {Harris}, {Jenkins}, {Jensen}, {Latham}, {Law}, {Lund}, {Mann}, {Massey}, {Murgas}, {Ricker}, {Relles}, {Rowden}, {Schwarz}, {Schlieder}, {Shporer}, {Seager}, {Srdoc}, {Torres}, {Twicken}, {Vanderspek}, {Winn}, \& {Ziegler}}]{OddoLTTparams}
{Oddo}, D., {Dragomir}, D., {Brandeker}, A., {et~al.} 2023, \aj, 165, 134, \dodoi{10.3847/1538-3881/acb4e3}

\bibitem[{Osborn {et~al.}(2021)Osborn, Armstrong, Cale, Brahm, Wittenmyer, Dai, Crossfield, Bryant, Adibekyan, Cloutier, Collins, Delgado Mena, Fridlund, Hellier, Howell, King, Lillo-Box, Otegi, Sousa, Stassun, Matthews, Ziegler, Ricker, Vanderspek, Latham, Seager, Winn, Jenkins, Acton, Addison, Anderson, Ballard, Barrado, Barros, Batalha, Bayliss, Barclay, Benneke, Berberian, Bouchy, Bowler, Briceño, Burke, Burleigh, Casewell, Ciardi, Collins, Cooke, Demangeon, Díaz, Dorn, Dragomir, Dressing, Dumusque, Espinoza, Figueira, Fulton, Furlan, Gaidos, Geneser, Gill, Goad, Gonzales, Gorjian, Günther, Helled, Henderson, Henning, Hogan, Hojjatpanah, Horner, Howard, Hoyer, Huber, Isaacson, Jenkins, Jensen, Jordán, Kane, Kidwell, Kielkopf, Law, Lendl, Lund, Matson, Mann, McCormac, Mengel, Morales, Nielsen, Okumura, Osborn, Petigura, Plavchan, Pollacco, Quintana, Raynard, Robertson, Rose, Roy, Reefe, Santerne, Santos, Sarkis, Schlieder, Schwarz, Scott, Shporer, Smith, Stibbard, Stockdale, Strøm, Twicken, Tan,
  Tanner, Teske, Tilbrook, Tinney, Udry, Villaseñor, Vines, Wang, Weiss, West, Wheatley, Wright, Zhang, \& Zohrabi}]{Osborn_2021}
Osborn, A., Armstrong, D.~J., Cale, B., {et~al.} 2021, Monthly Notices of the Royal Astronomical Society, 507, 2782–2803, \dodoi{10.1093/mnras/stab2313}

\bibitem[{Paragas {et~al.}(2025)Paragas, Knutson, Hu, Ehlmann, Alemanno, Helbert, Maturilli, Zhang, Iyer, \& Rossman}]{Paragas_Knutson_Hu_Ehlmann_Alemanno_Helbert_Maturilli_Zhang_Iyer_Rossman_2025}
Paragas, K., Knutson, H.~A., Hu, R., {et~al.} 2025, The Astrophysical Journal, 981, 130, \dodoi{10.3847/1538-4357/ada9eb}

\bibitem[{Pass {et~al.}(2025)Pass, Charbonneau, \& Vanderburg}]{Pass_Charbonneau_Vanderburg_2025}
Pass, E.~K., Charbonneau, D., \& Vanderburg, A. 2025, \dodoi{10.48550/arXiv.2504.01182}

\bibitem[{{Pass} {et~al.}(2023{\natexlab{a}}){Pass}, {Winters}, {Charbonneau}, {Balkanski}, {Lewis}, {Lally}, {Bean}, {Cloutier}, \& {Eastman}}]{2023AJ....166..171P}
{Pass}, E.~K., {Winters}, J.~G., {Charbonneau}, D., {et~al.} 2023{\natexlab{a}}, \aj, 166, 171, \dodoi{10.3847/1538-3881/acf561}

\bibitem[{{Pass} {et~al.}(2023{\natexlab{b}}){Pass}, {Winters}, {Charbonneau}, {Balkanski}, {Lewis}, {Lally}, {Bean}, {Cloutier}, \& {Eastman}}]{PassLTTparams}
---. 2023{\natexlab{b}}, \aj, 166, 171, \dodoi{10.3847/1538-3881/acf561}

\bibitem[{Patel {et~al.}(2024)Patel, Brandeker, Kitzmann, de~la Roche, Bello-Arufe, Heng, Valdés, Persson, Zhang, Demory, Bourrier, Deline, Ehrenreich, Fridlund, Hu, Lendl, Oza, Alibert, \& Hooton}]{Patel_Brandeker_Kitzmann_de_la_Roche_Bello-Arufe_Heng_Valdes_Persson_Zhang_Demory_et_al._2024}
Patel, J.~A., Brandeker, A., Kitzmann, D., {et~al.} 2024.
\newblock \url{http://arxiv.org/abs/2407.12898}

\bibitem[{{Perryman}(2018)}]{2018exha.book.....P}
{Perryman}, M. 2018, {The Exoplanet Handbook}

\bibitem[{{Poggiali} {et~al.}(2021){Poggiali}, {Brucato}, {Dotto}, {Ieva}, {Barucci}, \& {Pajola}}]{2021Icar..35414040P}
{Poggiali}, G., {Brucato}, J.~R., {Dotto}, E., {et~al.} 2021, \icarus, 354, 114040, \dodoi{10.1016/j.icarus.2020.114040}

\bibitem[{Rackham {et~al.}(2018)Rackham, Apai, \& Giampapa}]{Rackham_Apai_Giampapa_2018}
Rackham, B.~V., Apai, D., \& Giampapa, M.~S. 2018, The Astrophysical Journal, 853, 122, \dodoi{10.3847/1538-4357/aaa08c}

\bibitem[{Rackham {et~al.}(2019)Rackham, Apai, \& Giampapa}]{Rackham_Apai_Giampapa_2019}
---. 2019, The Astronomical Journal, 157, 96, \dodoi{10.3847/1538-3881/aaf892}

\bibitem[{Rackham \& de~Wit(2024)}]{Rackham_deWit_2024}
Rackham, B.~V., \& de~Wit, J. 2024, The Astronomical Journal, 168, 82, \dodoi{10.3847/1538-3881/ad5833}

\bibitem[{Rackham {et~al.}(2023)Rackham, Espinoza, Berdyugina, Korhonen, MacDonald, Montet, Morris, Oshagh, Shapiro, Unruh, Quintana, Zellem, Apai, Barclay, Barstow, Bruno, Carone, Casewell, Cegla, Criscuoli, Fischer, Fournier, Giampapa, Giles, Iyer, Kopp, Kostogryz, Krivova, Mallonn, McGruder, Molaverdikhani, Newton, Panja, Peacock, Reardon, Roettenbacher, Scandariato, Solanki, Stassun, Steiner, Stevenson, Tregloan-Reed, Valio, Wedemeyer, Welbanks, Yu, Alam, Davenport, Deming, Dong, Ducrot, Fisher, Gilbert, Kostov, López-Morales, Line, Močnik, Mullally, Paudel, Ribas, \& Valenti}]{Rackham_Espinoza_Berdyugina_Korhonen_MacDonald_Montet_Morris_Oshagh_Shapiro_Unruh_etal._2023}
Rackham, B.~V., Espinoza, N., Berdyugina, S.~V., {et~al.} 2023, 2, 148–206, \dodoi{10.1093/rasti/rzad009}

\bibitem[{Rajpurohit {et~al.}(2019)Rajpurohit, Allard, Rajpurohit, Sharma, Teixeira, Mousis, \& Kamlesh}]{Rajpurohit_Allard_Rajpurohit_Sharma_Teixeira_Mousis_Kamlesh_2019}
Rajpurohit, A.~S., Allard, F., Rajpurohit, S., {et~al.} 2019, Astronomy \& Astrophysics, 622, C1, \dodoi{10.1051/0004-6361/201833500e}

\bibitem[{Rasmussen {et~al.}(2023)Rasmussen, Currie, Hagee, van Buchem, Malik, Savel, Brogi, Rauscher, Meadows, Mansfield, Kempton, Desert, Wardenier, Pino, Line, Parmentier, Seifahrt, Kasper, Brady, \& Bean}]{rasmussen2023nondetectionironhighresolutionemission}
Rasmussen, K.~C., Currie, M.~H., Hagee, C., {et~al.} 2023, A Non-Detection of Iron in the First High-Resolution Emission Study of the Lava Planet 55 Cnc e.
\newblock \doarXiv{2308.10378}

\bibitem[{Redfield {et~al.}(2024)Redfield, Batalha, Benneke, Biller, Espinoza, France, Konopacky, Kreidberg, Rauscher, \& Sing}]{Redfield_Batalha_Benneke_Biller_Espinoza_France_Konopacky_Kreidberg_Rauscher_Sing_2024}
Redfield, S., Batalha, N., Benneke, B., {et~al.} 2024, \dodoi{10.48550/arXiv.2404.02932}

\bibitem[{Rochon {et~al.}(2025)Rochon, Étienne Artigau, Weisserman, Dang, Doyon, Cadieux, \& Cloutier}]{rochon2025reanalysiseclipseslhs1140}
Rochon, A., Étienne Artigau, Weisserman, D., {et~al.} 2025, Reanalysis of the eclipses of LHS 1140 c: No evidence of an atmosphere and implications for the internal structure of the planet.
\newblock \doarXiv{2510.11397}

\bibitem[{{Rodrigo} \& {Solano}(2020)}]{2020sea..confE.182R}
{Rodrigo}, C., \& {Solano}, E. 2020, in XIV.0 Scientific Meeting (virtual) of the Spanish Astronomical Society, 182

\bibitem[{{Rodrigo} {et~al.}(2012){Rodrigo}, {Solano}, \& {Bayo}}]{2012ivoa.rept.1015R}
{Rodrigo}, C., {Solano}, E., \& {Bayo}, A. 2012, {SVO Filter Profile Service Version 1.0}, IVOA Working Draft 15 October 2012, \dodoi{10.5479/ADS/bib/2012ivoa.rept.1015R}

\bibitem[{{Rodrigo} {et~al.}(2024){Rodrigo}, {Cruz}, {Aguilar}, {Aller}, {Solano}, {G{\'a}lvez-Ortiz}, {Jim{\'e}nez-Esteban}, {Mas-Buitrago}, {Bayo}, {Cort{\'e}s-Contreras}, {Murillo-Ojeda}, {Bonoli}, {Cenarro}, {Dupke}, {L{\'o}pez-Sanjuan}, {Mar{\'\i}n-Franch}, {de Oliveira}, {Moles}, {Taylor}, {Varela}, \& {Rami{\'o}}}]{2024A&A...689A..93R}
{Rodrigo}, C., {Cruz}, P., {Aguilar}, J.~F., {et~al.} 2024, \aap, 689, A93, \dodoi{10.1051/0004-6361/202449998}

\bibitem[{{Sarmento} {et~al.}(2021){Sarmento}, {Rojas-Ayala}, {Delgado Mena}, \& {Blanco-Cuaresma}}]{SARMENTO2021A&A...649A.147S}
{Sarmento}, P., {Rojas-Ayala}, B., {Delgado Mena}, E., \& {Blanco-Cuaresma}, S. 2021, \aap, 649, A147, \dodoi{10.1051/0004-6361/202039703}

\bibitem[{Seager(2010)}]{48d96982-3471-3bab-9b6f-f8b21bafd6fb}
Seager, S. 2010, Exoplanet Atmospheres: Physical Processes (Princeton University Press).
\newblock \url{http://www.jstor.org/stable/j.ctvcm4gvv}

\bibitem[{{Seager} \& {Deming}(2009)}]{2009ApJ...703.1884S}
{Seager}, S., \& {Deming}, D. 2009, \apj, 703, 1884, \dodoi{10.1088/0004-637X/703/2/1884}

\bibitem[{{Shields} {et~al.}(2016){Shields}, {Ballard}, \& {Johnson}}]{induction_heating_2016PhR...663....1S}
{Shields}, A.~L., {Ballard}, S., \& {Johnson}, J.~A. 2016, \physrep, 663, 1, \dodoi{10.1016/j.physrep.2016.10.003}

\bibitem[{{Spaargaren} {et~al.}(2020){Spaargaren}, {Ballmer}, {Bower}, {Dorn}, \& {Tackley}}]{2020A&A...643A..44S}
{Spaargaren}, R.~J., {Ballmer}, M.~D., {Bower}, D.~J., {Dorn}, C., \& {Tackley}, P.~J. 2020, \aap, 643, A44, \dodoi{10.1051/0004-6361/202037632}

\bibitem[{Tamburo {et~al.}(2018)Tamburo, Mandell, Deming, \& Garhart}]{Tamburo_2018}
Tamburo, P., Mandell, A., Deming, D., \& Garhart, E. 2018, The Astronomical Journal, 155, 221, \dodoi{10.3847/1538-3881/aabd84}

\bibitem[{Tenthoff {et~al.}(2024)Tenthoff, Wohlfarth, Wöhler, Zieba, \& Kreidberg}]{Tenthoff_Wohlfarth_Wohler_Zieba_Kreidberg_2024}
Tenthoff, M., Wohlfarth, K., Wöhler, C., Zieba, S., \& Kreidberg, L. 2024, Reflectance and Emission Modelling of Airless Exoplanets No. EPSC2024-649, \dodoi{10.5194/epsc2024-649}

\bibitem[{Teske {et~al.}(2025)Teske, Wallack, Piette, Dang, Lichtenberg, Plotnykov, Pierrehumbert, Postolec, Boucher, McGinty, Peng, Valencia, \& Hammond}]{teske2025thickvolatileatmosphereultrahot}
Teske, J.~K., Wallack, N.~L., Piette, A. A.~A., {et~al.} 2025, A Thick Volatile Atmosphere on the Ultra-Hot Super-Earth TOI-561 b.
\newblock \doarXiv{2509.17231}

\bibitem[{{Treiman} {et~al.}(2021){Treiman}, {Filiberto}, \& {Vander Kaaden}}]{2021PSJ.....2...43T}
{Treiman}, A.~H., {Filiberto}, J., \& {Vander Kaaden}, K.~E. 2021, Planetary Science Journal, 2, 43, \dodoi{10.3847/PSJ/abd546}

\bibitem[{Tsuji(2002)}]{Tsuji_2002}
Tsuji, T. 2002, The Astrophysical Journal, 575, 264, \dodoi{10.1086/341262}

\bibitem[{{Tsuji} \& {Nakajima}(2014)}]{2014PASJ...66...98T}
{Tsuji}, T., \& {Nakajima}, T. 2014, \pasj, 66, 98, \dodoi{10.1093/pasj/psu078}

\bibitem[{{Tsuji} \& {Nakajima}(2016)}]{2016PASJ...68...13T}
---. 2016, \pasj, 68, 13, \dodoi{10.1093/pasj/psv119}

\bibitem[{{Tsuji} {et~al.}(2015){Tsuji}, {Nakajima}, \& {Takeda}}]{2015PASJ...67...26T}
{Tsuji}, T., {Nakajima}, T., \& {Takeda}, Y. 2015, \pasj, 67, 26, \dodoi{10.1093/pasj/psu160}

\bibitem[{Vanderspek {et~al.}(2019)Vanderspek, Huang, Vanderburg, Ricker, Latham, Seager, Winn, Jenkins, Burt, Dittmann, Newton, Quinn, Shporer, Charbonneau, Irwin, Ment, Winters, Collins, Evans, Gan, Hart, Jensen, Kielkopf, Mao, Waalkes, Bouchy, Marmier, Nielsen, Ottoni, Pepe, Ségransan, Udry, Henry, Paredes, James, Hinojosa, Silverstein, Palle, Berta-Thompson, Crossfield, Davies, Dragomir, Fausnaugh, Glidden, Pepper, Morgan, Rose, Twicken, Villaseñor, Yu, Bakos, Bean, Buchhave, Christensen-Dalsgaard, Christiansen, Ciardi, Clampin, Lee, Deming, Doty, Jernigan, Kaltenegger, Lissauer, McCullough, Narita, Paegert, Pal, Rinehart, Sasselov, Sato, Sozzetti, Stassun, \& Torres}]{Vanderspek_2019}
Vanderspek, R., Huang, C.~X., Vanderburg, A., {et~al.} 2019, The Astrophysical Journal Letters, 871, L24, \dodoi{10.3847/2041-8213/aafb7a}

\bibitem[{{Vidotto} {et~al.}(2014){Vidotto}, {Gregory}, {Jardine}, {Donati}, {Petit}, {Morin}, {Folsom}, {Bouvier}, {Cameron}, {Hussain}, {Marsden}, {Waite}, {Fares}, {Jeffers}, \& {do Nascimento}}]{induction_heating_2014MNRAS.441.2361V}
{Vidotto}, A.~A., {Gregory}, S.~G., {Jardine}, M., {et~al.} 2014, \mnras, 441, 2361, \dodoi{10.1093/mnras/stu728}

\bibitem[{Virtanen {et~al.}(2020)Virtanen, Gommers, Oliphant, Haberland, Reddy, Cournapeau, Burovski, Peterson, Weckesser, Bright, {van der Walt}, Brett, Wilson, Millman, Mayorov, Nelson, Jones, Kern, Larson, Carey, Polat, Feng, Moore, {VanderPlas}, Laxalde, Perktold, Cimrman, Henriksen, Quintero, Harris, Archibald, Ribeiro, Pedregosa, {van Mulbregt}, \& {SciPy 1.0 Contributors}}]{2020SciPy-NMeth}
Virtanen, P., Gommers, R., Oliphant, T.~E., {et~al.} 2020, Nature Methods, 17, 261, \dodoi{10.1038/s41592-019-0686-2}

\bibitem[{Wachiraphan {et~al.}(2024)Wachiraphan, Berta-Thompson, Diamond-Lowe, Winters, Murray, Zhang, Xue, Morley, Rosario-Franco, \& Duvvuri}]{wachiraphan2024thermalemissionspectrumnearby}
Wachiraphan, P., Berta-Thompson, Z.~K., Diamond-Lowe, H., {et~al.} 2024, The Thermal Emission Spectrum of the Nearby Rocky Exoplanet LTT 1445A b from JWST MIRI/LRS.
\newblock \doarXiv{2410.10987}

\bibitem[{Wohlfarth {et~al.}(2023)Wohlfarth, Wöhler, Hiesinger, \& Helbert}]{Wohlfarth_Wohler_Hiesinger_Helbert_2023}
Wohlfarth, K., Wöhler, C., Hiesinger, H., \& Helbert, J. 2023, Astronomy \& Astrophysics, 674, A69, \dodoi{10.1051/0004-6361/202245343}

\bibitem[{Xue {et~al.}(2024)Xue, Bean, Zhang, Mahajan, Ih, Eastman, Lunine, Mansfield, Coy, Kempton, Koll, \& Kite}]{Xue_Bean_Zhang_Mahajan_Ih_Eastman_Lunine_Mansfield_Coy_Kempton_et_al._2024}
Xue, Q., Bean, J.~L., Zhang, M., {et~al.} 2024.
\newblock \url{http://arxiv.org/abs/2408.13340}

\bibitem[{Xue {et~al.}(2025{\natexlab{a}})Xue, Zhang, Coy, Brady, Ji, Bean, Radica, Seifahrt, Sturmer, Luque, Basant, Brown, Das, Kasper, Piaulet-Ghorayeb, Kempton, \& Kite}]{Xue_Zhang_Coy_Brady_Ji_Bean_Radica_Seifahrt_Sturmer_Luque_etal._2025}
Xue, Q., Zhang, M., Coy, B.~P., {et~al.} 2025{\natexlab{a}}, \dodoi{10.48550/arXiv.2508.12516}

\bibitem[{Xue {et~al.}(2025{\natexlab{b}})Xue, Zhang, Coy, Brady, Ji, Bean, Radica, Seifahrt, Sturmer, Luque, Basant, Brown, Das, Kasper, Piaulet-Ghorayeb, Kempton, \& Kite}]{xue2025jwstrockyworldsddt}
---. 2025{\natexlab{b}}, The JWST Rocky Worlds DDT Program reveals GJ 3929b to likely be a bare rock.
\newblock \doarXiv{2508.12516}

\bibitem[{Zahnle \& Catling(2017)}]{Zahnle_2017}
Zahnle, K.~J., \& Catling, D.~C. 2017, The Astrophysical Journal, 843, 122, \dodoi{10.3847/1538-4357/aa7846}

\bibitem[{Zaini {et~al.}(2012)Zaini, van~der Meer, \& van~der Werff}]{Zaini_van}
Zaini, N., van~der Meer, F., \& van~der Werff, H. 2012, Remote Sensing, 4, 987–1003, \dodoi{10.3390/rs4040987}

\bibitem[{Zhang {et~al.}(2024)Zhang, Hu, Inglis, Dai, Bean, Knutson, Lam, Goffo, \& Gandolfi}]{Zhang_Hu_Inglis_Dai_Bean_Knutson_Lam_Goffo_Gandolfi_2024}
Zhang, M., Hu, R., Inglis, J., {et~al.} 2024, The Astrophysical Journal Letters, 961, L44, \dodoi{10.3847/2041-8213/ad1a07}

\bibitem[{Zhuang {et~al.}(2023)Zhuang, Zhang, Ma, Jiang, Yang, Milliken, \& Hsu}]{Zhuang_Zhang_Ma_Jiang_Yang_Milliken_Hsu_2023}
Zhuang, Y., Zhang, H., Ma, P., {et~al.} 2023, Icarus, 391, 115346, \dodoi{10.1016/j.icarus.2022.115346}

\bibitem[{Zieba {et~al.}(2022)Zieba, Zilinskas, Kreidberg, Nguyen, Miguel, Cowan, Pierrehumbert, Carone, Dang, Hammond, Louden, Lupu, Malavolta, \& Stevenson}]{Zieba_Zilinskas_Kreidberg_Nguyen_Miguel_Cowan_Pierrehumbert_Carone_Dang_Hammond_et_al._2022}
Zieba, S., Zilinskas, M., Kreidberg, L., {et~al.} 2022, Astronomy \& Astrophysics, 664, A79, \dodoi{10.1051/0004-6361/202142912}

\bibitem[{Zieba {et~al.}(2023)Zieba, Kreidberg, Ducrot, Gillon, Morley, Schaefer, Tamburo, Koll, Lyu, Acuña, Agol, Iyer, Hu, Lincowski, Meadows, Selsis, Bolmont, Mandell, \& Suissa}]{Zieba_Kreidberg_Ducrot_Gillon_Morley_Schaefer_Tamburo_Koll_Lyu_Acuna_et_al._2023}
Zieba, S., Kreidberg, L., Ducrot, E., {et~al.} 2023, Nature, 620, 746–749, \dodoi{10.1038/s41586-023-06232-z}

\bibitem[{Zilinskas {et~al.}(2022)Zilinskas, van Buchem, Miguel, Louca, Lupu, Zieba, \& van Westrenen}]{thermal_inversion_Zilinskas_2022}
Zilinskas, M., van Buchem, C. P.~A., Miguel, Y., {et~al.} 2022, Astronomy \& Astrophysics, 661, A126, \dodoi{10.1051/0004-6361/202142984}

\end{thebibliography}



\end{document}